\documentclass{iopjournal}
\usepackage[utf8]{inputenc}
\usepackage[english]{babel}
\usepackage[T1]{fontenc}
\usepackage{amsmath}
\usepackage{amsthm}
\usepackage{hyperref}
\usepackage[numbers,sort&compress]{natbib}
\usepackage{subfigure}
\usepackage[capitalize]{cleveref}
\usepackage{amsmath,amssymb}
\usepackage{multirow}
\usepackage{bm}
\usepackage{physics}
\usepackage{lipsum}
\usepackage{pifont}
\usepackage[acronym]{glossaries}
\glsdisablehyper
\usepackage{tikz}
\usepackage{titlesec} 
\usepackage{nicefrac}

\newcommand{\cmark}{\ding{51}}%
\newcommand{\xmark}{\ding{55}}%

\newtheorem{lemma}{Lemma}
\newtheorem{theorem}{Theorem}
\newtheorem*{theorem*}{Theorem}
\newtheorem{definition}{Definition}

\crefname{section}{section}{sections}
\Crefname{section}{Section}{Sections}
\Crefname{appsec}{Appendix}{Appendices}
\crefname{appsubsec}{appendix}{appendices}
\Crefname{appsubsec}{Appendix}{Appendices}
\crefname{figure}{figure}{figures}
\Crefname{figure}{Figure}{Figures}
\crefname{table}{table}{tables}
\Crefname{table}{Table}{Tables}
\crefname{equation}{equation}{equations}
\Crefname{equation}{Equation}{Equations}

\begin{document}

\newacronym{qec}{QEC}{quantum error correction}
\newacronym{varqec}{VarQEC}{variational quantum error correction}
\newacronym{vqc}{VQC}{variational quantum circuit}
\newacronym{cptp}{CPTP}{completely positive trace preserving}
\newacronym{ml}{ML}{machine learning}
\newacronym{rl}{RL}{reinforcement learning}
\newacronym{ai}{AI}{artificial intelligence}
\newacronym{eftqc}{eFTQC}{early fault-tolerant quantum computing}
\newacronym{rea}{REA}{randomized entangling ansatz}
\newacronym{qpu}{QPU}{quantum processing unit}
\newacronym{qldpc}{qLDPC}{quantum low-density parity-check}


\title{Learning Encodings by Maximizing State Distinguishability: Variational Quantum Error Correction}

\author{Nico Meyer$^{1,2,*}$\orcid{0000-0002-5463-5437}, Christopher Mutschler$^1$\orcid{0000-0001-8108-0230}, Andreas Maier$^2$\orcid{0000-0002-9550-5284} and Daniel D. Scherer$^1$\orcid{0000-0003-0355-4140}}

\affil{$^1$Fraunhofer IIS, Fraunhofer Institute for Integrated Circuits IIS, Nuremberg, Germany}

\affil{$^2$Pattern Recognition Lab, Friedrich-Alexander-University Erlangen-Nuremberg, Erlangen, Germany}

\affil{$^*$Author to whom any correspondence should be addressed.}

\email{nico.meyer@iis.fraunhofer.de}


\begin{abstract}
    Quantum error correction is crucial for protecting quantum information against decoherence. Traditional codes like the surface code require substantial overhead, making them impractical for near-term, early fault-tolerant devices.
    We propose a novel objective function for tailoring error correction codes to specific noise structures by maximizing the distinguishability between quantum states after a noise channel, ensuring efficient recovery operations. We formalize this concept with the distinguishability loss function, serving as a machine learning objective to discover resource-efficient encoding circuits optimized for given noise characteristics. We implement this methodology using variational techniques, termed variational quantum error correction (VarQEC).
    Our approach yields codes with desirable theoretical and practical properties and surpasses standard codes of comparable size under structured noise. We also provide proof-of-concept demonstrations on IBM and IQM hardware devices, highlighting the practical relevance of our procedure.
\end{abstract}

\section{Introduction}

Quantum computing promises to solve complex problems intractable for classical computers, such as cryptographic analysis~\cite{shor1999polynomial}, simulation~\cite{georgescu2014quantum,daley2022practical}, optimization~\cite{abbas2024challenges}, and machine learning~\cite{biamonte2017quantum,meyer2022survey}. However, maintaining quantum state coherence amidst noise~\cite{unruh1995maintaining} is a significant challenge. \Gls{qec} is essential to protect quantum information from errors due to environmental interactions and imperfect quantum operations~\cite{shor1995scheme,steane1996error}.

Traditional \gls{qec} codes aim to detect and correct arbitrary quantum errors up to a certain weight, like e.g.\ the surface code~\cite{fowler2012surface}. Recent experiments demonstrate that noise suppression increases exponentially with code size~\cite{acharya2024quantum}, confirming theoretical predictions~\cite{shor1996fault}. However, these codes require substantial overhead in additional qubits and complex recovery operations, posing a barrier for near-term devices and the upcoming \gls{eftqc} era~\cite{katabarwa2024early}, where resources are limited and error rates are high.

Thus, reducing resource requirements is crucial, e.g., by tailoring codes to specific noise structures of hardware platforms~\cite{aliferis2008fault,tuckett2018ultrahigh}. Alternatively, one can relax perfect correction conditions and use approximate \gls{qec} codes with less overhead~\cite{leung1997approximate,schumacher2002approximate}. Recent advancements explore integrating \gls{ai} techniques into quantum computing and \gls{qec}~\cite{alexeev2024artificial}. While much literature focuses on \gls{ml}-based decoders for error correction~\cite{torlai2017neural,bausch2020quantum}, there is growing interest in leveraging \gls{ai} to design better quantum codes and encoding strategies~\cite{fosel2018reinforcement,olle2024simultaneous}.

\begin{figure}[t]
    \centering
    \includegraphics[width=0.55\textwidth]{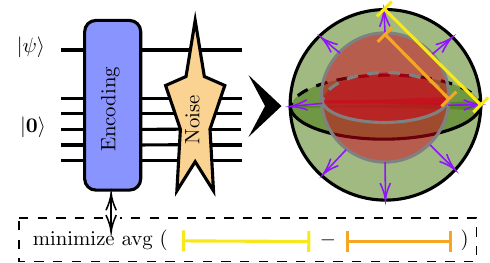}
    \caption{\label{fig:header}We propose a procedure that tailors quantum error correction encodings to arbitrary noise channels.
    In our setup, a pure state $\ket{\psi}$ is encoded into a larger ancilla system, and a noise channel then acts on the involved wires. We illustrate this with pairs of states $\ket{\psi_0} = \ket{0}$ and $\ket{\psi_1}=\ket{+}$, which in the noise-free case can be visualized on the top and right side of a unit sphere--the Bloch sphere in the single-qubit setup. Under noise, e.g., a combination of depolarizing and amplitude damping processes, the sphere shrinks and the states are drawn inwards. We quantify this by computing the trace distance between the original states (yellow line) and subtracting the trace distance between their noisy counterparts (orange line). To optimize the encoding for arbitrary states, we minimize the average distinguishability loss over pairs of states drawn from a two-design, thereby encouraging minimal sphere shrinkage (shown by blue arrows) and enabling high-fidelity recovery. For proof of principle, we implement this procedure as a variational quantum algorithm using a parameterized encoding ansatz, calling it variational quantum error correction (VarQEC).}
\end{figure}

\medskip
\noindent
\textbf{Main Contributions of this Work.}
We claim and summarize the main contributions of our paper as follows:
\renewcommand{\labelenumi}{\Roman{enumi}.}
\begin{enumerate}
    \item We introduce the novel \emph{distinguishability loss} function for learning approximate \gls{qec} codes. This loss aims to maximize the trace distance between pairs of quantum states after passing through a noise channel, ensuring effective recovery. We define worst-case and average-case variants and present a fast approximation using a two-design ensemble for efficient computation and optimization.
    \item We derive theoretical properties of the distinguishability loss, specifically proving its relationship to the fidelity after recovery. This demonstrates that minimizing the distinguishability loss leads to improved error correction performance. Furthermore, we connect our method to established metrics in \gls{qec} like the code distance.
    \item We incorporate our approach into a variational quantum algorithm, termed \glsentrylong{varqec}. This also integrates elements from the QVECTOR approach~\cite{johnson2017qvector} for learning recovery operations. We provide an end-to-end implementation to encourage reproducibility and future extensions.
    \item We conduct extensive empirical evaluations to demonstrate the efficiency and effectiveness of the \gls{varqec} procedure across various noise models and system sizes. Our evaluations analyze the stability and noise resilience of the resulting codes. We show that our method achieves lower information loss and higher recovery fidelity than traditional small stabilizer codes under structured noise, while matching performance in the absence of structure.
    \item We deploy the learned encoding operations on actual IBM and IQM quantum hardware. These experiments demonstrate consistent improvements over purely physical realizations and provide empirical evidence of the practical applicability of \gls{varqec} codes, highlighting their promising noise suppression capabilities as system size increases.
\end{enumerate}
In summary, our work advances the field of quantum error correction by introducing a novel, adaptable approach for learning error-correcting codes tailored to specific noise channels. Leveraging the distinguishability loss within a variational framework, we reduce the overhead associated with traditional \gls{qec} codes. Our method is particularly well-suited for the era of \gls{eftqc}~\cite{katabarwa2024early}, where resource efficiency and adaptability to hardware constraints are critical.

\medskip
\noindent
The remainder of this paper is organized as follows: In \cref{sec:related}, we position our work within the literature on \gls{ai}-enhanced \gls{qec}. \Cref{sec:background} summarizes the background on noise channels, distance measures for quantum states, and \gls{qec}. In \cref{sec:method}, we formally introduce the distinguishability loss, formulate the \gls{varqec} algorithm, and derive theoretical properties of the learned codes. The empirical results in \cref{sec:empirical} demonstrate the effectiveness of the \gls{varqec} codes compared to standard \gls{qec} codes. In \cref{sec:hardware}, we deploy code instances on IBM and IQM hardware systems to underline the practical relevance of our procedure. We identify current limitations of the \gls{varqec} procedure and outline future improvements and extensions in \cref{sec:future_work}. Finally, in \cref{sec:summary}, we provide an overview and discuss our results in the context of \gls{qec} and \gls{eftqc}.

\section{\label{sec:related}Related Work}

\Gls{ai} is drawing increasing interest in the domain of quantum computing in general~\cite{alexeev2024artificial}, and \gls{qec} in particular~\cite{wang2024artificial}. Much of the existing literature is, however, focused on developing \gls{ml}-based decoders~\cite{torlai2017neural,bausch2024learning}. In the following, we briefly summarize the existing literature on \gls{ai} for \gls{qec} beyond decoders, and highlight the connection to and delimitation from our work.

\medskip\noindent
\textbf{Classical Machine Learning for \gls{qec}.}
The integration of \gls{ai} into \gls{qec} began around 2018 with the pioneering work of F\"osel et al.~\cite{fosel2018reinforcement}, which employed \gls{rl}. In this approach, a reward function was constructed from several components, including the \emph{recoverable quantum information}, \emph{protection reward}, and \emph{recovery reward}. This methodology was used to train encoding and recovery circuits on systems with up to four physical qubits and one logical qubit, primarily focusing on bit-flip noise. This notion of recoverable quantum information is related to our proposed distinguishability loss, both being based on the trace distance. However, we argue in \cref{subsec:d_loss} that only the latter constitutes a loss function that is efficient to evaluate for complex noise structures. Furthermore, while the recoverable quantum information was restricted to stabilizer codes, our formulation extends to non-additive codes.
   
A modified \gls{rl} routine for \gls{qec} has recently been proposed, formulating reward functions based on the Knill-Laflamme conditions~\cite{olle2024simultaneous,olle2025scaling}. These works show that noise-aware RL agents can simultaneously discover codes and encoders tailored to a prescribed noise model and hardware connectivity, and scale this discovery process to larger Clifford circuits. Aspects of fault tolerance have also been incorporated by learning flag gadgets~\cite{zen2024quantum}. \gls{rl} has further been studied for modifying and combining existing codes~\cite{nautrup2019optimizing,mauron2024optimization,su2025discovery,yanay2026learning}, autonomous QEC~\cite{zeng2023approximate,yin2025discovering,park2025enhancing}, and the discovery of gadgets for low-density parity-check (qLDPC) codes~\cite{he2025discovering,doherty2026fast}. Additionally, \gls{rl} has been applied to the discovery of logical gates within specific experimental setups~\cite{ivanova2024discovery}.  Compared to \gls{rl}-based code-discovery frameworks, which perform a combinatorial search over discrete Clifford circuits guided by a reward signal, our \gls{varqec} scheme uses a gradient-based optimization of a trace-distance-based loss over arbitrary circuits and provides analytic guarantees (bounds between distinguishability loss and worst-case fidelity and a distance-like parameter). In this sense, the \gls{rl} and variational approaches are complementary: RL can explore large discrete architectures and fault-tolerant gadgets, while VarQEC offers a differentiable objective with provable properties for small to medium-sized codes.

Beyond \gls{rl}, evolutionary approaches have been employed to engineer codes and circuits~\cite{tandeitnik2024evolving,webster2025engineering}, and various other \gls{ml} approaches have been proposed for QEC-related tasks~\cite{guerrero2025game,seksaria2025correcting,chengyu2026bayesian}. A particular focus is also being put on \gls{ml}-enhanced decoding using supervised and unsupervised learning techniques~\cite{bausch2020quantum,wang2023automated,zeng2025neural,liu2025noise}, including neural-network and belief-propagation decoders~\cite{torlai2017neural}. In contrast to these decoder-centric approaches, our work focuses on \emph{learning encoders} and coherent \emph{measurement-free} recovery unitary, so no syndrome extraction or classical post-processing decoder is required in our setting. In principle, however, \gls{varqec} encodings could be combined with standard or \gls{ml}-based decoders whenever a stabilizer or syndrome structure is available, but this lies beyond the scope of the present work.

\medskip\noindent
\textbf{Variational Techniques for \gls{qec}.}
Variational methods have gained attention in the context of \gls{qec} due to their flexibility and adaptability to quantum hardware. The first proposal to formulate \gls{qec} as a variational algorithm was proposed in 2017 by Johnson et al.~\cite{johnson2017qvector}, dubbed QVECTOR. In this approach, the fidelity after recovery is used as a loss function to simultaneously train the free parameters of encoding and recovery circuits. The authors demonstrate the effectiveness of the approach in low-probability amplitude and phase damping noise scenarios, using systems with up to five physical qubits and a single logical qubit. However, we observed that the concept of simultaneously learning encoding and recovery is limited to rather small code instances and simple low-probability noise structures, due to training not converging in more complex scenarios. A modification to the QVECTOR approach proposes to enhance trainability by employing a Wasserstein version of the fidelity~\cite{zoratti2023improving}. However, in our evaluations, this improvement did not manifest. In contrast to QVECTOR, our proposed technique modularizes the training of the encoding and recovery operations, allowing for more flexibility and the consideration of complex noise structures.

A related approach by Cao et al.~\cite{cao2022quantum} is also variational and learning-based, but it optimizes parameterized encoders using cost functions derived directly from the Knill–Laflamme error-correction and detection conditions on a prescribed error set. In practice, their algorithm is used primarily as a code‑discovery tool: given target code parameters or a noise channel and a hardware connectivity graph, it searches for symmetric, asymmetric, and channel‑adaptive codes (including non‑stabilizer codes). Our work is similar in spirit, in that we also variationally learn encodings tailored to a given noise model. However, the objective and training strategy are fundamentally different. We introduce a trace‑distance–based distinguishability loss over pairs of logical input states and use it to derive analytic bounds on worst‑case fidelity after recovery as well as a notion of (approximate) code distance. On top of this, we realize a concrete two‑stage pipeline in which this distinguishability‑based VarQEC step for the encoder is followed by a separate, variationally trained recovery stage. Conceptually, this shifts the focus from satisfying (approximate) Knill–Laflamme operator equalities for a prescribed error set to preserving the geometry of the logical state space under the full noise channel, i.e., keeping pairs of logical states distinguishable, which leads to different guarantees and training behavior. In this sense, the two frameworks are complementary approaches to variational quantum error correction.

Other works have investigated the use of convolutional quantum neural networks~\cite{cong2019quantum} and quantum autoencoders for \gls{qec}~\cite{locher2023quantum,rathi2024quantum,lin2025design}. In some instances, these methods function more as denoising algorithms than as active \gls{qec} strategies. Different techniques aim to generate codes by formulating the task of \gls{qec} as finding the ground state of a Hamiltonian~\cite{cao2022quantum,xu2021variational,wang2022automated}. However, these approaches assume full direct access to the error operators and stabilizers. Additionally, work has been conducted on finding logical operations for given codes via variational techniques~\cite{chen2022automated,meyer2026logical}.

\medskip\noindent
\textbf{Delimitation to Large-Scale Stabilizer Codes.}
Traditional large-scale fault-tolerant architectures are based on stabilizer codes such as surface codes~\cite{fowler2012surface} and, more recently, \gls{qldpc} codes~\cite{bravyi2024high}, which are designed to achieve high distance and favorable thresholds under generic local noise, typically with simple, translation-invariant structure. In contrast, the VarQEC codes studied in this work are in general \emph{non-additive} codes without a compact stabilizer description or regular lattice geometry; they are intentionally kept small and tailored to a specific physical noise channel. Our goal is therefore not to compete with surface or \gls{qldpc} codes in the asymptotic, large-distance regime, but to provide complementary, hardware-adapted building blocks for \gls{eftqc}.

\section{\label{sec:background}Background}

As mentioned before, quantum systems are highly susceptible to errors due to decoherence and interactions with the environment~\cite{unruh1995maintaining}. To build reliable quantum computers, it is crucial to understand how quantum information is affected by noise and how to correct for the introduced errors. In this section, we introduce the underlying concepts on noise channels, distance measures for quantum states, and error correction that go into the \gls{varqec} procedure we introduce in this paper. Supplementary details can be found in \cref{app:background_definitions}.

\subsection{Quantum Noise Channels}

Quantum noise channels are mathematical models that describe how quantum states evolve in the presence of noise and decoherence. A pure $n$-qubit quantum state is typically described using Dirac notation as $\ket{\psi} \in \mathcal{H}^{\otimes n}$, where $\mathcal{H}$ denotes the Hilbert space, and it holds $\expval{\psi|\psi} = 1$. Using the density matrix formalism, one can express the same state as $\rho = \ket{\psi} \bra{\psi}$, which is of rank one for pure states. More general \emph{mixed states} represent a statistical ensemble of pure states and can be described by a density matrix
\begin{align}
    \rho = \sum_i p_i \ket{\psi_i} \bra{\psi_i},
\end{align}
where $p \geq 0$ and $\sum_i p_i = 1$. In the case of mixed states, the rank of $\rho$ is greater than one.

Quantum operations can either be of unitary or non-unitary nature. A unitary transformation $U$ acts on a state as $U \rho U^{\dagger}$, preserving purity and ensuring reversibility of the operation. However, real-world quantum systems are open and therefore interact with the environment, leading to non-unitary processes that introduce noise and decoherence. These processes are described by \gls{cptp} maps, which can be represented in the \emph{operator sum} representation as
\begin{align}
    \mathcal{N}(\rho) = \sum_{k} E_k \rho E_{k}^{\dagger},
\end{align}
where the $E_k$ are Kraus operators satisfying $\sum_k E_{k}^{\dagger} E_k = I$, which ensures trace preservation. This description is also often referred to as \emph{Kraus representation}~\cite{kraus1971general}.

A common example of a quantum noise channel is the \emph{depolarizing channel}, which models a state deteriorating to a fully mixed state with a certain probability. Single-qubit depolarizing noise can be described as
\begin{align}
    \mathcal{N}(\rho) = (1-p) \rho + \frac{p}{3} X \rho X^{\dagger} + \frac{p}{3} Y \rho Y^{\dagger} + \frac{p}{3} Z \rho Z^{\dagger},
\end{align}
where $p$ is the depolarization probability and $X$, $Y$, and $Z$ are the Pauli operators representing bit-flip, bit-flip and phase-flip, and phase-flip errors, respectively~\cite{nielsen2010quantum}. General single-qubit Pauli noise can be described as
\begin{align}
    \label{eq:general_pauli}
    \mathcal{N}(\rho) = (1-p) \rho + p_X X \rho X^{\dagger} + p_Y Y \rho Y^{\dagger} + p_Z Z \rho Z^{\dagger},
\end{align}
with overall noise strength $p = p_X + p_Y + p_Z$. Arbitrary quantum errors in multi-qubit systems can be expanded as linear combinations of tensor products of Pauli operators. Consequently, correcting bit-flip and phase-flip errors on each qubit inherently addresses more general multi-qubit error scenarios~\cite{gottesman1997stabilizer}. In the context of stabilizer-based \gls{qec}, this can be understood as syndrome measurements effectively projecting general errors into Pauli errors~\cite{knill1997theory}.

Throughout this work, we consider several families of quantum channels acting on the $n$ physical qubits. In many practical scenarios, and in most of this paper, errors are assumed to act independently on each qubit, leading to a composite noise model
\begin{align}
    \mathcal{N}^{\otimes n} = \bigotimes_{j=1}^{n} \mathcal{N}_j,
\end{align}
where $\mathcal{N}_j$ is a single-qubit noise channel acting on qubit $j$. For these single-qubit channels, we use Pauli-type noise (depolarizing, asymmetric depolarizing, bit-flip, and phase-damping), as well as non-unital channels such as amplitude damping and thermal relaxation; in the latter case, the channel parameters are chosen to match the physical device by fitting to coherence times $T_1$ and $T_2$. To probe the impact of correlated noise and the fault resilience of the encoding circuits, we additionally consider simple correlated two-qubit depolarizing channels. The explicit Kraus representations of all noise channels are given in \cref{subapp:noise_channels}, and the experiments involving correlated noise are discussed in \cref{subsec:stability_resilience}.




\subsection{Distance Measures for Quantum States}

Quantifying the difference between quantum states is essential for the task of quantum error correction. Measures of state similarity or distinguishability help in evaluating the performance of quantum algorithms and error correction codes~\cite{nielsen2010quantum}.

The fundamental metric we employ in this paper is the \emph{trace distance}, which for two quantum states $\rho$ and $\sigma$ is defined as 
\begin{align}
    T(\rho,\sigma) = \frac{1}{2} \| \rho-\sigma \|_1,
\end{align}
where $\| A \|_1 = \mathrm{Tr} \sqrt{A^t A}$ denotes the trace norm of the operator $A$~\cite{fuchs1999cryptographic}. The trace distance quantifies the distinguishability between quantum states and is bounded between $0$ and $1$, with $0$ indicating identical states and $1$ representing completely distinguishable (orthogonal) states. These bounds and operational meaning, as well as the fact that it defines a full metric, contribute to the definition of a stable loss function in \cref{sec:method}. The trace distance is invariant under unitary transformations, i.e., \
\begin{align}
    T(U \rho U^{\dagger}, U \sigma U^{\dagger}) = T(\rho,\sigma).
\end{align}
It furthermore satisfies the data processing inequality~\cite{uhlmann1977relative}, meaning it cannot increase under \gls{cptp} maps such as noise channels:
\begin{align}
    T(\mathcal{N}(\rho), \mathcal{N}(\sigma)) \leq T(\rho,\sigma).
\end{align}
These properties make the trace distance a reliable measure when evaluating the effect of quantum operations on state distinguishability.

In this paper, we formulate the goal of a \gls{qec} code to encode information in such a way that the trace distance between different encoded states remains large, facilitating better correction operations. This approach is inspired by the \emph{quantum relative entropy}~\cite{vedral2002role}, which measures the informational divergence between two states. Its theoretical connection to the \emph{recovery map}~\cite{petz1986sufficient} provides guarantees on the existence of successful recovery operations. However, the quantum relative entropy is not a real metric, violating symmetry and the triangle inequality. Furthermore, it is unbounded and challenging to compute, making it less practical for defining a loss function.

Another important measure is the \emph{fidelity}~\cite{jozsa1994fidelity} between quantum states $\rho$ and $\sigma$, defined as:
\begin{align}
    F(\rho,\sigma) = \left( \mathrm{Tr} (\sqrt{\sqrt{\rho} \sigma \sqrt{\rho}})\right)^2
\end{align}
It quantifies the overlap between the states, with values ranging from 0 (orthogonal states) to 1 (identical states). The fidelity is invariant under unitary transformations and non-decreasing under \gls{cptp} maps; however, it is not a full metric because it violates the triangle inequality~\cite{nielsen2010quantum}. While a metric property is not strictly required for training a loss function with gradient descent, it significantly enhances both stability and interpretability~\cite{liang2024foundations}, which is why we focus on the trace distance in our work.

\subsection{Quantum Error Correction}

\Glsentrylong{qec} is a framework designed to protect quantum information from errors due to decoherence and imperfect quantum operations~\cite{shor1995scheme,steane1996error}. By encoding quantum information into entangled states of multiple qubits, \gls{qec} allows for the detection and correction of certain types of errors without measuring and thereby collapsing the quantum information itself.

The first and crucial step of such an error correction procedure is to find an encoding operation $U_{\mathrm{enc}}$ that maps the $k$-qubit pure state into a larger composite $n$-qubit system, creating redundancy that can protect against errors~\cite{calderbank1996good}. The encoding operation produces the \emph{encoded} state using the original state $\rho$ and ancilla qubits initialized to $\ket{0}$:
\begin{align}
    \label{eq:encoding}
    \rho_L = U_{\mathrm{enc}} \left( \rho \otimes \ket{0} \bra{0}^{\otimes n-k} \right) U_{\mathrm{enc}}^{\dagger}.
\end{align}
Here, the $n-k$ ancilla qubits are \emph{encoding ancillas}: they are introduced once at the beginning of the protocol, remain part of the $n$-qubit register throughout encoding, noise, and recovery, and are only traced out in the final decoding step.

After the state passes through some noise channel $\mathcal{N}$, the resulting state is mixed and can be described as
\begin{align}
    \tilde{\rho}_L = \mathcal{N}(\rho_L).
\end{align}
Afterwards, a recovery operation $U_{\mathrm{rec}}$ has to be applied to obtain the corrected logical state, potentially using and successively discarding an additional recovery register~\cite{gottesman1997stabilizer}:
\begin{align}
    \label{eq:recovery}
    \hat{\rho}_L = \mathrm{Tr}_{r} \left( U_{\text{rec}} \left( \tilde{\rho}_L \otimes \ket{0} \bra{0}^{\otimes r} \right)
    U_{\text{rec}}^{\dagger} \right).
\end{align}
Hereby, $r$ is the number of \emph{recovery ancillas}, and $\mathrm{Tr}_{r}$ denotes the partial trace over these qubits. Conceptually, this step corresponds to the recovery: in standard, measurement-based \gls{qec} the additional ancillas are used for syndrome extraction via stabilizer measurements and then discarded~\cite{steane1996error}. In our measurement-free setting, the $r$ ancillas play an analogous role as an entropy sink during recovery~\cite{nielsen2010quantum,heussen2024measurement}: they are freshly initialized, coupled coherently to the noisy encoded state, accumulate error information, and are then traced out, thereby removing entropy from the logical subsystem without intermediate measurements. 

Once the protection should be lifted, e.g.\ when the state has been transmitted and should be measured, the decoding operation can be applied as
\begin{align}
    \hat{\rho} = \mathrm{Tr}_{n-k} \left( U_{\text{enc}}^{\dagger} \hat{\rho}_L U_{\text{enc}} \right),
\end{align}
where $\mathrm{Tr}_{n-k}$ traces out the original encoding ancilla qubits.

The overall error correction procedure $\mathcal{R}$ aims to satisfy $(\mathcal{R} \circ \mathcal{N}) (\rho) \propto \rho$ for all $\rho$ whose support is in the code space $\mathcal{C}$. Alternative formulations, such as the Knill-Laflamme conditions, provide criteria for when a set of errors can be perfectly corrected by a \gls{qec} code~\cite{knill1997theory}. In this ideal case, there exists a recovery operation $\mathcal{R}$ such that, for all states $\rho$ supported on $\mathcal{C}$ and for all error patterns in a chosen correctable set\textemdash{}typically all physical errors acting non-trivially on at most $t$ qubits\textemdash{}the recovered state $\hat{\rho}$ obeys
\begin{align}
    F(\rho, \hat{\rho}) = 1,
\end{align}
and we refer to this regime as \emph{exact} \gls{qec}. In practice, and in line with the framework of approximate error correction~\cite{beny2010general,schumacher2002approximate}, our proposed routine mostly operates in a relaxed setting where one only requires the fidelity between the recovered state $\hat{\rho}$ and the original state $\rho$ to be close to $1$, i.e.\
\begin{align}
    F(\rho, \hat{\rho}) \geq 1 - \delta,
\end{align}
with $\delta$ being a small positive error threshold~\cite{beny2010general}. We refer to this regime as \emph{approximate} \gls{qec}. In the experimental section, we discuss that some instances of the learned code reduce $\delta$ to values that are numerically indistinguishable from zero, so that these codes behave as practically exact for the noise models and parameter regimes considered.

An arbitrary \gls{qec} code can be characterized by its code parameters
\begin{align}
    ((n, K = 2^k, d)),
\end{align}
where $n$ is the number of physical qubits to encode $k$ logical qubits (general \emph{non-additive} codes do not restrict $k=\log_2 K$ to a whole number, but in this work we do not consider such codes). Furthermore, $d$ is the \emph{code distance}~\cite{calderbank1997quantum}, which gives the minimum number of qubits that need to be affected by an error operation to transform one valid codeword into another one. In other words, it is the minimum number of non-trivial single-qubit Pauli errors that break the code. A distance-$d$ \gls{qec} code can detect up to $d-1$ errors and correct up to $\lfloor \frac{d-1}{2} \rfloor$ errors, i.e.\ $t = \lfloor (d-1)/2 \rfloor$ in the exact-QEC setting above. Therefore, a larger code distance implies a higher capability to correct errors. However, in general, a larger code distance requires a larger number of physical qubits, which is also quantified by the \emph{code rate} $\frac{k}{n}$. When talking about codes with no provable or unknown distance, as for most \gls{varqec} instances, we simplify the notation to
\begin{align}
    ((n, K)).
\end{align}
The special case of \emph{additive} \gls{qec} codes, usually also called \emph{stabilizer} codes, is indicated by the modified notation
\begin{align}
    [[n, k, d]].
\end{align}
In typical \gls{qec} procedures, multiple recovery cycles are necessary because new errors can arise again after the previous recovery step. As our work is mostly concerned with the encoding operation, and not so much the recovery, we focus on only a single recovery step in the following. By tailoring the encoding operations to specific noise channels, we aim to enhance the effectiveness of the codes and reduce complexity.

\section{\label{sec:method}Learning Error Correction Codes}

In this section, we introduce our method for learning \gls{qec} codes by maximizing state distinguishability. We define a loss function based on the trace distance between quantum states, realize the method as a variational quantum algorithm, discuss how to perform error correction using recovery operations, and analyze the properties of the learned codes. Supplementary details can be found in \cref{app:properties_dloss,app:learning_recovery}.

\begin{figure}[t]
    \centering
    \subfigure[Effect of a noise channel on a single physical qubit visualized on the Bloch sphere. The shrinking of the sphere and the shift of the center towards the $\ket{0}$-state can, e.g.\ be caused by depolarizing combined plus amplitude damping noise~\cite{gyongyosi2012properties}. The blue line marks the distance of optimal $\ket{0}$ and $\ket{+}$ states, the red one marks the distance after the noise channel. For the single-qubit setup, one can equivalently use the trace or the Euclidean distance, yielding $T(\ket{0},\ket{+}) = \frac{1}{\sqrt{2}}$.\label{subfig:bloch_sphere_1}]
    {
        \includegraphics[width=0.29\textwidth]{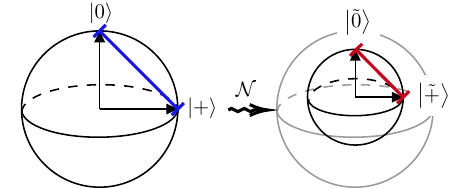}
    }
    \quad
    \subfigure[Effect of a composite noise channel on an encoded multi-qubit system. While these are not Bloch spheres, for the purpose of visualizing the trace distance, the undistorted sphere still has unit length. This follows from the invariance of the trace distance under unitary operations, consequently $T(\ket{0}_{L},\ket{1}_{L}) = \frac{1}{\sqrt{2}}$.\label{subfig:bloch_sphere_2}]
    {
        \includegraphics[width=0.29\textwidth]{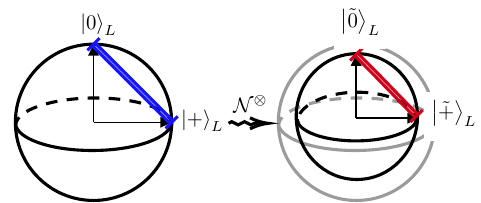}
    }
   \quad
    \subfigure[Effect of a composite noise channel on an encoded multi-qubit system. Instead of considering the entire sphere, we restrict ourselves to the effects on a spherical two-design. More concretely, we form a two-design of the underlying physical single-qubit state with $\mathcal{S} = \left\{ \ket{0},\ket{1},\ket{+},\ket{-},\ket{+i},\ket{-i} \right\}$.\label{subfig:bloch_sphere_3}]
    {
        \includegraphics[width=0.29\textwidth]{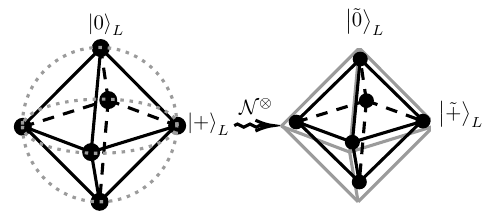}
    }
    \caption{\label{fig:bloch}Effect of noise on a physical state (a) and an encoded logical state (b). Under the assumption of a sophisticated encoding operation, we can satisfy $T(\ket{\tilde{0}},\ket{\tilde{1}}) < T(\ket{\tilde{0}}_L,\ket{\tilde{1}}_L)$. This allows us to reduce the lost trace distance $\Delta_{T}(\ket{0},\ket{1})$. In practice, we have to minimize this measure over all possible state pairs, or alternatively over all combinations of the spherical two-design in (c).}
\end{figure}

\subsection{\label{subsec:d_loss}Lost Trace and Distinguishability Loss}

To begin with, we define the \emph{lost trace distance} between two pure quantum states $\rho$ and $\sigma$ under a noise channel $\mathcal{N}$ as
\begin{align}
    \label{eq:lost_trace}
    \Delta_{T}(\rho,\sigma;\mathcal{N}{\color{gray};\Theta}) = T(\rho,\sigma) - T(\mathcal{N}(\rho_L),\mathcal{N}(\sigma_L)),
\end{align}
where $\rho_L$ and $\sigma_L$ are prepared using the potentially parameterized encoding operation $U_{\mathrm{enc}}{\color{gray}(\Theta)}$ as described in \cref{eq:encoding}. Due to the trace distance’s invariance under both tensoring with pure states and unitary transformations, we can also express the first term as $T(\rho_L,\sigma_L)$, but the former is usually more straightforward to compute. This measure is also indicated in \cref{fig:header} and its components are explicitly marked in \cref{fig:bloch}.

The motivation behind this definition is to find an encoding operation that minimizes the loss of distinguishability between quantum states after a noise channel. By keeping the lost trace distance as small as possible, we ensure that sufficient information is preserved for the subsequent recovery operation. Based on this, we define the \emph{distinguishability loss}. Since we do not know in advance which states we need to encode, the encoding operation should work effectively for arbitrary pure states. We are usually concerned with the worst-case scenario, defining the \emph{worst-case distinguishability loss} as
\begin{align}
    \label{eq:dloss_worst}
    \overline{\mathcal{D}}(\mathcal{N}{\color{gray};\Theta}) = \max_{\rho,\sigma} \Delta_{T} (\rho,\sigma;\mathcal{N}{\color{gray};\Theta}).
\end{align}
For practical optimization, evaluating the worst-case loss is challenging because it involves a maximization over an infinite set of state pairs. In principle, it is possible to use techniques from semi-definite programming~\cite{skrzypczyk2023semidefinite}, but it is hard to obtain guarantees for the efficiency of finding the solution. Moreover, we empirically found that the worst-case formulation also has the drawback that employing it as a loss function leads to severe instabilities in the training procedure, which aligns with findings from classical machine learning~\cite{hastie2009elements,bassily2020stability}. To address this, we define the \emph{average-case distinguishability loss as}
\begin{align}
    \mathcal{D}(\mathcal{N}{\color{gray};\Theta}) = \int_{\rho} \int_{\sigma} \Delta_{T}(\rho,\sigma;\mathcal{N}{\color{gray};\Theta}) \,d\nu(\rho) \,d\nu(\sigma),
\end{align}
where $\,d\nu$ denotes the integration over the Haar measure of pure quantum states~\cite{sommers2004statistical}. However, computing this integral is again impractical due to the continuous nature of the state space. Therefore, we use a two-design approximation~\cite{ambainis2007quantum} on the Hilbert space $\mathcal{H}^{\otimes k}$, where $k$ is the number of data qubits. This approximation uses a finite set $\mathcal{S}$ of states that for a two-design~\cite{dankert2009exact}, allowing to approximate the average-case loss as
\begin{align}
    \label{eq:avg_approx_d_loss}
    \mathcal{D}_{\mathcal{S}}(\mathcal{N}{\color{gray};\Theta}) = \frac{1}{\left| \mathcal{S} \right|^2} \sum_{\rho \in \mathcal{S}} \sum_{\sigma \in \mathcal{S}} \Delta_{T}(\rho,\sigma;\mathcal{N}{\color{gray};\Theta}).
\end{align}
For example, for a single data qubit we can choose $\mathcal{S} = \lbrace \ket{0}, \ket{1}, \ket{+}, \ket{-}, \ket{+i}, \ket{-i} \rbrace$, which reduces the number of evaluations to $\frac{\left| \mathcal{S} \right|^2 - \left| \mathcal{S} \right|}{2}$, accounting for the symmetry $\Delta_{T}(\rho,\sigma;\mathcal{N})=\Delta_{T}(\sigma,\rho;\mathcal{N})$ and $\Delta_{T}(\rho,\rho;\mathcal{N})=0$. While this approximation is not exact due to up to fourth-order moments in the loss function, it holds that $\mathcal{D}_{\mathcal{S}}(\mathcal{N}) \approx \mathcal{D}(\mathcal{N})$, which is sufficient for using the measure as a loss function.
The concept of quantum $t$-designs is more formally introduced and analyzed regarding approximation quality in \cref{subapp:dloss_two_design}. In short, the two-design approximation introduces a noise-dependent bias with low variance, effectively just shifting the loss function.

Conceptually, our distinguishability loss is related to the \emph{recoverable quantum information} introduced in Ref.~\cite{fosel2018reinforcement}, as both are built from the trace distance. The recoverable quantum information is defined as $\mathcal{R}_Q(t) = \frac{1}{2} \min_{\vec{n}} \left\| \hat{\rho}_{\vec{n}}(t) - \hat{\rho}_{-\vec{n}}(t) \right\|_1$, where $\hat{\rho}_{\vec{n}}$ describes the evolution under initial pure state $\vec{n}$, and $\hat{\rho}_{-\vec{n}}$ under the initially orthogonal state, respectively. For the special case of a single logical qubit under isotropic unital noise, which guarantees that the worst-case violation is caused by antipodal pairs, both measures coincide, concretely $\overline{\mathcal{D}}$ is proportional to $1-R_Q$. However, the distinguishability loss constitutes a more general and practically different object: It naturally extends to multiple logical qubits per patch, and as shown in \cref{subsec:properties} $\overline{\mathcal{D}}$ establishes analytic bounds on the fidelity after recovery, one of the standard figures of merit in \gls{qec}. Furthermore, while $\mathcal{R}_Q$ is defined as a discrete reward function only, the average-case proxy of the distinguishability loss $\mathcal{D}$ is smooth and differentiable in the encoding parameters, and can therefore be optimized directly with gradient-based techniques. Importantly, the recoverable quantum information assumes unital noise, and the efficient evaluation of $\mathcal{R}_Q$ has been tied to either isotropic channels, or more general Pauli noise under restrictions to CHZ circuits~\cite{nielsen2010quantum}. In contrast, the distinguishability loss is defined for arbitrary \gls{cptp} maps, and its two-design approximation $\mathcal{D}_\mathcal{S}$ ensures efficient evaluation. In this sense, the distinguishability loss generalizes the intuition behind the recoverable quantum information, and establishes an efficiently computable objective for learning noise-tailored encodings under arbitrary noise.

\subsection{Variational Quantum Error Correction}

\begin{figure}[t]
  \centering
  \includegraphics[width=0.6\linewidth]{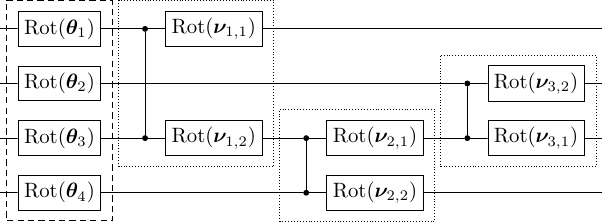}
  \caption{\label{fig:rea_small}An instance of the \emph{\glsentrylong{rea}} (REA) on $4$ qubits. It consists of an initial block of parameterized single-qubit rotations $3$ blocks of successive entangling operations. The entangling blocks themselves consist of a CZ gate and a parameterized single-qubit rotation on the two involved qubits. The control and target for each two-qubit block are determined uniformly at random, i.e.\ the exact instance of an ansatz is defined by the random seed that was used for generating it. A more general version of this ansatz is detailed in \cref{fig:randomized_entangling_ansatz}.}
\end{figure}

The underlying concept we propose is agnostic to the specific optimization method used. One could, for instance, employ the distinguishability loss as a reward function in a reinforcement learning framework that incrementally constructs the encoding circuit~\cite{fosel2018reinforcement,olle2024simultaneous}, or encode it into the fitness function of an evolutionary algorithm~\cite{rubinstein2001evolving,tandeitnik2024evolving}. 

In this work, for a proof-of-principle demonstration of the concept, we choose to formulate the task as a variational quantum algorithm~\cite{cerezo2021variational}. We first select a variational ansatz $U_{\mathrm{enc}}(\Theta)$ for the encoding unitary, with trainable parameters $\Theta$. Therefore, we utilize a newly developed \emph{\glsentrylong{rea}} (REA), which features an initial single-qubit layer, followed by randomly placed two-qubit blocks. An exemplary instance of such a layout is shown in \cref{fig:rea_small}. The motivation for this ansatz is to allow searching for well-suited shallow circuit instances by iterating over layouts with varying two-qubit placements. Furthermore, a \gls{rea} can easily be made hardware-efficient by restricting to two-qubit gates that allow for native realization on the hardware. The parameters can be optimized using either gradient-free or gradient-based optimization methods. We employ the quasi-Newton limited-memory Broyden–Fletcher–Goldfarb–Shanno (L-BFGS) algorithm~\cite{liu1989limited} due to its fast convergence for medium-sized models. However, conceptually, other optimization techniques like standard gradient descent can also be used. To enhance performance, in future work, one can also focus on a more structured approach for choosing optimal circuit architectures, using e.g.\ differentiable architecture search~\cite{zhang2022differentiable} or incrementally growing the ansatz~\cite{grimsley2019adaptive}.

\begin{figure}[t]
  \centering
    \includegraphics[width=0.99\textwidth]{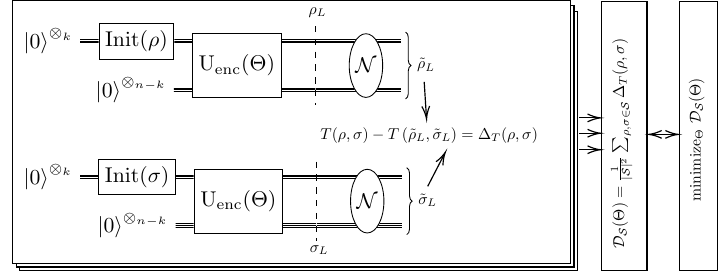}
  \caption{Pipeline of learning \gls{varqec} encodings with the distinguishability loss (adapted and extended from~\cite{meyer2025variational}). Pairs of $k$-qubit states from a two-design $\mathcal{S}$ are prepared and encoded into an $n$-qubit Hilbert space with the encoding unitary $U_{\text{enc}}(\Theta)$. The resulting encoded states are subjected to the noise channel $\mathcal{N}$, before the trace distance between the disturbed states is evaluated. The difference to the baseline trace distance is computed and denoted as the \emph{lost trace distance} $\Delta_T(\rho,\sigma;\mathcal{N};\Theta)$. This information is averaged over all state pairs $\rho,\sigma\in\mathcal{S}$, giving the two-design approximation to the average-case distinguishability loss $\mathcal{D}_{\mathcal{S}}(\mathcal{N};\Theta)$. This objective is minimized with a suitable technique, in this paper instantiated as a variational quantum algorithm utilizing gradient-based optimization. The converged encoding constitutes a $((n,2^k))$ code -- either for exact or approximate \gls{qec}, depending on the setup -- finetuned for the noise channel $\mathcal{N}$.} 
  \label{fig:encoding_pipeline}
\end{figure}

\def\app#1#2{%
  \mathrel{%
    \setbox0=\hbox{$#1\sim$}%
    \setbox2=\hbox{%
      \rlap{\hbox{$#1\propto$}}%
      \lower1.1\ht0\box0%
    }%
    \raise0.25\ht2\box2%
  }%
}
\def\approxprop{\mathpalette\app\relax}

For training the parameterized encoding ansatz $U_{\mathrm{enc}}(\Theta)$, we employ the distinguishability loss as a training objective. As argued above, optimizing the worst-case loss directly is difficult. Therefore, we use the approximate average-case loss as an effective and stable proxy:
\begin{align}
    \min_{\Theta} \mathcal{D}_{\mathcal{S}}(\mathcal{N};\Theta)
\end{align}
While it trivially holds that $\mathcal{D}_{\mathcal{S}}(\mathcal{N};\Theta) \approx \mathcal{D}(\mathcal{N};\Theta) \leq \overline{\mathcal{D}}(\mathcal{N};\Theta)$, in practice, we observe $\mathcal{D}_{\mathcal{S}}(\mathcal{N};\Theta) \approxprop \overline{\mathcal{D}}(\mathcal{N};\Theta)$. In all considered scenarios, this relation was sufficient to guarantee a good validation performance quantified by the worst-case loss $\overline{\mathcal{D}}(\mathcal{N};\Theta)$.

Due to its resemblance to a variational quantum algorithm, we denote this realization of the training procedure as \emph{\glsentrylong{varqec} (VarQEC)}~\footnote{
\textbf{Historical note}: The acronym ``VarQEC'' was first introduced by Cao et al.~\cite{cao2022quantum} for a variational, Knill–Laflamme‑based, channel‑adaptive code‑discovery framework. Only after releasing the first version of this work did we realize that we had independently adopted the same acronym for our distinguishability‑based scheme, and we therefore explicitly acknowledge their prior use of the term. In this paper, VarQEC denotes a complementary and conceptually different framework in which encodings are learned by minimizing a trace‑distance‑based distinguishability loss over logical state pairs, rather than by optimizing Knill–Laflamme cost functions on a prescribed error set; throughout, unless explicitly stated otherwise, VarQEC refers to this distinguishability‑based approach. We further delimit and contrast our framework with that of Cao et al.~\cite{cao2022quantum} in \cref{sec:related}.}. 
The resulting circuits are dubbed \gls{varqec} codes, and grouped by their code parameters $((n,2^k))$. While in \cref{subsec:properties} we also derive a methodology to quantify the code distance $d$ of \gls{varqec} codes, in general, and by construction, the \gls{varqec} procedure does not explicitly enforce the ability to recover information under specific error terms. Rather, the distinguishability loss is used to automatically focus on specific error terms to minimize the overall loss of information. The described training procedure is conceptually illustrated in \cref{fig:encoding_pipeline}.

\subsection{Generating Recovery Operations}

While the main focus of our paper is the encoding operation, it is also important to construct fitting recovery operations $U_{\mathrm{rec}}$ for practical error correction. Therefore, we adopt a learning-based method inspired by the QVECTOR procedure~\cite{johnson2017qvector} to obtain optimal parameters $\Phi$ for a variational ansatz $U_{\mathrm{rec}}(\Phi)$:

For training the recovery operation, one defines the \emph{average-case fidelity loss} between the original state $\rho$ and the recovered state $\hat{\rho}$ as
\begin{align}
    \label{eq:fidelity_loss_average}
    \mathcal{F}(\mathcal{N};\Theta,\Phi) = 1 - \int_{\rho} F(\rho,\hat{\rho}) \,d\nu(\rho),
\end{align}
where again $\rho$ are Haar-random states. Similar to the distinguishability loss, we can also define a worst-case version
\begin{align}
    \overline{\mathcal{F}}(\mathcal{N};\Theta,\Phi) = 1 - \min_{\rho} F(\rho,\hat{\rho}).
\end{align}
Interestingly, the two-design approximation of the average-case loss
\begin{align}
    \label{eq:fidelity_loss_worst}
    \mathcal{F}_{\mathcal{S}}(\mathcal{N};\Theta,\Phi) = 1 - \frac{1}{\left| \mathcal{S} \right|} \sum_{\rho \in \mathcal{S}} F(\rho,\hat{\rho}),
\end{align}
is exact, i.e.\ $\mathcal{F} = \mathcal{F}_{\mathcal{S}}$, as the loss function involves at most second-order moments~\cite{ambainis2007quantum}. In the original QVECTOR approach, one trains both the encoding and recovery operations with this loss, i.e.\ minimizes w.r.t.\ both parameter sets $\Theta$ and $\Phi$. As we have already obtained an encoding operation by training for the distinguishability loss, we focus solely on training the recovery unitary:
\begin{align}
    \label{eq:fidelity_loss_simplified}
    \min_{\Phi} \mathcal{F}_{\mathcal{S}}(\mathcal{N};\Theta,\Phi).
\end{align}
In our setting, \cref{eq:fidelity_loss_simplified} is used to learn a \emph{single-round}, coherent, measurement-free recovery operation in the sense of \cref{eq:recovery}: a unitary $U_{\mathrm{rec}}(\Phi)$ acts once after the noise channel $\mathcal{N}$ on the $n$ encoded qubits, optionally together with a fresh register of $r$ recovery ancillas initialized in $\ket{0}^{\otimes r}$, and the recovery ancillas are then traced out. There are no intermediate syndrome measurements or classical feed-forward steps in this work. The analysis of fault tolerance and the extension to multiple recovery rounds is left for future work.

In the experimental section, we demonstrate that our proposed two-step procedure, i.e.\ first optimizing the encoding and then the recovery operation, is superior to the end-to-end QVECTOR method. Moreover, the distinguishability loss also allows for determining sophisticated encoding operations, completely abstracted from the recovery technique.

\medskip
\noindent
While our training of recovery operations in \cref{eq:fidelity_loss_simplified} is only slightly modified from the QVECTOR procedure~\cite{johnson2017qvector}, the overall \gls{varqec} framework is substantially more general and should not be viewed as a merely incremental extension. In QVECTOR, encoding and recovery are optimized simultaneously with respect to an average-case fidelity loss, tightly coupling the learned encoding to a particular recovery ansatz. In contrast, \gls{varqec} first learns an encoding by minimizing the distinguishability loss $\overline{\mathcal{D}}$ via its average-case two-design proxy $\mathcal{D}_{\mathcal{S}}$, which both are based on the trace distance. This procedure comes with analytic guarantees relating $\overline{\mathcal{D}}$ to the worst-case fidelity after recovery, and an approximate code distance, as discussed in \cref{subsec:properties}. Only in a second, modular step, we train the recovery operation using a QVECTOR-like fidelity objective, so that QVECTOR effectively appears as one possible instantiation of the recovery stage on top of our novel encoding-centric objective.

From an architectural perspective, this separation enables using the same learned encodings with different coherent (or, in future work, potentially also measurement-based) recovery strategies, which is not supported by the original end-to-end QVECTOR formulation. Empirically, we find that this two-stage, distinguishability-driven approach yields encodings that match or surpass established codes and consistently outperform QVECTOR-style end-to-end training, where fidelity-only optimization often fails to converge to useful solutions (see \cref{subsec:recovery}).

\subsection{\label{subsec:properties}Properties of Learned Codes}

As elaborated above, by construction, the VarQEC procedure does not explicitly enforce the ability to recover information from a prescribed error set (e.g.\ all Pauli errors up to a certain weight). Instead, the distinguishability loss naturally guides the encoding to focus on the most detrimental errors for the given noise channel, minimizing the overall loss of information. In this sense, and in line with the framework of approximate \gls{qec}, \gls{varqec} codes should in general be classified as approximate quantum error correction codes.

\paragraph{Diagnosing code distance via distinguishability loss.}

Even though we do not impose Knill–Laflamme conditions during training, certain \gls{varqec} encoders behave as exact codes. To formalize this, we use the notion of a \emph{potential code distance} $d^*$ defined purely in terms of distinguishability loss under Pauli noise; full details and examples are given in \cref{subapp:dloss_code_distance}.

Operationally, for a given encoding, we probe all Pauli errors $E$ of weight $\mathrm{wt}(E)=w$ by applying the corresponding Pauli channel $\mathcal{N}_p(\rho) = (1-p)\rho + p\,E\rho E^\dagger$ for $0 \le p \le 1$ and evaluating the distinguishability loss $D_S(\mathcal{N}_p)$ on a two-design. If $D_S(\mathcal{N}_p)=0$ (up to numerical accuracy) for all $p$ and for all Pauli errors with $\mathrm{wt}(E)\le w$, and if there exists at least one Pauli error with $\mathrm{wt}(E)=w+1$ for which $D_S(\mathcal{N}_p)>0$, we say that the code has potential code distance $d^* = w+1$. Intuitively, vanishing distinguishability loss for all weight $\le w$ Pauli errors means that \emph{all} such errors are detectable: they do not reduce the trace distance between any pair of logical codewords. Conversely, a weight-$(w+1)$ Pauli error with non-zero loss implies that not all such errors are detectable, so the distance cannot exceed $w+1$.

At the level of the code space, this detection-based notion is equivalent to the usual definition of distance: detecting all Pauli errors of weight at most $w$ and not all of weight $w+1$ is precisely the statement that the code has distance $d=w+1$. Combined with a suitable recovery, this yields the standard operational interpretation that up to $\lfloor (d-1)/2\rfloor$ errors drawn from the span of weight $\le\lfloor (d-1)/2\rfloor$ Paulis can be corrected with unit fidelity. In \cref{subapp:dloss_code_distance}, we verify this equivalence numerically for several standard codes with known distances.

Applying this diagnostic to \gls{varqec} encodings, we find that codes with parameters $((n\geq 5,2))$ trained on \emph{symmetric} single-qubit depolarizing noise exhibit a potential code distance $d^* = 3$, i.e.\ they detect all one- and two-qubit Pauli errors and fail to detect at least some weight-3 Pauli errors. In particular, the $((5,2))$ VarQEC encoders behave as distance-3 codes, matching the $[[5,1,3]]$ perfect code~\cite{laflamme1996perfect}. Since no $[[5,1,d>3]]$ code exists (by the quantum Hamming bound), these encodings are distance-optimal for $n=5,~k=1$ in the usual sense.

The potential code distance is defined via such \emph{probe Pauli channels} and is therefore independent of the noise model used during training. In this work, we only evaluate distances explicitly for \gls{varqec} codes trained on symmetric depolarizing noise, where the traditional Hamming distance is the natural performance metric and where we can directly compare to distance-optimal stabilizer codes such as $[[5,1,3]]$. For strongly biased Pauli noise (e.g.\ $p_Z \gg p_X$) or non-Pauli channels (e.g.\ amplitude damping), the standard distance treats all Pauli error types equally and often ceases to be predictive for the relevant performance trade-offs: a code that aggressively suppresses dominant $Z$-errors at the price of tolerating slightly more $X$-errors may have the same or even smaller distance but be strictly preferable in practice. In these regimes, we therefore regard code distance as a secondary, diagnostic quantity, and use the distinguishability loss—together with its fidelity guarantees described below—as our primary, noise-aware figure of merit.

\paragraph{Theoretical guarantees for fidelity after recovery.}

As further motivation for training with the distinguishability loss, we now summarize the theoretical relationship between distinguishability loss and worst-case fidelity after recovery.
Full statements and proofs are given in \cref{subapp:dloss_fidelity_bounds}; here we state the results informally.

\begin{theorem*}[Bounds on fidelity loss, informal]
    Then the worst-case distinguishability loss $\overline{\mathcal{D}}$ under a noise channel $\mathcal{N}$ and the worst-case fidelity loss $\overline{\mathcal{F}}$ after recovery are related by two-sided bounds:
    \begin{align}
        \label{eq:bound}
        1 - \left( 1 - \overline{\mathcal{D}}(\mathcal{N}) \right)^2
        \;\geq\;
        \overline{\mathcal{F}}(\mathcal{N})
        \;\geq\;
        \overline{\mathcal{D}}(\mathcal{N})^2,
    \end{align}
    In particular, there exists a recovery operation for which the worst-case fidelity loss is upper-bounded by the distinguishability loss on the left-hand side. Furthermore, the worst-case fidelity loss for any\textemdash{}in particular, including the optimal\textemdash{}recovery operation, is lower-bounded by the distinguishability loss on the right-hand side.
\end{theorem*}

For a fixed encoding and noise channel $\mathcal{N}$, the inequality in \cref{eq:bound} shows that the worst-case distinguishability loss $\overline{\mathcal{D}}$ completely determines the interval of admissible worst-case fidelity losses $\overline{\mathcal{F}}$ after recovery. The upper bound expresses that there always exists at least one, possibly non-unique, recovery operation whose worst-case fidelity loss does not exceed $1 - \left( 1 - \overline{\mathcal{D}}(\mathcal{N}) \right)^2$. The lower bound expresses a fundamental limitation: no recovery operation, including the optimal one, can reduce the worst-case fidelity loss below $\overline{\mathcal{D}}(\mathcal{N})^2$. Consequently, a smaller worst-case distinguishability loss allows for a smaller worst-case fidelity loss—and hence a higher worst-case fidelity—after some suitable recovery, while a larger distinguishability loss implies a worse achievable fidelity loss for \emph{all} possible recoveries.

This provides a general theoretical framework and strong motivation for using the distinguishability loss as a design objective: it is efficiently approximable via its two-design proxy, compatible with gradient-based training, and comes with explicit, noise-model–independent guarantees on the best possible recovery performance, irrespective of whether the code is stabilizer or non-additive. In practice, the upper bound in \cref{eq:bound} is typically not tight; our variationally trained recovery circuits achieve substantially smaller fidelity losses than the guaranteed worst-case values. In \cref{subsec:recovery}, we empirically confirm that encodings with lower distinguishability loss tend to yield higher worst-case and average-case fidelities after recovery.

\section{\label{sec:empirical}Empirical Evaluations}

\begin{figure}[t]
    \centering
    \subfigure[\label{subfig:encoding_depolarizing}Depolarizing noise with noise strength $p=0.1$.]{
        \includegraphics[width=0.47\linewidth]{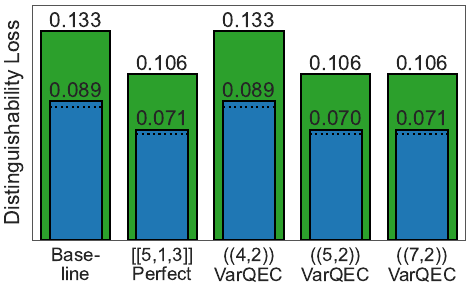}
    }\quad
    \subfigure[\label{subfig:encoding_asymmetric_depolarizing}Asymmetric depolarizing noise with noise strength $p=0.1$ and asymmetry $c=0.5$ following \cref{eq:noise_asymmetry}.]{
        \includegraphics[width=0.47\linewidth]{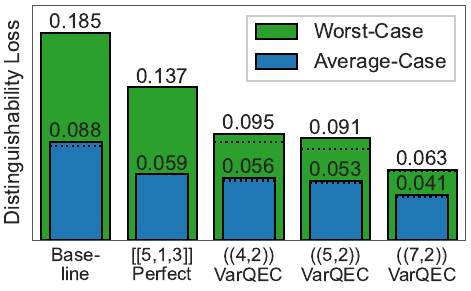}
    }
    \caption{\label{fig:encoding}Distinguishability loss for different \gls{qec} codes on (a) depolarizing and (b) asymmetric depolarizing noise. All bars visualize the average-case loss $\mathcal{D}$ and worst-case loss $\overline{\mathcal{D}}$. Additionally, the two-design approximations $\mathcal{D}_{\mathcal{S}}, \overline{\mathcal{D}}_{\mathcal{S}}$ are visualized as dotted lines if the approximation was not exact. For both noise models, the baseline of an unencoded qubit, the $[[5,1,3]]$ \emph{perfect} code~\cite{laflamme1996perfect}, and \gls{varqec} codes with $n=4$, $n=5$, and $n=7$ physical qubits are compared. All \gls{varqec} ansätze are trained for $10$ epochs and contain $(n-1)\cdot(n-2)$ two-qubit blocks. Results on additional noise models can be found in \cref{fig:encoding_extended}, an extension to two data qubits is discussed in \cref{fig:multi_qubit}.}
\end{figure}

In this section, we present empirical evaluations of the proposed \gls{varqec} method. We, for now, focus on training and validation in simulation, which we will complement with deployment of actual quantum devices in \cref{sec:hardware}. The employed ansätze and the experimental setup are detailed in \cref{app:implementation_setup}. We assess the performance of \gls{varqec} codes under symmetric and asymmetric depolarizing noise in the following, and defer considerations of various other noise models and setups with multiple logical qubits to \cref{app:experiments_extended}.

\subsection{\label{subsec:encoding}Distinguishability Loss of Encodings}

We begin by evaluating the distinguishability loss of different \gls{qec} and \gls{varqec} codes when subjected to depolarizing noise. In addition, we simulate the same noise on an unencoded qubit as a baseline (see \cref{subapp:dloss_baseline} for analytical justification). \Cref{subfig:encoding_depolarizing} illustrates the distinguishability loss for this baseline, the $[[5,1,3]]$ perfect code~\cite{laflamme1996perfect}, and \gls{varqec} codes with parameters $((4,2))$, $((5,2))$ and $((7,2))$. For all instances, the worst-case loss $\overline{\mathcal{D}}$, as well as the average-case loss $\mathcal{D}$, is displayed. Additionally, we indicate the two-design approximation values if they deviate from the ground-truth. We note that these values are themselves estimated with $1000$ Haar-random states, which is, however, sufficient for an accuracy of $10^{-4}$ (see \cref{subapp:dloss_two_design}). In this context, we have to remember that the $[[5,1,3]]$ perfect code is the smallest code that can correct for arbitrary single-qubit errors~\cite{laflamme1996perfect}. Therefore, it is expected that it serves as a kind of lower bound for the specific setup of symmetric depolarizing noise. Indeed, the smaller \gls{varqec} code with only $n=4$ physical qubits cannot improve the loss beyond the unencoded baseline of $\overline{\mathcal{D}}=0.133$. However, if we consider \gls{varqec} models with $n\geq5$ qubits, these perform on par with the loss of $\overline{\mathcal{D}} = 0.106$ of the perfect \gls{qec} code. In that respect, we consider the slightly improved average-case performance of the $((5,2))$ code with $\mathcal{D}=0.070$ only an artifact of the evaluation procedure, and not an actual performance improvement. We empirically confirm that the $((5,2))$ \gls{varqec} code exhibits a code distance of $d=3$, allowing for correcting single-qubit errors with a high fidelity (see \cref{subapp:dloss_code_distance}). Moreover, in \cref{subsec:recovery} we demonstrate that these \gls{varqec} codes allow for practically exact error correction on depolarizing noise. Overall, the learned encoding is on par with the optimal classical code for this setup.

We extend our evaluation to asymmetric depolarizing noise in \cref{subfig:encoding_asymmetric_depolarizing}. This noise channel represents general Pauli noise in the sense of \cref{eq:general_pauli}, instantiated by a bias parameter $c = \frac{\log p_Z}{\log p_X} = \frac{\log p_Z}{\log p_Y}$~\cite{olle2024simultaneous}. For a total noise strength of $p=0.1$ and an asymmetry factor of $c=0.5$, this yields the probability of an isolated phase-flip is $p_z \approx 0.085$, while the other two probabilities are $p_x = p_y \approx 0.0075$. In the presence of this asymmetry, the $[[5,1,3]]$ perfect code can no longer be assumed to be optimal. Indeed, already the \gls{varqec} code with $n=4$ qubits significantly improves the worst-case distinguishability loss. Moreover, the loss values decrease further when increasing the number of physical qubits to $n=5$ and $n=7$. 

The results demonstrate that the \gls{varqec} procedure can produce codes that are competitive on noise structures where optimal \gls{qec} codes are known, and develop novel codes for more complex noise models.

\subsection{\label{subsec:stability_resilience}Stability and Fault Resilience}

\begin{figure}[t]
  \centering
  \subfigure[\label{subfig:stability_noise_asymmetry}Asymmetric depolarizing noise with a noise strength of $p=0.1$ and asymmetry $c$ as denoted on the $x$-axis. For $c<1$ the noise landscape is dominated by $Z$ errors, while for $c>1$ mainly $X$ and $Y$ errors occur, compare \cref{eq:noise_asymmetry}.]{
    \includegraphics[width=0.47\linewidth]{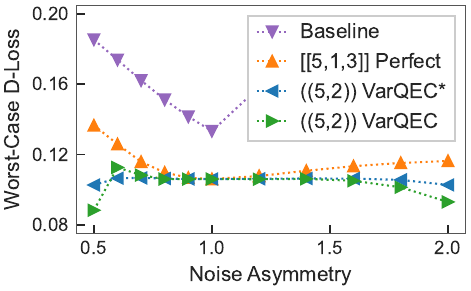}
  }\quad
  \subfigure[\label{subfig:stability_noise_resilience}Performance of noisy encoding operations on asymmetric depolarizing noise with a noise strength of $p=0.1$ and asymmetry $c=0.5$. In contrast to the other experiments, the gates in the encoding themselves are assumed to be non-ideal. To simulate this, correlated depolarizing noise of strength as denoted on the $x$-axis is added after each two-qubit operation, compare \cref{eq:correlated_depolarizing}.]{
    \includegraphics[width=0.47\linewidth]{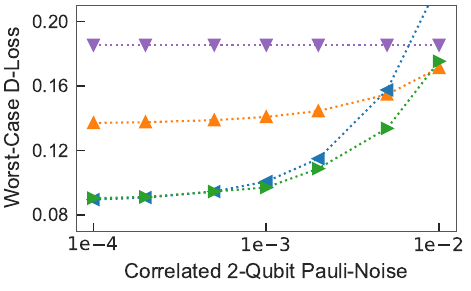}
  }
  \caption{\label{fig:stability}Stability and resilience of standard \gls{qec} and \gls{varqec} codes in setups of (a) varying noise asymmetry and (b) non-ideal encoding operations. All data points represent the respective worst-case distinguishability loss $\overline{\mathcal{D}}$. The $((5,2))^*$ \gls{varqec} code is always the same \emph{static} one trained on symmetric depolarizing noise with noise strength $p=0.1$ and under the assumption of ideal encoding operations.}
\end{figure}

We further analyze the stability and resilience of the \gls{varqec} codes and \gls{varqec} procedure in general in \cref{fig:stability}. The experiments in \cref{subfig:stability_noise_asymmetry} on varying asymmetry levels highlight the adaptability of the \gls{varqec} procedure. For all setups from $c=0.5$, i.e.\ mostly phase-flips, to $c=2.0$, i.e.\ mostly bit-flips and combined bit- and phase-flips, all codes exhibit a distinguishability loss significantly beyond the unencoded baseline. The \emph{static} version of the $((5,2))$ \gls{varqec} code (marked by $((5,2)^*$), i.e.\ the code learned on the symmetric setup $c=1.0$ from \cref{subfig:encoding_depolarizing}, outperforms the $[[5,1,3]]$ perfect \gls{qec} code especially on setups with high asymmetries. Moreover, as one could expect, the improvement when training the \gls{varqec} codes explicitly on the respective noise structure also increases with larger asymmetries. In this context, we refer to the codes noted by $((5,2))$ as \emph{dynamically} trained codes.

In a more realistic scenario, one has to assume that the encoding operations themselves are non-ideal, which is typically considered within the context of fault-tolerant quantum computing~\cite{shor1996fault}. However, this fault-tolerance analysis typically requires some highly formalized understanding of the involved encoding circuit structures, which has not been established yet for the \gls{varqec} codes. Therefore, we approach this topic from a more empirical perspective, and therefore dub this analysis as the \emph{fault-resilience} of the encodings. In \cref{subfig:stability_noise_resilience}, we do this by adding correlated depolarizing noise (see \cref{eq:correlated_depolarizing}) of increasing strength after each two-qubit operation in the encoding. This is motivated by the observation that on current hardware systems, most error-prone operations are these involving multiple qubits. While single-qubit fidelity is already quantified around $99.99\%$, two-qubit gate fidelity is just recently approaching the infamous \emph{three 9's}~\cite{mckay2023benchmarking,decross2024computational}. We do not claim that these gate fidelities can be directly related to the noise strength in \cref{subfig:stability_noise_resilience}. However, one can see that for a two-qubit gate fidelity beyond a certain threshold, the \gls{varqec} encodings are reasonably fault-resilient. Moreover, for a wide range of error rates, they outperform the (also not fault-tolerant) encoding of the $[[5,1,3]]$ perfect code, which only gradually catches up due to shallower circuit depth. Simulating the two-qubit errors already during the training procedure and therefore dynamically tailoring the codes further improves the results. The successful deployment of \gls{varqec} encodings on quantum hardware we conduct in \cref{sec:hardware} underlines these fault-resilience properties.

In \cref{subapp:restricted_connectivity_resilience} we analyze fault-resilience under the assumption of restricted connectivity, which further strengthens the case for the \gls{varqec} encodings. It is left for future work to enrich the approach with guarantees on actual fault-tolerance, e.g.\ by sufficiently extending the learning-based approach to simultaneously learn flag-gadgets~\cite{chamberland2018flag,zen2024quantum}.

\subsection{\label{subsec:recovery}Fidelity Loss with Recovery}

\begin{figure}[t]
  \centering
  \subfigure[\label{subfig:recovery_depolarizing}Depolarizing noise with noise strength $p=0.1$.]{
    \includegraphics[width=0.47\linewidth]{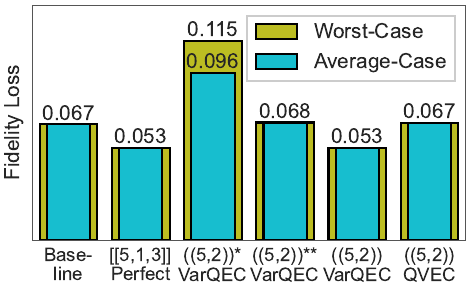}
  }\quad
  \subfigure[\label{subfig:recovery_asym_depolarizing}Asymmetric depolarizing noise with noise strength $p=0.1$ and asymmetry $c=0.5$ following \cref{eq:noise_asymmetry}.]{
    \includegraphics[width=0.47\linewidth]{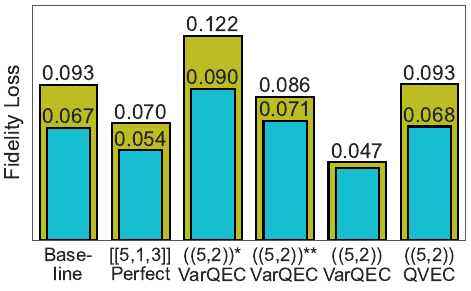}
  }
  \caption{Fidelity loss for different \gls{qec} codes and respective recovery operations on (a) depolarizing and (b) asymmetric depolarizing noise. All bars visualize the average-case loss $\mathcal{F}$ and worst-case loss $\overline{\mathcal{F}}$. For both noise models, the baseline of an unencoded qubit and the $[[5,1,3]]$ \emph{perfect} code are visualized. Additionally, we compare three instances of \gls{varqec} codes, where $((5,2))^{*}$ employs a random encoding, $((5,2))^{**}$ employs an encoding that has converged to a locally optimal distinguishability loss, and the encoding of $((5,2))$ has converged to a global optimum. The recovery operations are successively trained following \cref{eq:fidelity_loss_simplified}. For the rightmost model, the encoding and recovery operations have been trained end-to-end with the QVECTOR approach~\cite{johnson2017qvector} as described in \cref{app:learning_recovery}. The recovery ansätze are trained using the fidelity loss for $50$ epochs and contain $200$ two-qubit blocks. As we consider a single-recovery scenario, no separate recovery register was used, i.e.\ the recovery operation acts solely on the $n$-qubit encoded and disturbed state~\cite{johnson2017qvector}. The distinguishability loss for all encodings is summarized in \cref{tab:recovery_dloss}. Results on additional noise models can be found in \cref{fig:recovery_extended}.}
  \label{fig:recovery}
\end{figure}

Finally, we analyze the achievable fidelities after recovery for the noise channels and models we already considered in \cref{subsec:encoding}. This is quantified using the worst-case and average-case fidelity loss from \cref{eq:fidelity_loss_worst,eq:fidelity_loss_average}, which is evaluated over a two-design. In \cref{fig:recovery} we compare the unencoded baseline and $3$ code types: First, the $[[5,1,3]]$ perfect \gls{qec} code with its static encoding and recovery operation. Second, three versions of a \gls{varqec} code with $n=5$ physical qubits, one denoted as $((5,2))^{*}$ with a randomly initialized untrained encoding, one referred to as $((5,2))^{**}$ where the distinguishability loss has converged to a local optimum, and the $((5,2))$ \gls{varqec} code from \cref{fig:encoding} where the encoding produces a presumably globally optimal distinguishability loss. The recovery operation is trained successively following \cref{eq:fidelity_loss_simplified}. Third, a code with $n=5$ physical qubits where the encoding is not pre-trained, but encoding and recovery operations are simultaneously learned with the QVECTOR approach~\cite{johnson2017qvector}, for details refer to \cref{app:learning_recovery}. We note that as we consider the single-recovery scenario, we do not employ additional recovery qubits for the variational models~\cite{johnson2017qvector}. We list the respective distinguishability losses of the encodings and the bounds on the fidelity for the \gls{varqec} models following \cref{the:lower_bound,the:upper_bound} in \cref{tab:recovery_dloss}.

\begin{table}[t]
    \centering
    \begin{tabular}{cc|cccccc}
         & & Base- & {$[[5,1,3]]$} & {$((5,2))$}$^*$ & {$((5,2))$}$^{**}$ & {$((5,2))$} & {$((5,2))$} \\
         & & line & Perfect & VarQEC & VarQEC & VarQEC & QVEC
         \\
         \hline
         \multirow{4}{*}{symmetric depolarizing} & $\overline{\mathcal{D}}=$ & ~~$0.133$~~ & ~~$\pmb{0.106}$~~ & ~~$0.215$~~ & ~~$0.133$~~ & ~~$\pmb{0.106}$~~ & ~~$0.133$~~ \\
         & $\overline{\mathcal{F}} \leq$ & & & ~~$0.216$~~ & ~~$0.249$~~ & ~~$0.201$~~ & \\
         & $\overline{\mathcal{F}} =$ & ~~$0.067$~~ & ~~$\pmb{0.053}$~~ & ~~$0.115$~~ & ~~$0.068$~~ & ~~$\pmb{0.053}$~~ & ~~$0.067$~~ \\
         & $\overline{\mathcal{F}} \geq$ & & & ~~$0.013$~~ & ~~$0.018$~~ & ~~$0.011$~~ & \\
        \hline
         \multirow{4}{*}{asymmetric depolarizing} & $\overline{\mathcal{D}}=$ & $0.185$ & $0.137$ & $0.237$ & $0.161$ & $\pmb{0.091}$ & $0.185$ \\ 
         & $\overline{\mathcal{F}} \leq$ & & & ~~$0.228$~~ & ~~$0.296$~~ & ~~$0.174$~~ & \\
         & $\overline{\mathcal{F}} =$ & ~~$0.093$~~ & ~~$0.070$~~ & ~~$0.122$~~ & ~~$0.086$~~ & ~~$\pmb{0.047}$~~ & ~~$0.093$~~ \\
         & $\overline{\mathcal{F}} \geq$ & & & ~~$0.015$~~ & ~~$0.026$~~ & ~~$0.008$~~ & \\
         
    \end{tabular}
    \caption{\label{tab:recovery_dloss}Worst-case distinguishability loss $\overline{\mathcal{D}}$ exhibited by the encodings for the models in \cref{fig:recovery}. For the baseline and $[[5,1,3]]$ code, this is evaluated on the unencoded baseline and static encoding operation, respectively. For the \gls{varqec} codes, this corresponds to an untrained encoding for $((5,2))^{*}$, to a local optimum for $((5,2))^{**}$, and to a global optimum for $((5,2))$. For the $((5,2))$ QVECTOR code, the same ansatz with $12$ two-qubit blocks as for the \gls{varqec} codes is employed, but training was only implicitly conducted via the fidelity loss. Additionally, we again report the worst-case fidelity loss $\overline{\mathcal{F}}$ (third line) and the respective upper (second line) and lower bounds (fourth line) following \cref{the:lower_bound,the:upper_bound} for the \gls{varqec} models.}
\end{table}

The results in \cref{subfig:recovery_depolarizing} on symmetric depolarizing noise demonstrate that a lower distinguishability loss of the encoding correlates with a lower fidelity loss, i.e.\ a higher fidelity after recovery. The results on the untrained \gls{varqec} encoding and the one converged to a local optimum are below or just at the unencoded baseline, which further underlines the importance of the training procedure. We highlight that the optimal $((5,2))$ \gls{varqec} code achieves, up to numerical accuracy, an identical fidelity of $1-0.053 = 0.947$ as the $[[5,1,3]]$ perfect code, which underlines the potential code distance of $d=3$ of this code and the interpretation as practically exact error correction. The $((5,2))$ QVECTOR code is not successful in improving the fidelity beyond the unencoded baseline of $1-0.067 = 0.933$. We can observe qualitatively similar results on the asymmetric depolarizing noise channel in \cref{subfig:recovery_asym_depolarizing}. Moreover, here the $((5,2))$ \gls{varqec} code guarantees a worst-case fidelity of $1 - 0.047 = 0.953$, which significantly improves upon the $[[5,1,3]]$ \gls{qec} code with a fidelity of only $1 - 0.070 = 0.930$. Again, the QVECTOR approach is not able to improve the fidelity beyond the unencoded baseline.

Regarding the QVECTOR results, we want to note that we employed the same ansätze as for the best \gls{varqec} codes. Consequently, the underlying models in principle are guaranteed to be expressive enough to also produce the same competitive results. However, for the considered setups, the end-to-end training procedure based purely on the fidelity loss is unable to produce sophisticated codes. The results in \cref{tab:recovery_dloss} indicate that this is mostly because the resulting encoding operations do not sufficiently protect the state information before application of the noise channel. This does not contradict the results from the original QVECTOR work~\cite{johnson2017qvector}, as we consider more complex and stronger noise channels and employ ansätze with significantly fewer parameters.

Beyond the overall fidelity figures, it is instructive to inspect the effective logical noise channels induced by the encoding–noise–recovery–decoding cycle. Using the process tomography procedure described in Ref.~\cite{meyer2026learning}, we fit a single-qubit Pauli channel to the logical output for the baseline, the $[[5,1,3]]$ perfect code, and the $((5,2))$ \gls{varqec} code. The resulting total logical error probabilities $p$ and relative Pauli error weights $\nicefrac{p_X}{p}$, $\nicefrac{p_Y}{p}$, and $\nicefrac{p_Z}{p}$ are summarized in \cref{tab:effective_channel}. For symmetric depolarizing noise, both the $[[5,1,3]]$ and $((5,2))$ \gls{varqec} encodings produce essentially identical effective logical channels: they reduce the total error rate from $p=0.10$ to $p\approx 0.0795$ while preserving a perfectly symmetric $p_X,p_Y,p_Z$ distribution, consistent with the distance-$3$ behavior and matching worst-case fidelities in \cref{subfig:recovery_depolarizing}. For initially asymmetric depolarizing noise, the $[[5,1,3]]$ code only partially symmetrizes the noise, whereas the $((5,2))$ \gls{varqec} code produces an effectively almost isotropic logical channel. Importantly, the overall logical error probability is reduced to $p\approx 0.057$, which is significantly lower than $p\approx 0.081$ for the standard \gls{qec} code. This channel-level analysis corroborates the fidelity results in \cref{subfig:recovery_asym_depolarizing} and illustrates how the learned \gls{varqec} encoding adapts to the biased noise in a way that the standard stabilizer code does not.

\begin{table}[t]
    \centering
    \begin{tabular}{cc|cccc}
          & \multirow{2}{*}{\textbf{code}} & \textbf{noise strength} & \multicolumn{3}{c}{\textbf{noise proportion}} \\
            & & $p$ & ~~~$\nicefrac{p_X}{p}$~~~ & ~~~$\nicefrac{p_Y}{p}$~~~ & ~~~$\nicefrac{p_Z}{p}$~~~ \\
            \hline
            \multirow{3}{*}{symmetric depolarizing} & Baseline & $0.1000$ & $\nicefrac{1}{3}$ & $\nicefrac{1}{3}$ & $\nicefrac{1}{3}$ \\
             & $[[5,1,3]]$ Perfect & $0.0795$ & $\nicefrac{1}{3}$ & $\nicefrac{1}{3}$ & $\nicefrac{1}{3}$ \\
            & $((5,2))$ VarQEC & $0.0795$ & $\nicefrac{1}{3}$ & $\nicefrac{1}{3}$ & $\nicefrac{1}{3}$ \\
            \hline
            \multirow{3}{*}{asymmetric depolarizing} & Baseline & $0.1000$ & $0.070$ & $0.070$ & $0.860$ \\
             & $[[5,1,3]]$ Perfect & $0.0812$ & $0.438$ & $0.438$ & $0.124$ \\
            & $((5,2))$ VarQEC & $0.0569$ & $0.339$ & $0.339$ & $0.322$
    \end{tabular}
    \caption{\label{tab:effective_channel}Effective single-logical-qubit Pauli noise channels induced by the encoding–noise–recovery–decoding operation for symmetric and asymmetric depolarizing physical noise with strength $p=0.1$. For the $[[5,1,3]]$ code, we use the canonical encoding and recovery, and for the $((5,2))$ \gls{varqec} code, the learned encoder–recovery pair. The column $p$ reports the total logical error probability of the fitted Pauli channel, while $p_X/p$, $p_Y/p$, and $p_Z/p$ give the relative contributions of logical $X$, $Y$, and $Z$ errors, respectively. The characterization is performed using a variant of process tomography described in Ref.~\cite{meyer2026learning}.}
\end{table}

\medskip
\noindent
In summary, our empirical evaluations demonstrate that \gls{varqec} codes, trained using the distinguishability loss, are competitive with, and\textemdash{}in scenarios of structured noise\textemdash{}can outperform traditional small-scale stabilizer codes. The codes offer stable performance across multiple noise models and are furthermore fault-resilient, assuming a sufficiently large physical operation fidelity.

\section{\label{sec:hardware}Hardware Experiments}

In this section, we extend our experiments to actual \glsentrylong{qpu}s (QPUs) to validate the practical applicability of our proposed \gls{varqec} codes. We deploy our models on the 156-qubit \texttt{ibm\_marrakesh} device featuring a \texttt{Heron r2} chip provided by the IBM Quantum services~\cite{IBMQuantum}. All \gls{varqec} encoders used in these experiments are trained in classical simulation using noise models fitted to the device characteristics; on hardware, we deploy these pre-trained encoders and collect measurement data. The experiments involve running multiple logical patches in parallel on different regions of the chip. We compare the unencoded baseline to the logical patches by reconstructing the relevant output density matrices with mitigated state tomography~\cite{kanazawa2023qiskit} and computing the distinguishability loss from these reconstructed states. Details on the experimental procedure and the reconstruction method are summarized in \cref{subapp:hardware_setup}. The implementation of the deployment procedure on hardware and the raw results are provided as noted in the data availability statement.

\begin{figure}[t]
  \centering
  \subfigure[\label{subfig:experiment_delay}The $((4,2))$ \gls{varqec} code is trained in simulation for a noise duration of $10.0\mu\text{s}$. We deploy the model on the QPU for delays varying from $2.5\mu\text{s}$ to $10.0\mu\text{s}$. The bar plots visualize the distribution of the worst-case distinguishability loss over multiple logical patches, with orange lines denoting the median and outliers exceeding $1.5$ times the interquartile range marked separately.]{
    \includegraphics[width=0.47\linewidth]{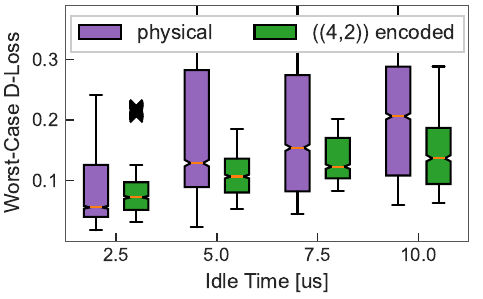}
  }\quad
  \subfigure[\label{subfig:experiment_wires}\gls{varqec} codes with increasing number of physical qubits, i.e.\ $((3,2))$, $((4,2))$, and $((5,2))$ are trained in simulation for a noise duration of $10.0\mu\text{s}$. To account for additional sources of errors, like noise introduced during the encoding operation itself and measurement delays, the delay is reduced to $5.0\mu\text{s}$ for hardware deployment. The violin plots visualize the distribution of the worst-case distinguishability loss over multiple logical patches.]{
    \includegraphics[width=0.47\linewidth]{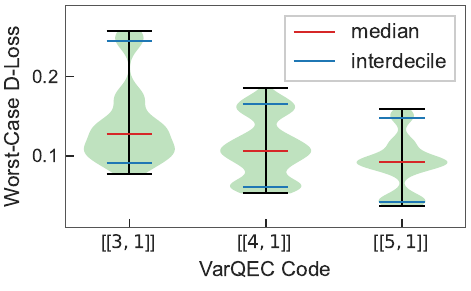}
  }
  \caption{\label{fig:experiment}Experiment on the \texttt{ibm\_marrakesh} device~\cite{IBMQuantum}, which features a 156-qubit \texttt{Heron r2} chip. We induce thermal relaxation noise by applying a delay on all wires for a specific duration, assuming a median $T_1=180\mu\text{s}$ and $T_2=120\mu\text{s}$. The \gls{varqec} codes are trained in classical simulation assuming these noise hyperparameters. Evaluation is conducted on hardware simultaneously on $8$ logical patches distributed uniformly over the QPU. We successively apply cluster re-sampling to enhance the granularity of data points~\cite{cameron2008bootstrap,meyer2025benchmarking}. Additional details on the experimental procedure and the reconstruction of the worst-case distinguishability loss $\overline{\mathcal{D}}$ via mitigated state tomography can be found in the appendix. The predictions of the distinguishability loss from the simulation can be found in \cref{subfig:experiment_simulation_ibmq}.
  }
\end{figure}

To control noise, we idle the qubits for a set duration, thereby inducing thermal relaxation noise; mitigation techniques such as dynamic decoupling are disabled for all experiments. We identify the median properties of the \texttt{ibm\_marrakesh} qubits to be $T_1 = 180\mu\text{s}$ and $T_2 = 120\mu\text{s}$, and, in our simulations, we model decoherence using single-qubit thermal relaxation (amplitude- and phase-damping) channels whose parameters are fitted to these reported coherence times. This captures the dominant Markovian incoherent noise on the device, but does not attempt to model gate-dependent drifts, leakage, or non-Markovian effects. We adapt our ansatz directly to the heavy-hex lattice of the device, selecting CZ gates as the two-qubit gates and parameterized $\text{R}_{\text{zxz}} = R_z R_x R_z$ gates as single-qubit rotations to align with the native gate set. We train models on $n=3$, $n=4$, and $n=5$ qubits in simulation with a noise duration of $t=10.0\mu\text{s}$.

In \cref{subfig:experiment_delay}, we present results for several delay values ranging from $t=2.5\mu\text{s}$ to $t=10.0\mu\text{s}$ for a $((4,2))$ \gls{varqec} code trained at $t=10.0\mu\text{s}$. We observe that the baseline physical distinguishability loss increases sharply with increasing delay duration, while the distinguishability loss for encoded patches increases much more slowly, demonstrating the effectiveness of our encoding in preserving quantum information over time. Since our training in simulation assumes ideal encoding operations, we reduce the noise duration to $t=5.0\mu\text{s}$ in the subsequent experiment shown in \cref{subfig:experiment_wires} to partially account for additional sources of error such as noise introduced during the encoding operations themselves and measurement delays. We also train and deploy $((3,2))$ and $((5,2))$ \gls{varqec} codes. The most important observation is that with an increasing number of physical qubits, the distinguishability loss decreases, indicating improved error correction performance with larger codes. We did not explicitly report hardware results for the $[[5,1,3]]$, as after compilation to the restricted topology, the encoding circuit gets quite deep. Initial trials revealed that execution and successive tomography on hardware yield close-to-random measurement outcomes, due to accumulating noise.

Reliably predicting and quantifying scalability behavior from this limited data is challenging and is left for future work, as it requires extensive access to hardware platforms. In principle, these results could be improved by refining the noise model of the device, i.e.\ incorporating two-qubit correlated noise directly into the training procedure. We analyze the experiments and the results on the individual patches in more detail in \cref{subapp:hardware_ibmq}. Furthermore, we extend our experiments to a $20$-qubit IQM Resonance device~\cite{IQMResonance} in \cref{subapp:hardware_iqm}, where we demonstrate similar results. While the hardware limits qubit stability to shorter durations, its uniform qubit performance is beneficial for our procedure, since our training relies on consistent noise levels across all qubits. Overall, these experiments demonstrate the practicality and adaptability of our approach across multiple existing hardware platforms.

\section{\label{sec:future_work}Discussion of Future Extensions}

Throughout this paper, we have demonstrated that the \gls{varqec} procedure is highly flexible and adaptable to various noise channels. The resulting codes exhibit desirable properties, are stable, and are noise-resilient. This effectiveness has also been demonstrated on actual quantum hardware. At the same time, our simulations and hardware experiments rely on relatively simple, though physically motivated, noise models: independent and correlated Pauli channels, amplitude damping, and thermal relaxation derived from device $T_1/T_2$ parameters. These models capture important qualitative features (e.g.\ non-unitality, bias, and some two-qubit correlations), but they do not constitute a full description of real hardware noise, which can be strongly gate-dependent, non-Markovian, and include leakage. Conceptually, \gls{varqec} only requires an oracle that implements the channel $\mathcal{N}$ and can therefore, in future work, also be extended to complex, more realistic noise structures.

\subsection{Computational Complexity and Scalability Limitation}

Despite these promising results, we acknowledge that there are open problems and limitations associated with our procedure. These challenges are not unique to our proposal but are present in most \gls{ai}-based approaches to \gls{qec}. We focus particularly on the open tasks surrounding the encoding operations, as recovery operations are not the primary target of this work.

From an algorithmic point of view, the dominant cost per training step is the evaluation of the two-design-based distinguishability loss $D_S(\mathcal{N};\Theta)$ in \cref{eq:avg_approx_d_loss}. For $k$ logical qubits, an exact unitary 2-design on the logical Hilbert space has cardinality $|S| = \mathcal{O}(4^k)$. In this work we restrict ourselves to $k \leq 2$ and use small spherical two-designs with $|S| = 6$ for $k = 1$ and $|S| = 16$ for $k = 2$ (compare \cref{subapp:dloss_two_design}). The loss $D_S$ uses all ordered pairs $(\rho,\sigma) \in S \times S$, so the cost of a single loss evaluation scales as
\begin{align}
    \mathrm{cost}(\mathcal{D}_{\mathcal{S}}) = \mathcal{O}(4^{2k} \cdot 4^n),
\end{align}
where the second term denotes the cost of evolving the encoding circuit under noise for one specific input state using \emph{density matrix} simulation. The overall classical training cost scales approximately linearly in the number of optimization steps times the cost of one loss evaluation; in practice, the number of L-BFGS steps per training run is on the order of $10^2$–$10^3$.

We interpret this analysis and summarize the main limitations as follows:

\renewcommand{\labelenumi}{\roman{enumi}.}
\begin{enumerate}
    \item The evaluation of the two-design approximation scales exponentially with the number of logical qubits as $4^k$.
    \item The current formulation requires simulating the underlying quantum circuits on a classical computer, with a cost that is exponential in the number of physical qubits $n$, restricting the approach to system sizes that are classically tractable.
    \item Instantiating the method as a variational algorithm may introduce the problem of barren plateaus for increased circuit size and depth~\cite{mcclean2018barren}, making training challenging.
\end{enumerate}

\subsection{Scaling by Modularization}

For the above reasons, we do not propose VarQEC as a standalone solution for large monolithic codes, but rather as a tool for learning small, noise-tailored patches that can be used as building blocks in larger architectures (e.g.\ via concatenation or patch-based logical qubits as discussed below). Within this regime of small logical patches, the per-step training cost remains well controlled.

\textbf{Code concatenation} offers the possibility to compose a larger QEC code from several smaller codes, enhancing the overall code distance and error correction capability~\cite{knill1996concatenated}. This concept has to be extended to the approximate \gls{qec} setup, focusing more on performance under specific noise models rather than on achieving a specific code distance. By using smaller, modular codes, we can mitigate the exponential scaling issue (limitation i) and reduce the computational resources required for simulation (limitation ii). Additionally, working with smaller circuits helps alleviate the barren plateau problem (limitation iii). In follow-up work~\cite{meyer2026learning}, we make this concrete by combining small, noise-tailored VarQEC patches with standard stabilizer codes in a level-wise, noise-aware concatenation pipeline. There, we introduce a fidelity-only, two-design-based estimator for the Pauli-diagonal part of the effective logical channel at each concatenation level, which avoids full process tomography and thus points to a scalable route for characterizing and selecting codes without performing state tomography on large systems.

\textbf{Patch-based logical qubits} eliminate the necessity to realize a large number of logical qubits within a single code patch. This concept is especially present in surface codes~\cite{fowler2012surface}, where single-qubit logical patches are interacted via techniques like lattice surgery~\cite{horsman2012surface}. Developing similar procedures for VarQEC codes ensures that each patch remains small enough for efficient classical training and, where desired, small-scale tomographic validation, keeping the overhead for evaluating the loss function essentially constant (limitation i). In related work on learning logical operations~\cite{meyer2026logical}, we show how to co-design encodings and logical gate sets between such patches, providing concrete examples of shallow, hardware-compatible logical gates acting between small VarQEC-encoded blocks.

\textbf{Training directly on quantum hardware}, inspired by the QVECTOR approach~\cite{johnson2017qvector}, would eliminate the necessity to simulate the involved circuits (limitation ii). A scalable implementation would, however, have to avoid full state tomography when evaluating the distinguishability loss on the device. Possible routes include estimating easier-to-measure loss proxies such as the Hilbert-Schmidt distance~\cite{fuchs1999cryptographic}, or using approximate or factorized tomography on small blocks of qubits instead of reconstructing the full global state. Even though such factorized schemes only approximate the true many-qubit state, the resulting loss signal may still be sufficiently informative to guide the optimization of the encoder.

\textbf{Barren Plateau-free Ans\"atze} have recently been proposed in the literature~\cite{park2024hamiltonian,deshpande2024dynamic}, resolving the potential trainability issues (limitation iii).
While there is debate about whether the absence of barren plateaus implies classical simulability~\cite{cerezo2023does}, this is less problematic for encoding operations in \gls{qec}. In particular, encoding operations for stabilizer codes are purely Clifford operations. Alternatively, we can modify the loss function to serve as an \gls{rl} reward function and develop fast tableau-based simulators, similar to methods used with the Knill-Laflamme conditions~\cite{olle2024simultaneous}. However, finding such efficient ways to approximate the trace distance with sufficient accuracy remains an open question. Apart from that, in the small-patch regime (up to about $20$ physical qubits per patch), barren plateaus are unlikely to be a bottleneck in practice.

\medskip
\noindent
In summary, we identified several avenues to address the current challenges in the \gls{varqec} approach, particularly concerning the distinguishability loss and scalability. Our overall strategy is to keep individual VarQEC patches small so that classical simulation and variational training remain tractable and barren plateaus are unlikely to arise, and embed these patches into larger architectures via code concatenation, patch-based logical qubits, and, in the longer term, hardware-in-the-loop variants with proxy losses. In practice, we expect a hybrid strategy\textemdash{}combining proxy losses, approximate or factorized tomography on moderate-size patches, and the use of small VarQEC patches within concatenated and patch-based logical architectures\textemdash{}to be the most promising route towards scaling.

\glsresetall

\section{\label{sec:summary}Summary and Conclusion}

In this paper, we introduced a novel approach for learning \gls{qec} codes by minimizing an objective called the distinguishability loss. This procedure ensures that the distinguishability between quantum states, quantified by the trace distance, is preserved after passing through a noise channel, which is crucial for effective recovery operations. We formalized this relationship by deriving bounds and guarantees on the fidelity after recovery. Additionally, we introduced a method to evaluate the code distance of a \gls{qec} code purely based on the distinguishability loss. By minimizing the loss of trace distance over a two-design ensemble of states, we provide an efficient evaluation method suitable for incorporation into machine learning routines.

We incorporated the distinguishability loss function into a variational algorithm, termed \gls{varqec}. This procedure trains the parameters of a variational ansatz to tailor the encoding to the most detrimental errors of a given noise model. We also modified and integrated an established routine to construct a recovery operation, enabling a full \gls{qec} cycle. The extensive empirical evaluations highlight the flexibility and adaptability of the \gls{varqec} procedure. The learned \gls{varqec} codes are shown to be stable and noise-resilient, reducing the loss of information compared to similar-sized stabilizer codes under structured noise. The deployment of \gls{varqec} code instances on IBM and IQM hardware demonstrates the practical feasibility of the procedure.

While we acknowledge current limitations surrounding our proposed \gls{varqec} approach, particularly the scalability to larger code instances, we are confident that the potential for future improvements and extensions enables a sophisticated learning-based approach to \gls{qec}. Moreover, we note that the formulation as a variational algorithm is just one specific instance of a procedure based on the distinguishability loss. In principle, it can be incorporated into a wide range of AI-based techniques, including reinforcement learning and evolutionary procedures.

\medskip
\noindent
In conclusion, our work highlights the importance of overhead reduction and noise tailoring in advancing \gls{qec}. By leveraging the distinguishability loss, we provide a flexible framework for developing \gls{qec} codes optimized for specific noise characteristics and hardware constraints. This approach is well-suited for the era of early fault-tolerant quantum computing, where resource efficiency and adaptability are paramount. Future research can build upon this foundation to explore larger systems, implement fault-tolerant logical operations, and further integrate our approach with noise-tailored \gls{qec} strategies.

\ack{
    We acknowledge the use of IBM Quantum services for this work. The views expressed are those of the authors and do not reflect the official policy or position of IBM or the IBM Quantum team.
    
    \smallskip\noindent
    We acknowledge the use of IQM Resonance services for this work. The views expressed are those of the authors and do not reflect the official policy or position of IQM or the IQM Resonance team.
    
    \smallskip\noindent
    The authors gratefully acknowledge the scientific support and HPC resources provided by the Erlangen National High Performance Computing Center (NHR@FAU) of the Friedrich-Alexander-Universit\"at Erlangen-N\"urnberg (FAU). The hardware is funded by the German Research Foundation (DFG).

    \smallskip\noindent The authors used GPT-5.1 for language editing of the paper. All content was reviewed and edited by the authors, who take full responsibility for the final work.

    \medskip\noindent    
    Some results of this study were previously reported in condensed form in~\cite{meyer2025variational}; here we present the complete, significantly extended version.

}

\funding{
    The research was supported by the Bavarian Ministry of Economic Affairs, Regional Development and Energy with funds from the Hightech Agenda Bayern via the project BayQS, and partially by the German Federal Ministry of Research, Technology and Space, funding program Quantum Systems, via the project Q-GeneSys, grant number 13N17389.
    The research is also part of the Munich Quantum Valley (MQV), which is supported by the Bavarian state government with funds from the Hightech Agenda Bayern Plus.
}

\data{
    The pipeline for training the proposed \gls{varqec} codes, including the distinguishability loss function and all other required components, is available at \url{https://github.com/nicomeyer96/varqec}. The enclosed README provides setup details and instructions for reproducing the results of the paper. The scripts for employing trained codes on IBM and IQM hardware, as well as the raw results and analysis methods, are available separately at \url{https://github.com/nicomeyer96/varqec-experiment}. Further information and data are available upon reasonable request.
}

\appendix
\crefalias{section}{appsec}
\crefalias{subsection}{appsubsec}
\renewcommand{\thesection}{\Alph{section}}
\renewcommand{\thesubsection}{\thesection.\arabic{subsection}}
\titleformat{\section}
  {\normalfont\Large\bfseries}
  {Appendix \thesection:}{1em}{}
  
\section{\label{app:background_definitions}Background and Definitions}

This appendix summarizes some important definitions and properties used throughout this paper that are already established in the literature. We start in \cref{subapp:measures_states} by providing an overview of different measures on and between quantum states. The various quantum noise channels that are considered in this paper are formally introduced in \cref{subapp:noise_channels}.


\subsection{\label{subapp:measures_states}Measures on Quantum States}

There exist various measures on and between quantum states, of which the \emph{trace distance}, \emph{fidelity}, and partially the \emph{quantum relative entropy} are relevant to our work. In general, one can note an arbitrary pure $n$-qubit quantum state using the Dirac notation as
\begin{align}
    \ket{\psi} \in \mathcal{H}^{\otimes n},
\end{align}
where $H^{\otimes n}$ is the associated $2^n$-dimensional Hilbert space, and it holds $\expval{\psi|\psi} = 1$. Alternatively, one can use the density matrix notation, in which the same pure state can be expressed as
\begin{align}
    \rho = \ket{\psi} \bra{\psi}.
\end{align}
For pure states, this matrix has rank $1$, is idempotent, and also has trace $1$. A mixed state, which e.g.\ results from a noise channel being applied to an initially pure state, is a statistical ensemble of pure states. Formally, it is a convex combination of pure state projectors, i.e., \
\begin{align}
    \rho = \sum_i p_i \ket{\psi_i} \bra{\psi_i},
\end{align}
where $p_i \geq 0$ and $\sum_i p_i = 1$. A density matrix representing a mixed quantum state is hermitian, positive semi-definite, has trace $1$, and has a rank greater than $1$~\cite{nielsen2010quantum}. One can use different measures to quantify how different two states are, all with different properties and interpretations.

The \textbf{trace distance} between two quantum states $\rho,\sigma$ represents the maximum probability of distinguishing $\rho$ from $\sigma$ in a single measurement, and is defined as
\begin{align}
    T(\rho,\sigma) &= \frac{1}{2} \| \rho - \sigma \|_1 \\
    &= \frac{1}{2} \mathrm{Tr} \left[ \sqrt{(\rho-\sigma)^{\dagger}(\rho-\sigma)} \right] \\
    &= \frac{1}{2} \sum_{i=1}^{r} \left| \lambda_i \right|,
\end{align}
where $\lambda_i$ are the eigenvalues of $\rho - \sigma$, and $r$ it the rank. The trace distance satisfies the properties of a metric, and is bounded by $0 \leq T(\rho,\sigma) \leq 1$. A value of $0$ thereby indicates that both states are identical, while a value of $1$ implies that the states are orthogonal~\cite{fuchs1999cryptographic}. The trace distance is invariant under unitary operations $U$, i.e.\ $T(U\rho U^{\dagger}, U \sigma U^{\dagger}) = T(\rho,\sigma)$. Furthermore, it is contractive for CPTP maps $\Phi$, such as quantum noise channels, following the data processing inequality~\cite{uhlmann1977relative}, i.e.\ $T(\Phi(\rho), \Phi(\sigma)) \leq T(\rho,\sigma)$.

An alternate distance measure is the \textbf{fidelity}~\cite{jozsa1994fidelity}, which quantifies the closeness or overlap between quantum states $\rho$ and $\sigma$. For arbitrary quantum states, it is defined as
\begin{align}
    F(\rho,\sigma) &= \left( \mathrm{Tr} \left( \sqrt{\sqrt{\rho} \sigma \sqrt{\rho}} \right) \right)^2 \\
    &= \left(\sum_i \sqrt{\lambda_i}\right)^2,
\end{align}
where $\lambda_i$ are the eigenvalues of the positive semi-definite matrix $\sqrt{\rho} \sigma \sqrt{\rho}$. For pure states $\rho = \ket{\psi} \bra{\psi}$ and $\sigma = \ket{\phi} \bra{\phi}$ it simplifies to
\begin{align}
    F(\rho,\sigma) = \left| \expval{\psi | \phi} \right|^2.
\end{align}
It is not a full metric, as it violates the triangle inequality, but can be extended to the \emph{Bures distance}, which is a proper metric on the space of density matrices~\cite{bures1969extension}. The fidelity is bounded by $0 \leq F(\rho,\sigma) \leq 1$, where a value of $1$ implies equivalence of both states. The fidelity is also invariant under unitary operations and non-decreasing under \gls{cptp} maps~\cite{nielsen2010quantum}. The trace distance and the fidelity are in some ways contrasting with each other, as typically a high trace distance implies a low fidelity, and vice versa. More formally, they can be related via the \emph{Fuchs-van de Graaf inequalities}~\cite{fuchs1999cryptographic} as
\begin{align}
    1 - \sqrt{F(\rho,\sigma)} \leq T(\rho,\sigma) \leq \sqrt{1 - F(\rho,\sigma)},
\end{align}
where the upper bound becomes tight if at least one of the states is pure.

Lastly, there is also the \textbf{quantum relative entropy}, which measures the informational divergence of two quantum states as
\begin{align}
    S(\rho \| \sigma) = \mathrm{Tr}(\rho(\log \rho - \log \sigma)).
\end{align}
The value range of this measure is unbounded, i.e. $0 \leq S(\rho \| \sigma) \leq \infty$, with again $S(\rho \| \sigma) = 0$ if both states are identical~\cite{vedral2002role}. In general, the quantum relative entropy is not symmetric and also does not satisfy the triangle inequality. Similar to the trace distance, the quantum relative entropy is also invariant under unitary operations and satisfies an equivalent version of the data processing inequality~\cite{lindblad1975completely}. We do not explicitly use it in this paper, but mainly took its connections to the \emph{Petz recovery map}~\cite{petz1986sufficient} as inspiration for deriving explicit relations to the fidelity after (optimal) recovery. We argue in the main part of this paper why we can not directly employ this measure as a loss function, but instead base our objective on the trace distance. For completeness, one can relate the quantum relative entropy to the trace distance via \emph{Pinker's inequality}~\cite{csiszar2011information} by
\begin{align}
    S(\rho \| \sigma) \geq 2 \cdot T(\rho,\sigma)^2,
\end{align}
and to the fidelity by~\cite{hiai1991proper}
\begin{align}
    S(\rho \| \sigma) \geq - \ln F(\rho,\sigma).
\end{align}


\subsection{\label{subapp:noise_channels}Overview of Considered Noise Channels}

Most of the noise channels we consider are well-known from the literature; we just provide this summary for convenience. Additionally, we formally define an asymmetric depolarizing channel that was introduced in \cite{olle2024simultaneous}. To begin with, we need to recall that a noise channel can be described using the \emph{Kraus representation}
\begin{align}
    \mathcal{N}(\rho) = \sum_{i} K_i \rho K_{i}^{\dagger},
\end{align}
where the $K_i$ are the so-called \emph{Kraus matrices} with $\sum_i K_i^{\dagger} K_i = \mathbb{I}$. With that notion, in the following, we define the respective noise channels by the involved Kraus matrices. For that, let us first define the standard Pauli matrices for convenience:
\begin{align}
    I = \begin{bmatrix} 1 & 0 \\ 0 & 1 \end{bmatrix}, ~X = \begin{bmatrix} 0 & 1 \\ 1 & 0 \end{bmatrix}, ~Y = \begin{bmatrix} 0 & -i \\ i & 0 \end{bmatrix}, ~Z = \begin{bmatrix} 1 & 0 \\ 0 & -1 \end{bmatrix}
\end{align}
The arguably simplest noise channel we consider, which is inspired by classical noise, is the \textbf{bit-flip channel} $\mathcal{N}_{\text{bit}}$. With a probability of $p$ a bit-flip, i.e.\ an $X$-error occurs, while with a probability of $1-p$ the state remains unchanged. On the Bloch sphere of a single qubit, this can be visualized as shrinking along the $y$ and $z$ axes. The associated Kraus matrices are as follows:
\begin{align}
    K_0 = \sqrt{1-p} \cdot I, ~K_1 = \sqrt{p} \cdot X
\end{align}
Extending this to also allow for phase-flips and combinations of bit-flip and phase-flip, we arrive at the \textbf{depolarizing channel} $\mathcal{N}_{\text{dep}}$. In the standard notion, it is assumed that for an overall noise strength of $p$, a bit-flip, a phase-flip, and the combination thereof each happen with probability of $\frac{p}{3}$. This induces a uniform shrinking of the Bloch sphere along all axes. Again, we explicitly state the Kraus matrices:
\begin{align}
    \label{eq:noise_depolarizing}
    K_0 = \sqrt{1-p} \cdot I, ~K_1 = \sqrt{\frac{p}{3}} \cdot X, ~K_2 = \sqrt{\frac{p}{3}} \cdot Y, ~K_3 = \sqrt{\frac{p}{3}} \cdot Z
\end{align}
In addition to the uniform version, in this paper we also consider an \textbf{asymmetric depolarizing channel} $\mathcal{N}_{\text{adep}}$. While there is no unambiguous definition in literature, we resort to the instance defined in~\cite{olle2024simultaneous}. It introduces a bias parameter $c$, for which it holds
\begin{align}
    \label{eq:noise_asymmetry}
    c = \frac{\log p_Z}{\log p_X} = \frac{\log p_Z}{\log p_Y},
\end{align}
where $p_X$, $p_Y$, $p_Z$ denote the probability of $X$, $Y$, $Z$ errors, respectively. Overall, it still holds $p_x + p_y + p_z = p$, where for given $p$ and $c$ we can determine $p_X$ by solving
\begin{align}
    2 p_X + p_{X}^{c} - p = 0,
\end{align}
and consecutively determine $p_Y = p_X$ and $p_Z = p - 2 p_X$. For $c=1$, we get a uniform depolarizing channel, for $c < 1$, $Z$ errors are more likely than $X$ and $Y$ errors, and for $c > 1$, it is the other way around. For these cases, non-uniform shrinking of the Bloch sphere is induced, with the following Kraus matrices:
\begin{align}
    K_0 = \sqrt{1-p} \cdot I, ~K_1 = \sqrt{p_X} \cdot X, ~K_2 = \sqrt{p_Y} \cdot Y, ~K_3 = \sqrt{p_Z} \cdot Z
\end{align}
Another noise type we consider is described by the \textbf{amplitude damping channel} $\mathcal{N}_{\text{ad}}$, which captures the irreversible loss of energy from a quantum system to its environment, leading to spontaneous transitions from high-energy to low-energy states. Concretely, the damping probability $0 \leq \gamma \leq 1$ describes the likelihood of losing energy from the $\ket{1}$ state towards the $\ket{0}$ state, which itself remains stable under this noise model. This channel is \emph{non-unital}, i.e.\ the center of the Bloch sphere is moved from its origin. The Kraus matrices are:
\begin{align}
    K_0 = \begin{bmatrix} 1 & 0 \\ 0 & \sqrt{1 - \gamma} \end{bmatrix}, ~K_1 = \begin{bmatrix} 0 & \sqrt{\gamma} \\ 0 & 0 \end{bmatrix}
\end{align}
The somewhat complementary \textbf{phase damping channel} $\mathcal{N}_{\text{pd}}$ describes the loss of quantum coherence without energy dissipation. It affects the relative phase between quantum states, but leaves the probability of being in $\ket{0}$ and $\ket{1}$ states unchanged. The strength of the noise is similarly captured by a damping probability $0 \leq \gamma \leq 1$. It induces a shrinking of the Bloch sphere along the $x$ and $y$ axes, and is described by:
\begin{align}
    K_0 = \begin{bmatrix} 1 & 0 \\ 0 & \sqrt{1 - \gamma} \end{bmatrix}, ~K_1 = \begin{bmatrix} 0 & 0 \\ 0 & \sqrt{\gamma} \end{bmatrix}
\end{align}
We also frequently consider a \textbf{consecutive amplitude and phase damping channel} $\mathcal{N}_{\text{apd}}$. We realize this trivially by first applying an amplitude damping channel and afterwards a phase damping channel. While in principle one can use different noise strengths, in this work we employ the same damping parameter $\gamma$ for both channels. Additionally, in particular with regards to the conducted hardware experiments, we consider the \textbf{thermal relaxation channel} $\mathcal{N}_{\text{thr}}$, which describes how a qubit relaxes towards the thermal equilibrium due to interaction with the environment. In principle, it can also be modeled by a combination of amplitude and phase damping. However, it simplifies directly incorporating properties of the hardware device, in particular the $T_1$ and $T_2$ relaxation times, and the noise duration $t$.

In large parts of this paper, we assume that the respective noise channels independently act on the involved qubits. More formally, we use single-qubit noise channels $\mathcal{N}_j$ acting on qubit $j$. This allows the full noise channel on the $n$-qubit system to be expressed as
\begin{align}
    \mathcal{N} = \otimes_{j}^{n} \mathcal{N}_j,
\end{align}
where typically $N_j$ is the same for all qubits. However, we additionally analyze the resilience of the encoding operations to multi-qubit correlated noise. As the employed \gls{varqec} encoding ansätze consist of at most two-qubit interactions, we also restrict experiments to two-qubit correlated errors. In particular, we model a \textbf{correlated two-qubit depolarizing channel}, with strength $p$ and the following $4^2 = 16$ Kraus matrices:
\begin{align}
    \label{eq:correlated_depolarizing}
    K_0 = \sqrt{1-p} \cdot II, ~K_{1-15} = \sqrt{\frac{p}{15}} \cdot P ~~\text{with}~ P \in \lbrace IX,IY,IZ,XI,\cdots,ZZ \rbrace  
\end{align}
All noise channels represent CPTP maps, which implies that the trace distance underlies the data processing inequality as introduced in \cref{subapp:measures_states}, i.e.
\begin{align}
    T(\mathcal{N}(\rho), \mathcal{N}(\sigma)) \leq T(\rho,\sigma)
\end{align}
for arbitrary states $\rho$ and $\sigma$.


\section{\label{app:properties_dloss}Properties of Distinguishability Loss}

In this appendix, we analyze the properties of the introduced distinguishability loss from a more theoretical perspective. An empirical evaluation with the loss integrated in the introduced \gls{varqec} pipeline can be found in the main part and in \cref{app:experiments_extended}. For convenience, we re-state the definition of the distinguishability loss for a noise channel $\mathcal{N}$, both in the average-case formulation $\mathcal{D}$ and worst-case formulation $\overline{\mathcal{D}}$:
\begin{align}
    \label{eq:d_loss_ac_app}
    \mathcal{D}(\mathcal{N}{\color{gray};\Theta}) &= \int_{\rho} \int_{\sigma} T(\rho,\sigma) - T(\mathcal{N} (\rho_L), \mathcal{N} (\sigma_L)) \,d\nu(\rho) \,d\nu(\sigma) \\
    \label{eq:d_loss_wc_app}
    \overline{\mathcal{D}}(\mathcal{N}{\color{gray};\Theta}) &= \max_{\rho,\sigma} \left( T(\rho,\sigma) - T(\mathcal{N} (\rho_L), \mathcal{N} (\sigma_L)) \right)
\end{align}
Hereby, $T$ denotes the trace distance, and $\rho_L, \sigma_L$ are the encoded states corresponding to the pure initial states $\rho, \sigma$, respectively. These are produced by applying an encoding operation $U_{\mathrm{enc}}$ to the $k$-qubit state and additional $n-k$ ancilla qubits, initialized in state $\ket{0}$: 
\begin{align}
    \label{eq:logical_state}
    \rho_L = U_{\mathrm{enc}}{\color{gray}(\Theta)} \left( \rho \otimes \ket{0} \bra{0}^{n-k} \right) U_{\mathrm{enc}}^{\dagger}{\color{gray}(\Theta)}
\end{align}
Optionally, to allow for training, the encoding operation can depend on adjustable parameters $U_{\mathrm{enc}}(\Theta)$, e.g.\ variational rotation angles. To allow for a practically tractable evaluation, we define approximations based on a two-design $\mathcal{S}$ (see \cref{subapp:dloss_two_design} for details) which reads as:
\begin{align}
    \label{eq:d_loss_ac_approx_app}
    \mathcal{D}_{\mathcal{S}}(\mathcal{N}{\color{gray};\Theta}) &= \frac{1}{\left| \mathcal{S} \right|^2} \sum_{\rho,\sigma \in \mathcal{S}} T(\rho,\sigma) - T(\mathcal{N} (\rho_L), \mathcal{N} (\sigma_L)) \\
    \label{eq:d_loss_wc_approx_app}
    \overline{\mathcal{D}}_{\mathcal{S}}(\mathcal{N}{\color{gray};\Theta}) &= \max_{\rho,\sigma \in \mathcal{S}} \left( T(\rho,\sigma) - T(\mathcal{N} (\rho_L), \mathcal{N} (\sigma_L)) \right)
\end{align}
When using these loss functions as objectives for learning quantum error correction codes, which we dub as \gls{varqec} codes, the actual target should be to optimize the worst-case behavior. Throughout this paper, we employ the average-case formulation as a proxy for this objective, as it allows for faster and more stable training. Both, the worst-case and average-case formulation trivially are lower bounded by $0$ for the case that $T(\rho,\sigma) = T(\mathcal{N}(\rho),\mathcal{N}(\sigma))$ for all state pairs $\rho,\sigma$, i.e.\ there is no loss of information under noise channel $\mathcal{N}$.

In the following, we analyze different properties of this distinguishability loss. We start by analytically calculating the baseline for a single physical qubit under depolarizing noise in \cref{subapp:dloss_baseline}. One of the central statements of this paper, the connection between the distinguishability loss and the existence of high-fidelity recovery operations, is derived in \cref{subapp:dloss_fidelity_bounds}. In \cref{subapp:dloss_code_distance} we establish a direct connection between the value of the loss function and the distance of the respective code, and discuss the distinction between exact and approximate \gls{qec}. Finally, in \cref{subapp:dloss_two_design} we analyze the accuracy of the two-design approximations of the loss function.


\subsection{\label{subapp:dloss_baseline}Baseline for Depolarizing Noise}

We now derive the unencoded baseline for the distinguishability loss, i.e.\ the value of the measure for a physical state without an encoding operation. As an instance, we choose the depolarizing channel $\mathcal{N}_{\text{dep}}$, but in principle this analysis can also be extended to other noise models. Throughout most of the paper, we determine the unencoded baseline empirically, i.e.\ either by evaluation of two-designs or averaging over Haar-random states. The analytic findings in this section coincide with the empirical evaluations and therefore reinforce the correct functionality of our framework.

Let us start with an equivalent re-formulation of the depolarizing channel in \cref{eq:noise_depolarizing}, which interprets the noise as driving towards a maximally mixed state:
\begin{align}
    \mathcal{N}_{\text{dep}}(\rho) = (1-\gamma) \rho + \frac{\gamma}{2} \mathbb{I}
\end{align}
Hereby, $p$ and $\gamma$ are related by $\gamma = \frac{4}{3}p$. Using this formulation on pure states $\rho$, $\sigma$ yields:
\begin{align}
    \mathcal{N}_{\text{dep}}(\rho) &= (1 - \frac{4}{3}p) \rho + \frac{2}{3} p \cdot \mathbb{I} \\
    \mathcal{N}_{\text{dep}}(\sigma) &= (1 - \frac{4}{3}p) \sigma + \frac{2}{3} p \cdot \mathbb{I}
\end{align}
With the definition of the trace distance and the property that the trace is a linear map, the relation between the two resulting noisy states reads as
\begin{align}
    T(\mathcal{N}_{\text{dep}}(\rho),\mathcal{N}_{\text{dep}}(\sigma)) &= \frac{1}{2} \mathrm{Tr} \left\| \mathcal{N}_{\text{dep}}(\rho) - \mathcal{N}_{\text{dep}}(\sigma) \right| \\
    &= \left| 1 - \frac{4}{3}p \right| \cdot \frac{1}{2} \mathrm{Tr} \left| \rho - \sigma \right\| \\
    &= (1-\frac{4}{3}p) \cdot T(\rho,\sigma),
\end{align}
assuming that $p \leq \frac{3}{4}$. Plugging this into the definition of the lost trace distance from \cref{eq:lost_trace} yields
\begin{align}
    \label{eq:lost_trace_depoalrizing_1}
    \Delta_T(\rho,\sigma;\mathcal{N}_{\text{dep}}) &= T(\rho,\sigma) - (1-\frac{4}{3}p) \cdot T(\rho,\sigma) \\
    \label{eq:lost_trace_depoalrizing_2}
    &= \frac{4}{3}p \cdot T(\rho,\sigma).
\end{align}
This already allows us to derive the worst-case distinguishability loss for this setup. Using \cref{eq:dloss_worst} and the fact that the trace distance takes its maximum value of $1$ for orthogonal pure states, we get
\begin{align}
    \overline{\mathcal{D}}(\mathcal{N}_{\text{dep}}) &= \max_{\sigma,\rho} \Delta_T(\rho,\sigma;\mathcal{N}_{\text{dep}}) \\
    &= \frac{4}{3}p \cdot \max_{\rho,\sigma} T(\rho,\sigma) \\
    &= \frac{4}{3}p,
\end{align}
where $\rho, \sigma$ are Haar-random states. For a noise strength of $p=0.1$, this yields $\overline{\mathcal{D}}(\mathcal{N}_{\text{dep}}) \approx 0.133$, which coincides with the experimental predictions in \cref{subfig:encoding_depolarizing}.

Things are a bit more difficult for the average-case formulation. Here we first state and prove the following lemma that relates the average-case trace distance between pure Haar-random states solely to the dimension of the Hilbert space:

\begin{lemma}\label{lem:average_fidelity}
Let $T$ be the trace distance and let $\ket{\psi},\ket{\phi}$ be Haar-random states in a Hilbert space of dimension $d$. The average of trace distances between pairs of these states depends on $d$ as
    \begin{align}
        \int_{\ket{\psi}} \int_{\ket{\phi}} T(\ket{\psi},\ket{\phi}) \,d\nu(\ket{\psi}) \,d\nu(\ket{\phi}) = \frac{d-1}{d-\frac{1}{2}}.
    \end{align}
\end{lemma}
\begin{proof}
We use the relation of the trace distance to the fidelity, which for pure quantum states reads
    \begin{align}
        T(\ket{\psi},\ket{\phi}) = \sqrt{1 - \left| \expval{\psi|\phi} \right|^2}
    \end{align}
For independent Haar-random states, the squared overlap $F := \left| \expval{\psi|\phi} \right|^2$ follows a Beta distribution $B(1,d-1)$~\cite{zyczkowski2005average}, i.e.\ we have the probability density function
    \begin{align}
        P(F) = (d-1) (1-F)^{d-2}.
    \end{align}
Using this property, we finally get
    \begin{align}
        \int_{\ket{\psi}} \int_{\ket{\phi}} T(\ket{\psi},\ket{\phi}) \,d\nu(\ket{\psi}) \,d\nu(\ket{\phi}) 
        &= \int_{0}^{1} \sqrt{1-F} \cdot P(F) \,dF \\
        &= (d-1) \int_{0}^{1} (1-F)^{d-\frac{3}{2}} \,dF \\
        &= (d-1) \cdot B(1, d-\frac{1}{2}) \\
        &= \frac{d-1}{d-{\frac{1}{2}}}.
    \end{align}
\end{proof}
\noindent
With the use of \cref{eq:lost_trace_depoalrizing_1,eq:lost_trace_depoalrizing_2} and \cref{lem:average_fidelity} for the single-qubit case, i.e.\ a Hilbert space of dimension $d=2$, we can now continue:
\begin{align}
    \mathcal{D}(\Theta;\mathcal{N}_{\text{dep}}) &= \int_{\rho} \int_{\sigma}  \Delta_T(\rho,\sigma;\mathcal{N}_{\text{dep}}) \,d\nu(\rho) \,d\nu(\sigma)  \\
    &= \frac{4}{3}p \cdot \int_{\rho} \int_{\sigma} T(\rho,\sigma) \,d\nu(\rho) \,d\nu(\sigma) \\
    &= \frac{4}{3}p \cdot \frac{2}{3} \\
    &= \frac{8}{9}p
\end{align}
For a noise strength of $p=0.1$, we get an average-case distinguishability loss of $\mathcal{D}(\mathcal{N}_{\text{dep}}) \approx 0.089$, which also matches the empirical result in \cref{subfig:encoding_depolarizing}.


\subsection{\label{subapp:dloss_fidelity_bounds}Connection to Fidelity after Recovery}

The central goal of our paper is to learn resource-efficient encoding operations $U_{\mathrm{enc}}(\Theta)$ that protect quantum information against given noise channels $\mathcal{N}$. This ability is quantified by the worst-case distinguishability loss $\overline{\mathcal{D}}(\mathcal{N};\Theta)$. However, to perform successful quantum error correction, one also has to construct and perform a recovery operation $U_{\mathrm{rec}}(\Phi)$. How to learn a recovery operation we elaborate superficially in the main part of this paper, and discuss in more detail in \cref{app:learning_recovery}. In the end, the measure of interest is how well the recovered (and decoded) state aligns with the original state, which we quantify by the worst-case fidelity loss $\overline{\mathcal{F}}(\mathcal{N};\Theta,\Phi)$. In the empirical sections of this paper, we demonstrate for various noise models and setups that an encoding with lower distinguishability loss allows for a recovery operation with higher fidelity, i.e.\ lower fidelity loss. While it is not essential for the main statements of our paper, we think it is still valuable to derive a rigorous relation between these two quantities. Therefore, we state and prove the following theorems, which we also informally stated in a summarized fashion in the main part:

\begin{theorem}[Upper bound on fidelity loss]
    \label{the:upper_bound}
    Let $\mathcal{N}$ be a noise channel, for which $T(\ket{\psi},\mathcal{N}(\ket{\psi})) \leq \max_{\ket{\phi}} \left\{ T(\ket{\psi},\ket{\phi}) - T(\mathcal{N}(\ket{\psi}),\mathcal{N}(\ket{\phi})) \right\}$ holds for arbitrary pure states $\ket{\psi}$ (with $\ket{\phi} = \ket{\psi}_{\perp}$ one can see that this holds e.g.\ for arbitrary depolarizing and phase damping noise, or bit/phase-flip noise with strength $p \leq \frac{2}{3}$). Let furthermore $U_{\mathrm{enc}}$ be an encoding operation, and let $\rho,\sigma$ be arbitrary pure quantum states. Then there always exists an recovery operation $U_{\mathrm{rec}}$, for which the worst-case fidelity loss $\overline{\mathcal{F}}$ is upper-bounded by the worst-case distinguishability loss $\overline{\mathcal{D}}$ as:
    \begin{align}
        \overbrace{1 - \min_{\rho} F(\rho, U_{\mathrm{enc}}^{\dagger} \circ U_{\mathrm{rec}} \circ \mathcal{N} \circ U_{\mathrm{enc}} (\rho))}^{\overline{\mathcal{F}}(\mathcal{N})} \leq 1 - ( 1 - \overbrace{\max_{\rho,\sigma} (T(\rho,\sigma) - T(\mathcal{N} \circ U_{\mathrm{enc}} (\rho), \mathcal{N} \circ U_{\mathrm{enc}} (\sigma)))}^{\overline{\mathcal{D}}(\mathcal{N})} )^2
    \end{align}
    To maintain clarity in the notation, we do not explicitly denote partial traces. Operations involving the tracing out of certain subsystems (e.g.\ the recovery qubits) are understood implicitly.
\end{theorem}
\begin{proof}
    We start with a lower bound on the fidelity given by the Fuchs-van de Graaf inequality~\cite{fuchs1999cryptographic} for arbitrary states $\rho,\sigma$:
    \begin{align}
        1 - \sqrt{F(\rho, U_{\mathrm{enc}}^{\dagger} \circ U_{\mathrm{rec}} \circ \mathcal{N} \circ U_{\mathrm{enc}} (\rho))} \leq T(\rho, U_{\mathrm{enc}}^{\dagger} \circ U_{\mathrm{rec}} \circ \mathcal{N} \circ U_{\mathrm{enc}} (\rho))
    \end{align}
    We now rearrange the terms and use the invariance of the trace distance w.r.t.\ unitary operations:
    \begin{align}
        \sqrt{F(\rho, U_{\mathrm{enc}}^{\dagger} \circ U_{\mathrm{rec}} \circ \mathcal{N} \circ U_{\mathrm{enc}} (\rho))} &\geq 1 - T(\rho, U_{\mathrm{enc}}^{\dagger} \circ U_{\mathrm{rec}} \circ \mathcal{N} \circ U_{\mathrm{enc}} (\rho)) \\
        &= 1 - T(U_{\mathrm{enc}} (\rho), U_{\mathrm{enc}} \circ U_{\mathrm{enc}}^{\dagger} \circ U_{\mathrm{rec}} \circ \mathcal{N} \circ U_{\mathrm{enc}} (\rho)) \\
        &= 1 - T(U_{\mathrm{enc}} (\rho), U_{\mathrm{rec}} \circ \mathcal{N} \circ U_{\mathrm{enc}} (\rho))
    \end{align}
    With the existence of a trivial recovery operation $U_{\mathrm{rec}} = \mathbb{I}$, we can further simplify, use the above assumption on the error channel $\mathcal{N}$, and again employ the invariance of the trace distance:
    \begin{align}
        \sqrt{F(\rho, U_{\mathrm{enc}}^{\dagger} \circ U_{\mathrm{rec}} \circ \mathcal{N} \circ U_{\mathrm{enc}} (\rho))} &\geq 1 - T(U_{\mathrm{enc}} (\rho), \mathcal{N} \circ U_{\mathrm{enc}} (\rho)) \\
        &\geq 1 - \max_{\sigma} \left( T(U_{\mathrm{enc}} (\rho), U_{\mathrm{enc}} (\sigma)) - T(\mathcal{N} \circ U_{\mathrm{enc}} (\rho), \mathcal{N} \circ U_{\mathrm{enc}} (\sigma)) \right) \\
        &= 1 - \max_{\sigma} \left( T(\rho, \sigma) - T(\mathcal{N} \circ U_{\mathrm{enc}} (\rho), \mathcal{N} \circ U_{\mathrm{enc}} (\sigma)) \right)
    \end{align}
    Finally, as this statement holds for arbitrary pure $\rho$, we can also state it in the worst-case formulation and rearrange terms:
    \begin{align}
        \min_{\rho} \sqrt{F(\rho, U_{\mathrm{enc}}^{\dagger} \circ U_{\mathrm{rec}} \circ \mathcal{N} \circ U_{\mathrm{enc}} (\rho))} &\geq \min_{\rho} \left( 1 - \max_{\sigma} \left( T(\rho, \sigma) - T(\mathcal{N} \circ U_{\mathrm{enc}} (\rho), \mathcal{N} \circ U_{\mathrm{enc}} (\sigma)) \right) \right) \\
        &= 1 - \max_{\rho,\sigma} \left( T(\rho, \sigma) - T(\mathcal{N} \circ U_{\mathrm{enc}} (\rho), \mathcal{N} \circ U_{\mathrm{enc}} (\sigma)) \right)
    \end{align}
    Utilizing that both left and right sides of the equation can only take values in $[0,1]$, we just need to take the square and subtract both sides from one (which inverts the inequality) to arrive at the desired outcome.
\end{proof}
\noindent
In simpler words, \cref{the:upper_bound} states that decreasing the distinguishability loss for an encoding operation guarantees a lower fidelity loss, i.e.\ an improved fidelity after recovery. In practice, this bound is rarely tight, as this relation holds already purely due to the existence of a trivial recovery operation $U_{\mathrm{rec}} = \mathbb{I}$, as used in the proof. Therefore, usually it should be possible to find much better recovery operations than guaranteed by this \cref{the:upper_bound}. The experimental results in this paper, generated with the procedure for learning recovery operations described in \cref{app:learning_recovery}, confirm this assumption. However, we also empirically observe that a lower distinguishability loss allows for a higher-fidelity recovery operation.

Under the same requirements, one can also derive a lower bound on the fidelity loss, depending on the distinguishability loss, which reads as:
\begin{theorem}[Lower bound on fidelity loss]
    \label{the:lower_bound}
    Let $\mathcal{N}$ be a noise channel, for which $T(\ket{\psi},\mathcal{N}(\ket{\psi})) \leq \max_{\ket{\phi}} \left\{ T(\ket{\psi},\ket{\phi}) - T(\mathcal{N}(\ket{\psi}),\mathcal{N}(\ket{\phi})) \right\}$ holds for arbitrary pure states $\ket{\psi}$ (with $\ket{\phi} = \ket{\psi}_{\perp}$ one can see that this holds e.g.\ for arbitrary depolarizing and phase damping noise, or bit/phase-flip noise with strength $p \leq \frac{2}{3}$). Let furthermore $U_{\mathrm{enc}}$ be an encoding operation, and let $\rho,\sigma$ be arbitrary pure quantum states. Then for any recovery operation $U_{\mathrm{rec}}$, the worst-case fidelity loss $\overline{\mathcal{F}}$ is lower-bounded by the worst-case distinguishability loss $\overline{\mathcal{D}}$ as:
    \begin{align}
        \overbrace{1 - \min_{\rho} F(\rho, U_{\mathrm{enc}}^{\dagger} \circ U_{\mathrm{rec}} \circ \mathcal{N} \circ U_{\mathrm{enc}} (\rho))}^{\overline{\mathcal{F}}(\mathcal{N})} \geq  \overbrace{\max_{\rho,\sigma} (T(\rho,\sigma) - T(\mathcal{N} \circ U_{\mathrm{enc}} (\rho), \mathcal{N} \circ U_{\mathrm{enc}} (\sigma))}^{\overline{\mathcal{D}}(\mathcal{N})} )^2
    \end{align}
    To maintain clarity in the notation, we do not explicitly denote partial traces. Operations involving the tracing out of certain subsystems (e.g.\ the recovery qubits) are understood implicitly.
\end{theorem}
\noindent
As the proof follows an identical strategy, we omit stating it at this point. Complementary to the above, \cref{the:lower_bound} states that it is not only sufficient but also necessary to achieve a low distinguishability loss, to allow for a high-fidelity recovery operation. However, especially for small $\overline{\mathcal{D}}$, this bound is close to $1$, i.e.\ almost trivial.

Overall, both theorems establish an interesting analytic connection between the objective of minimizing the distinguishability loss, and minimizing the fidelity loss, i.e.\ achieving a high fidelity after recovery. Combined with the empirical evaluations and correlations we highlight throughout the paper, we are confident that our proposed procedure has its merits for quantum error correction.


\subsection{\label{subapp:dloss_code_distance}Code Distance and Approximate Error Correction}

We recall from the literature~\cite{gottesman2002introduction} that an error on an $n$-qubit system can be expressed as a tensor product of single-qubit Pauli operators
\begin{align}
    E = E_1 \otimes E_2 \otimes \cdots \otimes E_n,
\end{align}
where each $E_j$ is from $\lbrace I, X, Y, Z \rbrace$. The weight of an error operator $wt(E)$ is defined as the number of qubits on which $E$ acts non-trivially, i.e.
\begin{align}
    \label{eq:error_weight}
    wt(E) = \left| \lbrace j \mid E_j \neq I \rbrace \right|.
\end{align}
It is easy to see that the number of errors of weight $w$ is given by
\begin{align}
    \label{eq:number_paulis_weight}
    N_{w} = \binom{n}{w} \cdot 3^w,
\end{align}
where the first factor represents the number of unique configurations of $w$ qubits on which errors occur, and the second factor expresses the fact that $3$ non-identity Pauli errors can occur on each of these qubits. The distance $d$ of a code can be defined as the minimum weight of a non-trivial error term that transforms one valid codeword into another one. More formally, assume that a code can detect all weight-$w$ errors, but can not detect all errors with $wt(E) = w+1$. Then this code has a code distance of
\begin{align}
    d = w+1,
\end{align}
which by definition allows it to detect $d-1 = w$ errors and correct for 
\begin{align}
    t = \lfloor \tfrac{d-1}{2} \rfloor = \lfloor \tfrac{w}{2} \rfloor
\end{align}
errors, provided a suitable recovery operation exists. With the number of encoded logical qubits $k$ and the underlying number of physical qubits $n$, one typically uses the description
\begin{align}
    ((n,2^k,d)),
\end{align}
or $[[n,k,d]]$ for stabilizer codes, to classify the error correction capabilities of a code. In this work, we consider \gls{qec} codes with $n \leq 10$, $k \leq 2$, and $d \leq 4$, but in the main part, we also sketch potential improvements that could allow scaling to larger code parameters.

For some \gls{qec} codes, in particular stabilizer codes, there exist analytic but also numerical techniques to determine the code distance $d$ for a given encoding~\cite{terhal2015quantum,campbell2017roads,pryadko2023qdistrnd}. However, these results can not be directly utilized to determine the distance of \gls{varqec} codes. Therefore, we introduce the notion of the \emph{potential code distance} $d^{*}$, which can be determined purely based on the distinguishability loss and encodes the \emph{detection} properties of the code space:

\begin{definition}[Potential code distance]
    \label{def:pot_dist}
    Let $\mathcal{C}$ be a \gls{qec} code that encodes a $k$-qubit logical state into $n$ physical qubits. Assume that the distinguishability loss of this code satisfies
    \begin{align}
        \label{eq:pot_distance}
        \overline{\mathcal{D}}_{\mathcal{S}}(\mathcal{N}_p) = 0
    \end{align}
    for all noise strengths $0 \leq p \leq 1$ and all Pauli noise channels
    \begin{align}
        \label{eq:pauli_noise_channels}
        \mathcal{N}_p(\rho) = (1-p) \cdot \rho + p \cdot E \rho E^{\dagger},
    \end{align}
    where $wt(E) \leq w$. Furthermore, assume there is at least one noise channel $\mathcal{N}_p$ with $wt(E) = w+1$ for which $\overline{\mathcal{D}}_{\mathcal{S}}(\mathcal{N}_p) > 0$. Then we say that $\mathcal{C}$ has a potential code distance of $d^* := w+1$, which lets us describe the code with the parameters
    \begin{align}
        ((n, 2^k, d^*)).
    \end{align}
\end{definition}
\noindent
In the case of \cref{def:pot_dist}, it would also be possible to alternatively use the average-case code distinguishability loss $\mathcal{D}_{\mathcal{S}}$ instead of the worst-case formulation $\overline{\mathcal{D}}_{\mathcal{S}}$. We observed that in all cases the \emph{strongest signal} $\overline{\mathcal{D}}_{\mathcal{S}}(\mathcal{N}_p)$ is exhibited for $p=0.5$, i.e.\ for empirical evaluations we focus on this setup. 

By construction, $d^*$ is defined solely in terms of how the code space transforms under Pauli errors: $\overline{\mathcal{D}}_{\mathcal{S}}(\mathcal{N}_p)=0$ for all $wt(E)\le w$ means that \emph{all} such Pauli errors leave the trace distance between any pair of logical codewords unchanged, i.e.\ they are detectable in the usual sense. The notion of a potential code distance therefore quantifies the detection capability of an encoding (code space) without referring to any particular recovery map.

Especially given the \gls{varqec} codes, we also define a relaxed version of the potential code distance, which we dub the \emph{potential approximate code distance} $d^{*}_{\varepsilon}$:
\begin{definition}[Potential approximate code distance]
    \label{def:pot_approx_dist}
    Let $\mathcal{C}$ be an approximate quantum error correction code encoding a $k$-qubit logical state into $n$ physical qubits. Assume that the distinguishability loss of this code satisfies
    \begin{align}
        \label{eq:pot_approx_distance}
        \overline{\mathcal{D}}_{\mathcal{S}}(\mathcal{N}_p) \leq \varepsilon
    \end{align}
    for a small positive value of $\varepsilon$, all noise strengths $0 \leq p \leq 1$ and all Pauli noise channels
    \begin{align}
        \mathcal{N}_p(\rho) = (1-p) \cdot \rho + p \cdot E \rho E^{\dagger},
    \end{align}
    where $wt(E) \leq w$. Furthermore, assume there is at least one noise channel $\mathcal{N}_p$ with $wt(E) = w+1$ for which $\overline{\mathcal{D}}_{\mathcal{S}}(\mathcal{N}_p) > \varepsilon$. Then we say that $\mathcal{C}$ has a potential approximate code distance of $d^{*}_{\varepsilon} := w+1$, which lets us describe the code with the parameters
    \begin{align}
        ((n, 2^k, d^{*}_{\varepsilon})).
    \end{align}
\end{definition}

\begin{table}[t]
    \centering
    \begin{tabular}{c|cc|cc|cc|cc}
         & \multicolumn{2}{c|}{$1$ fault} & \multicolumn{2}{c|}{$2$ faults} & \multicolumn{2}{c|}{$3$ faults} & \multicolumn{2}{c}{$4$ faults} \\
         & number & detect & number & detect & number & detect & number & detect \\
         \hline
         $[[3,1,1]]$ & 9 & \xmark & & & & & & \\
         \hline
         $[[4,1,2]]$ & \multirow{2}{*}{12} & \cmark & \multirow{2}{*}{54} & \xmark & & & & \\
         ~$((4,2))${\footnotesize$^{\boldsymbol{a}}$~~} & & (\cmark) & & \xmark & & & & \\
         \hline
         $[[5,1,3]]$ & \multirow{2}{*}{15} & \cmark & \multirow{2}{*}{90} & \cmark & \multirow{2}{*}{270} & \xmark & & \\
         ~$((5,2))${\footnotesize$^{\boldsymbol{b}}$~~} & & (\cmark) & & (\cmark) & & \xmark & & \\
         \hline
         $[[7,1,3]]$ & \multirow{2}{*}{21} & \cmark & \multirow{2}{*}{189} & \cmark & \multirow{2}{*}{945} & \xmark & & \\
         ~$((7,2))${\footnotesize$^{\boldsymbol{b}}$~~} & & (\cmark) & & (\cmark) & & \xmark & & \\
         \hline
         $[[9,1,3]]$ & 27 & \cmark & 324 & \cmark & 2268 & \xmark & & \\
         \hline
         $[[10,1,4]]$ & 30 & \cmark & 405 & \cmark & 3240 & \cmark & 17010 & \xmark\\
    \end{tabular}
    \newline {\footnotesize{$^{\boldsymbol{a}}$ trained on amplitude damping noise with $\gamma=0.1$; $^{\boldsymbol{b}}$ trained on depolarizing noise with $p=0.1$;}}
    \caption{Evaluation of potential (approximative) code distance following \cref{def:pot_dist,def:pot_approx_dist} for different \gls{qec} and \gls{varqec} codes. We perform evaluations for $1$ to $4$ faults, corresponding to the respective error weights in \cref{eq:error_weight}. The \emph{number} column denotes the according count of possible error configurations from \cref{eq:number_paulis_weight} that are evaluated. A checkmark \cmark~for \emph{detect} indicates that \cref{eq:pot_distance} is satisfied for this code and error weight, for (\cmark) at least the weaker \cref{eq:pot_approx_distance} holds with $\varepsilon=0.002$. For \xmark~neither holds, which also indicates the potential (approximative) code distance of the respective code.}
    \label{tab:distance}
\end{table}

\noindent
Based on these definitions, we report empirical evaluations of several standard \gls{qec} codes, and two instances of \gls{varqec} codes in \cref{tab:distance}. In particular, we determine the potential code distance following \cref{def:pot_dist} for standard (non-approximative) codes with known code distance, and the potential approximate code distance for codes learned on depolarizing and amplitude damping noise.

For all codes with known code distance $d$, the evaluated potential distance $d^{*}$ coincides with $d$. This is expected: for Pauli noise, vanishing distinguishability loss $\overline{\mathcal{D}}_{\mathcal{S}}(\mathcal{N}_p)=0$ for all errors of weight $wt(E)\le w$ is equivalent to saying that \emph{all} such Pauli errors are detectable in the usual sense (they cannot map one logical codeword to another, nor reduce the trace distance between distinct logical states). Hence, at the level of error detection, the potential code distance $d^{*}$ defined via the distinguishability loss reproduces the standard code distance $d$:
\begin{align}
    ((n,2^k,d^*)) \quad\text{and}\quad ((n,2^k,d))
\end{align}
describe the same detection capability with respect to Pauli errors.

To turn this detection property into \emph{correction} in the usual operational sense (``the code can correct up to $\lfloor (d-1)/2\rfloor$ errors''), one additionally needs a recovery operation that actually implements this correction. For the stabilizer codes in \cref{tab:distance}, such recovery maps are known analytically from standard \gls{qec} theory. For \gls{varqec} codes, we do not assume a specific decoder a priori, but in the main text, we explicitly construct high-fidelity recovery unitaries variationally (\cref{subsec:recovery}) and verify that, for example, the $((5,2))$ VarQEC code on depolarizing noise behaves operationally like a distance-3 code when combined with its learned recovery.

Assuming the existence of a suitable recovery, the standard QEC result then implies that a code with (potential) distance $d^{*}$ can perfectly correct up to
\begin{align}
    t = \Big\lfloor \frac{d^*-1}{2} \Big\rfloor
\end{align}
errors drawn from the span of Pauli errors of weight $\le t$, i.e.\ there exists a recovery operation for which
\begin{align}
    \label{eq:correction_fidelity}
    \overline{\mathcal{F}}(\mathcal{N}_p) = 1
\end{align}
for all noise channels $\mathcal{N}_p(\rho) = (1-p) \cdot \rho + p \cdot E \rho E^{\dagger}$ where $E$ has a weight of $w \leq \lfloor \frac{d^*-1}{2} \rfloor$. We note that the results from \cref{the:upper_bound,the:lower_bound} are not directly applicable here, as these noise channels do not satisfy the stated requirements; instead, this statement relies on the standard Knill-Laflamme theory for Pauli error sets. Conceptually, the potential distance $d^*$ tells us that the \emph{code space} has the right detection properties; the existence and explicit construction of a recovery map is the additional step needed to obtain a full QEC code of distance $d^*$ in the operational sense.

Alternatively to this framework of \emph{perfect} error correction, there is also the notion of \emph{approximate} error correction~\cite{schumacher2002approximate,beny2010general}, which relaxes \cref{eq:correction_fidelity} to
\begin{align}
    \label{eq:correction_fidelity_approx}
    \overline{\mathcal{F}}(\mathcal{N}_p) \geq 1 - \delta
\end{align}
for some small positive value $\delta$. For completeness, we notice that both perfect and approximate error correction conditions can also be stated in terms of the \emph{Knill-Laflamme} conditions~\cite{knill1997theory}, but the notion using the fidelity is easier to interpret for our work. By construction, the proposed \gls{varqec} procedure does not explicitly enforce the ability to recover information under specific error terms. Rather, the distinguishability loss is used to automatically focus on specific error terms, to minimize the overall loss of information. Consequently, approximate error correction is the more fitting framework to analyze these codes. Similar to above, we hypothesize that a \gls{varqec} code with parameters $((n, 2^k, d^{*}_{\varepsilon}))$ allows for the correction of up to $\lfloor \frac{d^*_{\varepsilon}-1}{2} \rfloor$ errors with a high fidelity following \cref{eq:correction_fidelity_approx}, where $\delta$ correlates with $\varepsilon$, again provided a suitable recovery is available. Interestingly, the $((5,2))$ \gls{varqec} code that is analyzed in \cref{subfig:recovery_depolarizing} practically performs equivalent to the $[[5,1,3]]$ \gls{qec} code, i.e.\ we can assume $\delta$ being very close to $0$, or at least below numerical accuracy of the evaluation procedure. Consequently, it is reasonable to interpret this instance of a \gls{varqec}-code, together with its learned recovery, as being capable of performing practically exact error correction.


\subsection{\label{subapp:dloss_two_design}Approximation with Two-Designs}
To allow for an efficient evaluation of both the average-case distinguishability loss $\mathcal{D}$ and worst-case distinguishability loss $\overline{\mathcal{D}}$, we introduced two-design approximations $\mathcal{D}_{\mathcal{S}}$ and $\overline{\mathcal{D}}_{\mathcal{S}}$, respectively. For the exact definitions, please refer to the main part or \cref{eq:d_loss_ac_app,eq:d_loss_wc_app,eq:d_loss_ac_approx_app,eq:d_loss_wc_approx_app} in the appendix. The non-approximative instances are based on integrating or maximizing over pairs of Haar-random pure $k$-qubit states~\cite{sommers2004statistical}, respectively. Empirically, these operations have to be mimicked by averaging or maximizing over a finite set of $H$ states following a Haar-random distribution.

\begin{figure}[t]
  \centering
  \subfigure[Analysis for a single data qubit, i.e.\ $k=1$.]{
    \includegraphics[width=0.47\linewidth]{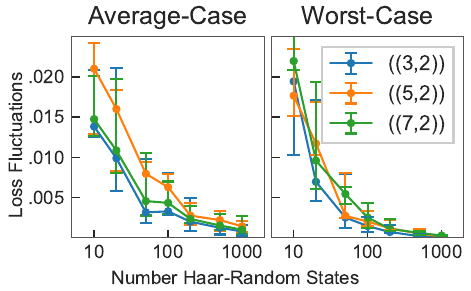}
  }\quad
  \subfigure[Analysis for two data qubits, i.e.\ $k=2$.]{
    \includegraphics[width=0.47\linewidth]{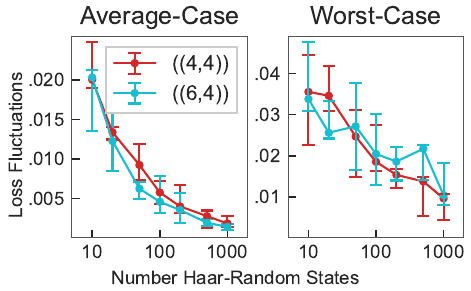}
  }
  \caption{Empirical analysis of how many Haar-random states are necessary to get a stable approximation of the ground-truth loss, which integrates or maximizes over an infinite number of states. The loss fluctuations denote the maximal difference between distinguishability loss values over $10$ instances of the same encoding, with different random seeds for generating the Haar-random states. The data points denote the median, while the error bars show the smallest/largest fluctuation over $5$ different encoding circuits. The plots refer to depolarizing noise with $p=0.1$, but similar results can be obtained for other noise models.}
  \label{fig:two_design_number}
\end{figure}

In \cref{fig:two_design_number}, we analyze how many such states are actually necessary for $k=1$ and $k=2$ to get a stable approximation to the ground-truth. In the main parts of this paper, we base our evaluations on $H=1000$ Haar-random states, which should guarantee a deviation from the ground truth of at most $0.001$ in the $k=1$ case and $0.01$ for $k=2$. To save computational resources, in the appendix we reduce this to $H=200$, which still should limit the deviations to about $0.005$.

Formally, one defines that $\mathcal{X}$ forms a quantum $t$-design, if the following equation holds for any polynomial $P_t$ of a most degree $t$ in the entries of $\rho$, as well as most degree $t$ in the complex conjugate of these entries~\cite{ambainis2007quantum}:
\begin{align}
    \label{eq:quantum_t_design_app}
    \frac{1}{S} \sum_{\rho_k \in \mathcal{X}} P_t(\rho_k) = \int_{\rho} P_t(\rho) \,d\nu(\rho)
\end{align}
On the right side of \cref{eq:quantum_t_design_app}, $\rho$ are Haar-random quantum states. In principle, the distinguishability loss contains degrees of up to order $4$, i.e.\ one would require a $4$-design to guarantee this equivalence. However, we evaluate below that in practice, using a two-design, which is much easier to construct and evaluate, allows for a sufficiently accurate approximation.

In general, the minimum size of a two-design for a $k$-qubit system is $\mathrm{dim}(\mathcal{H}^{\otimes k})^2 = 4^k$. One trivial two-design, which is, however, vastly exceeding this bound, is given by the Clifford group~\cite{dankert2009exact}. For a single-qubit system, this would allow for a two-design with four elements, i.e., the \emph{tetrahedron states}. However, for easier preparation of the initial states, we resort to the more common six-element spherical two-design 
\begin{align}
    \mathcal{S} = \lbrace \ket{0}, \ket{1}, \ket{+}, \ket{-}, \ket{+i}, \ket{-i} \rbrace.
\end{align}
This spherical two-design originating from a single qubit is also visualized in \cref{subfig:bloch_sphere_3}. For two data qubit, i.e.\ $k=2$, we use a formulation with $12$ product states and $4$ entangled states, reading as
\begin{align}
    \mathcal{S} =& \lbrace \ket{00}, \ket{01}, \ket{10}, \ket{11}, \\
    & ~\ket{++}, \ket{+-}, \ket{-+}, \ket{--}, \\
    & ~\ket{+i+i}, \ket{+i-i}, \ket{-i+i}, \ket{-i-i}, \\
    & ~\ket{00} + \ket{11}, \ket{00} - \ket{11}, \ket{01} + \ket{10}, \ket{01} - \ket{10} \rbrace,
\end{align}
where the normalization factors for the entangled states are understood implicitly. We acknowledge that this approximation still is untractable for large $k$, especially as we evaluate trace distance between all states in $\mathcal{S}$, which consequently scales as $(4^k)^2 = 8^k$. However, we want to emphasize that, in general, it is not desirable to encode an arbitrary number of logical qubits in a single code patch. We envision a modular realization, where several code patches, each implementing a limited number of logical qubits $k$, interact. An extreme example of this from existing \gls{qec} literature would be the use of lattice surgery~\cite{horsman2012surface} on logical surface code patches~\cite{fowler2012surface}, each encoding only a single logical qubit.

\begin{figure}[t]
  \centering
  \subfigure[\label{subfig:two_design_approximate_1}Analysis for a single data qubit, i.e.\ $k=1$.]{
    \includegraphics[width=0.47\linewidth]{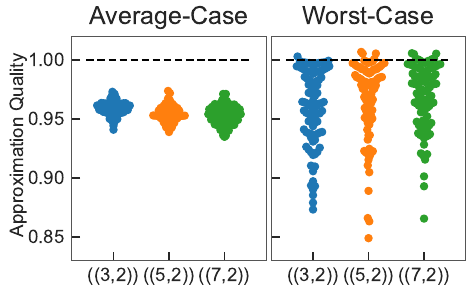}
  }\quad
  \subfigure[\label{subfig:two_design_approximate_2}Analysis for two data qubits, i.e.\ $k=2$.]{
    \includegraphics[width=0.47\linewidth]{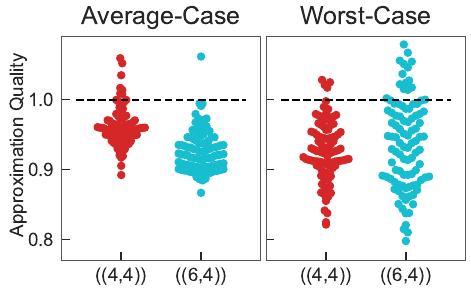}
  }
  \caption{Empirical analysis on how well the two-design formulations approximate the groundtruth of the distinguishability loss, which we quantify by $\mathcal{D} / \mathcal{D}_{\mathcal{S}}$ and $\overline{\mathcal{D}} / \overline{\mathcal{D}}_{\mathcal{S}}$, respectively. Hereby, we evaluate $100$ random encoding circuit instances for each setup, and estimate the ground truth with $H=200$ Haar-random states. The plots refer to a depolarizing noise channel with $p=0.1$.}
  \label{fig:two_design_approximate}
\end{figure}

In \cref{fig:two_design_approximate}, we analyze how well we can approximate the ground-truth distinguishability loss with the respective two-design formulations. By the definitions, it must, in principle, hold
\begin{align}
    \label{eq:bound_d_approx}
    \overline{\mathcal{D}} \geq \overline{\mathcal{D}}_{\mathcal{S}}.
\end{align}
This is because the state pair creating the largest loss of trace distance within the two-design must trivially be reflected in the consideration of all state pairs. For the average-case formulation, no such direct relation exists. Empirically, we quantify the approximation quality as $\mathcal{D} / \mathcal{D}_{\mathcal{S}}$ and $\overline{\mathcal{D}} / \overline{\mathcal{D}}_{\mathcal{S}}$, respectively. For both the $k=1$ setup in \cref{subfig:two_design_approximate_1} and the $k=2$ setup in \cref{subfig:two_design_approximate_2}, we can see that while there is some bias in average-case approximation, the variance of results is not too large. In contrast, for the worst-case approximation, the bias is smaller, but the variance is larger. The points above $1.0$, which in principle violate \cref{eq:bound_d_approx}, can be explained by the fact that the baseline is only estimated with a finite number of Haar-random states. Moreover, we do not need to care too much about the worst-case approximation quality, as for evaluating this loss we usually directly resort to $\overline{\mathcal{D}}$ instead of $\overline{\mathcal{D}}_{\mathcal{S}}$. For training, we employ the average-case approximation $\mathcal{D}_{\mathcal{S}}$, but as it is used as a trainable loss function, the approximate proportionality to the ground truth $\mathcal{D}$ is sufficient.

\begin{figure}[t]
  \centering
  \subfigure[Approximation with standard 2-design.]{
    \includegraphics[width=0.47\linewidth]{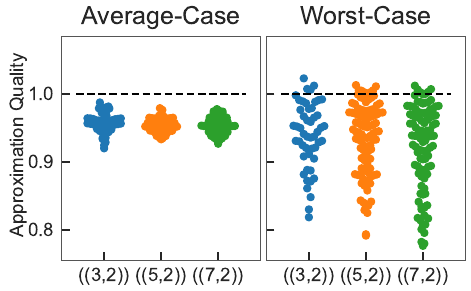}
  }\quad
  \subfigure[Approximation with weighted 2-design.]{
    \includegraphics[width=0.47\linewidth]{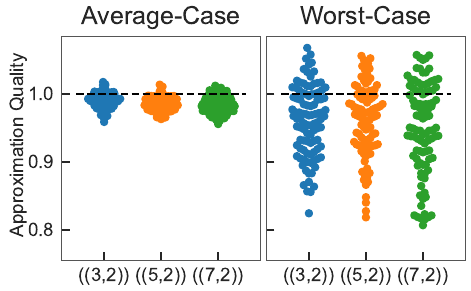}
  }
  \caption{Empirical analysis on how a weighted two-design can reduce the approximation bias, compared to a standard two-design on an amplitude damping channel with $\gamma=0.1$. Hereby, we evaluate $100$ random encoding circuit instances for each setup, and estimate the ground truth with $H=200$ Haar-random states. the weights are set as summarized in \cref{eq:weights_amplitude_damping}.}
  \label{fig:two_design_weighted}
\end{figure}

While not strictly necessary, there are also ways to reduce the bias, especially of the average-case formulation, without significantly reducing the variance. One possibility we briefly consider here is the use of a \emph{weighted two-design}~\cite{roy2007weighted}, which incorporates prior information about the noise channel. For this, we can extend the definitions of the distinguishability loss as
\begin{align}
    \mathcal{D}_{\mathcal{S}_w}(\mathcal{N}{\color{gray};\Theta}) &= \frac{1}{\sum_{\rho,\sigma \in \mathcal{S}_w} w_{\rho} \cdot w_{\sigma}} \sum_{\rho,\sigma \in \mathcal{S}} w_{\rho} \cdot w_{\sigma} \cdot \left( T(\rho,\sigma) - T(\mathcal{N} (\rho_L), \mathcal{N} (\sigma_L)) \right) \\
    \overline{\mathcal{D}}_{\mathcal{S}_w}(\mathcal{N}{\color{gray};\Theta}) &= \max_{\rho,\sigma \in \mathcal{S}} w_{\rho} \cdot w_{\sigma} \cdot \left( T(\rho,\sigma) - T(\mathcal{N} (\rho_L), \mathcal{N} (\sigma_L)) \right),
\end{align}
where the $w$'s are weights associated with the respective elements of the two-design. It is natural but not necessary to ensure that it holds $w_{\rho} \geq 0$ for all elements $\rho$ and $\sum_{\rho \in \mathcal{S}} w_{\rho} = \left| \mathcal{S} \right|$. With setting all $w_{\rho}=1$ the equations simplify to the original definitions,

In \cref{fig:two_design_weighted} we exemplarily consider an amplitude damping channel with damping parameter $\gamma=0.1$. With this channel, low-energy states like $\ket{0}$ are less or not affected at all, while high-energy states like $\ket{1}$ are disproportionally strongly affected. We reflect this in setting the weights as
\begin{align}
    \label{eq:weights_amplitude_damping}
    w_{\ket{0}} = 0.95, ~w_{\ket{1}} = 1.05, ~w_{\ket{+}} = 1.0, ~w_{\ket{-}} = 1.0, ~w_{\ket{+i}} = 1.0, ~w_{\ket{-i}} = 1.0.
\end{align}
The comparison between the pure two-design and the weighted two-design formulation underlines that this reduces the bias of the average-case distinguishability loss, without increasing the variance.


\section{\label{app:learning_recovery}Learning Recovery Operations with Fidelity Loss}

In this appendix, we sketch a procedure with which one can learn a recovery operation, given a \gls{varqec}-encoding already trained with the distinguishability loss. We emphasize that deriving such a recovery operation is not the primary scope of this work, but acknowledge that it is still helpful to consider and evaluate our technique within the scope of error correction. Therefore, the technique we describe in the following is to a large extent based on the work by Johnson et al.~\cite{johnson2017qvector}. There, the authors establish the QVECTOR approach, which can be used to train encoding and recovery operations for a given noise channel simultaneously from scratch, based on evaluations of the fidelity. We neither claim that this is the only method that could be modified to learn recovery operations given fixed encodings, nor do we claim that it is the most sophisticated technique for this task. Still, empirically, we get convincing results, and we outperform the original QVECTOR approach in a variety of setups.

\begin{figure}[t]
  \centering
  \includegraphics[width=0.99\textwidth]{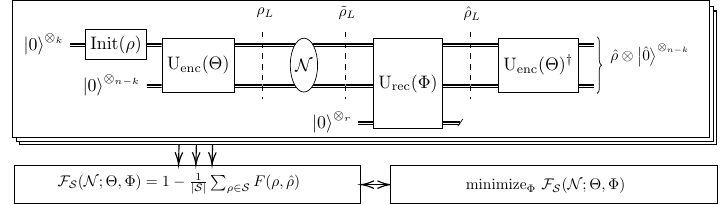}
  \caption{The adapted QVECTOR pipeline~\cite{johnson2017qvector} for training recovery operations $U_{\mathrm{rec}}(\Phi)$, under the premise that the encoding operations $U_{\mathrm{enc}}(\Theta)$ have already been trained with the distinguishability loss. The average-case fidelity loss $\mathcal{F}_{\mathcal{S}}$ is evaluated on states from a two-design $\mathcal{S}$ and is minimized w.r.t.\ the free parameters of the recovery operation. We note that at the end of the circuits it is not guaranteed that the state $\hat{\rho}$ on the $k$ data qubits is completely disentangled from the $n-k$ ancilla qubits, but for the single-recovery scenario, this can be neglected.}
  \label{fig:recovery_pipeline}
\end{figure}

To set the state, let us start with the definition of an encoded state $\rho_L$ encoded with the potentially parameterized encoding operation $U_{\mathrm{enc}}{\color{gray}(\Theta)}$ as described in \cref{eq:logical_state}. After a noise channel $\mathcal{N}$ acts on this state, the task is now to find a recovery operation $U_{\mathrm{rec}}{\color{gray}(\Phi)}$, which reverts the effect of $\mathcal{N}$ as much as possible. Especially for the multi-recovery setting, i.e.\ multiple consecutive recovery operations, this requires the use of additional \emph{recovery} qubits initialized to a known state, usually $\ket{0}^{\otimes r}$.  Most established \gls{qec} procedures use this register for measuring stabilizers and extracting syndromes~\cite{gottesman1997stabilizer}. Subsequently, classically controlled correction operations are applied to the encoded states. While it should be possible to adapt the presented \gls{varqec} routine within this framework, it more naturally aligns with the concept of \emph{measurement-free} or \emph{autonomous} \gls{qec}~\cite{lebreuilly2021autonomous,zeng2023approximate,heussen2024measurement}. After applying the recovery operation, the register is traced out, which in practice could be done by re-initializing to $\ket{0}^{\otimes r}$ or supplying new qubits for the next recovery cycle. Considering the single-recovery setup, the resulting state $\hat{\rho}_L$ reads as
\begin{align}
    \hat{\rho}_L = \mathrm{Tr}_{r} \left( U_{\text{rec}}{\color{gray}(\Phi)} \left( \mathcal{N}({\rho}_L) \otimes \ket{0} \bra{0}^{\otimes r} \right) U_{\text{rec}}^{\dagger}{\color{gray}(\Phi)} \right)
\end{align}
Consecutively, one can decode the state using the inverse of the encoding operation $U^{\dagger}_{\mathrm{enc}}{\color{gray}(\Theta)}$ and trace out the ancilla encoding qubits to get the final decodes state $\hat{\rho}$ as
\begin{align}
    \hat{\rho} = \mathrm{Tr}_{n-k} \left( U_{\text{enc}}^{\dagger}{\color{gray}(\Theta)} \hat{\rho}_L U_{\text{enc}}{\color{gray}(\Theta)} \right),
\end{align}
where $\rho$ should be close to $\hat{\rho}$ for a successful \gls{qec} cycle. This closeness can be quantified by the fidelity $F(\rho,\hat{\rho})$. This is the underlying idea of the QVECTOR approach, which uses variational ansätze for both the encoding and recovery operation, and defines the average-case \emph{fidelity loss} as
\begin{align}
    \mathcal{F}(\mathcal{N}{\color{gray};\Theta,\Phi}) = 1 - \int_{\rho} F(\rho,\hat{\rho}) \,d\nu(\rho),
\end{align}
where the states $\rho$ follow a Haar-random distribution. Similarly to the distinguishability loss, one can also define a worst-case version, which reads as
\begin{align}
    \overline{\mathcal{F}}(\mathcal{N}{\color{gray};\Theta,\Phi}) = 1 - \min_{\rho} F(\rho,\hat{\rho}).
\end{align}
We note that the original paper~\cite{johnson2017qvector} defines a slightly different formulation, in which the encoding qubits are not traced out, i.e.\ the underlying fidelity is
\begin{align}
    F(p \otimes \ket{0}^{n-k}, U^{\dagger}_{\mathrm{enc}}{\color{gray}(\Theta)}\hat{\rho}_L U_{\mathrm{enc}}{\color{gray}(\Theta)}).
\end{align}
However, this formulation is mainly motivated by the multi-recovery scenario, which we, for now, do not focus on in our work. Indeed, we experimented with both loss functions and did not observe a significant difference in performance for our setup. As the fidelity distance contains moments of at most order two, the two design formulations of the average-case loss are exact in this case:
\begin{align}
    \mathcal{F}_{\mathcal{S}}(\mathcal{N}{\color{gray};\Theta,\Phi}) &= 1 - \frac{1}{\left| \mathcal{S} \right|} \sum_{\rho \in \mathcal{S}} F(\rho,\hat{\rho}) = \mathcal{F}(\mathcal{N}{\color{gray};\Theta,\Phi})
\end{align}
The fidelity loss trivially converges to $0$ if the fidelity between original and recovered state is close to $1$, i.e.\ one can formulate and optimize the QVECTOR objective as
\begin{align}
    \label{eq:fidelity_loss_app}
    \min_{\Theta,\Phi} \mathcal{F}_{\mathcal{S}}(\mathcal{N};\Theta,\Phi),
\end{align}
which is used to simultaneously train encoding $U_{\mathrm{enc}}(\Theta)$ and recovery $U_{\mathrm{rec}}(\Phi)$. We slightly modify this setup, as initially we use our \gls{varqec} procedure based on the distinguishability loss to produce sophisticated encodings. The respective encoding ansätze are fixed after that, and we only train the recovery operation with the fidelity loss, which simplifies the second step to
\begin{align}
    \label{eq:fidelity_loss_simplified_app}
    \min_{\Phi} \mathcal{F}_{\mathcal{S}}(\Theta,\Phi).
\end{align}
This procedure is also sketched in \cref{fig:recovery_pipeline}. The end-to-end QVECTOR approach proves sophisticated enough for noise models like phase and amplitude damping with small damping parameters~\cite{johnson2017qvector}. However, it struggles with more complex noise models and larger noise rates. In the experimental section of this paper, we demonstrate that our two-step procedure succeeds on a variety of noise models and does so with a much shallower circuit, which is beneficial for fault-resilience and fault-tolerance.

\section{\label{app:implementation_setup}Implementation Details and Setup}

In this appendix, we summarize details regarding the actual experimental realization and implementation. We introduce a new type of ansatz we used for both the encoding and recovery operations in \cref{subapp:randomized_ansatz}. In \cref{subapp:implemented_pipeline} we highlight some implementation details and also discuss some algorithmic hyperparameters.


\subsection{\label{subapp:randomized_ansatz}Randomized Entangling Ansätze (REAs)}

In principle, the introduced \gls{varqec} procedure also allows for generating circuit ansätze from scratch. This could, e.g.\ be done by using the distinguishability loss as the reward function of a reinforcement learning agent~\cite{fosel2018reinforcement,olle2024simultaneous}, or encoding it into the fitness function of an evolutionary algorithm~\cite{rubinstein2001evolving,tandeitnik2024evolving}. However, for the proof-of-concept considerations of this paper, for simplicity we decided to formulate it as a variational quantum algorithm~\cite{cerezo2021variational} by fixing the structure $U_{\mathrm{enc}}$ and $U_{\mathrm{rec}}$ as a variational quantum circuit and purely training variational parameters which corresponds to rotation angles. To still have some flexibility and adaptability of the respective ansätze, we define the \gls{rea}. It is inspired by the simplified two-design ansatz~\cite{cerezo2021cost}, which incorporates single-qubit rotations and two-qubit control gates, but offers more flexibility.

\begin{figure}[t]
  \centering
  \includegraphics[width=\linewidth]{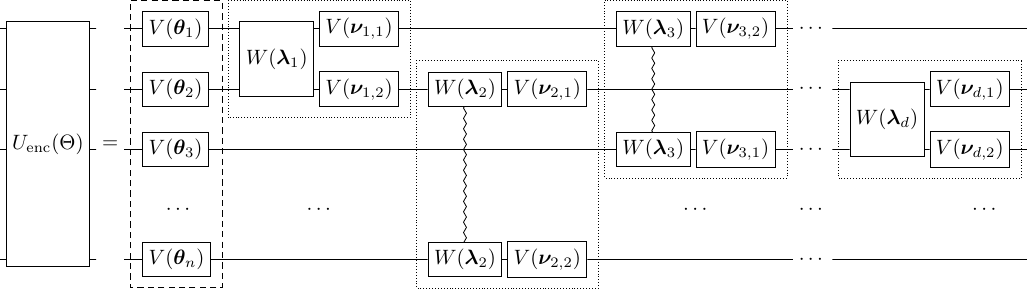}
  \caption{An instance of the \gls{rea} on $n$ qubits. It consists of an initial block of parameterized single-qubit gates $V$ and $d$ blocks of successive entangling operations. The entangling blocks themselves consist of an optionally parametrized two-qubit gate $W$, and parameterized single-qubit operations $V$ on the two involved qubits. The control and target for each two-qubit block are determined uniformly at random, i.e.\ the exact instance of an ansatz is defined by the random seed that was used for generating it.}
  \label{fig:randomized_entangling_ansatz}
\end{figure}

In \cref{fig:randomized_entangling_ansatz} we show an instance of such a \gls{rea} ansatz for the encoding operation. We note that the same ansatz can also be used for the recovery operations, potentially incorporating the additional recovery register. The general outline is defined by a parameterized single-qubit operation $V$ and an optionally also parameterized two-qubit operation $W$. Throughout this paper, we default to
\begin{align}
    V(\bm{\theta_j}) &= R_z(\theta_{j,3}) R_y(\theta_{j,2}) R_z(\theta_{j,1}) \\
    W(\bm{\lambda}_j) &= \mathrm{Controlled}\mathrm{-}V(\bm{\lambda}_j)
\end{align}
to stay as general as possible. However, to adapt to the native gate sets of the systems, we deviate from this in the hardware experiments. The details are laid out in the respective sections. The construction itself first places single-qubit operations parameterized by $\bm{\theta}$ on each contributing qubit. Subsequently, $d$ two-qubit blocks are placed, where control and target are each selected uniformly at random, but disjunct. These gates do not necessarily have to be parameterized, but in case they are, we refer to the parameters as $\bm{\lambda}$. After each of these blocks, we place two single-qubit operations parametrized by $\bm{\nu}$ on the involved qubits.

As the placement of two-qubit blocks is determined at random, the actual instance is determined by the random seed that was used for generating the placement. To generate a sophisticated ansatz, in particular a shallow one, throughout this paper, we use $100$ different seeds, train the respective \gls{varqec} circuit, and select the best performing one. We acknowledge that this procedure can not be expected to scale beyond a certain system size, as the number of ansätze one has to iterate over might increase proportionally to the problem complexity. However, we think it suffices as a proof-of-concept procedure to demonstrate that effective and shallow-depth encoding circuits can be found with the \gls{varqec} procedure. In particular, the parameter count is given as
\begin{align}
    \left| \bm{\theta} \right| &= n \cdot N_{V} \\
    \left| \bm{\lambda} \right| &= d \cdot N_{W} \\
    \left| \bm{\nu} \right| &= 2 \cdot d \cdot N_{V}
\end{align}
where $N_V$ and $N_W$ are the number of parameters per single-qubit and two-qubit operation, respectively. Alternative ways to perform architecture search in future work have already been outlined above, but one can also employ additional techniques from the literature, like e.g.\ differentiable architecture search~\cite{zhang2022differentiable} or incremental growing the ansatz~\cite{grimsley2019adaptive}.


\subsection{\label{subapp:implemented_pipeline}Hyperparameters and Python Module}

First, we want to note that all details on the used hyperparameters and settings can be extracted from the provided implementation and respective Readme file; links are provided in the data availability statement. In this section, we therefore only focus on the high-level functionalities and highlight the most important implementation details.

We implemented the \texttt{VarQEC} pipeline in \texttt{python v3.12}, but it should also be backward compatible with \texttt{python v3.10}. The underlying framework consists of about $4000$ lines of code and is embedded into the \texttt{PyTorch} cosmos~\cite{paszke2019pytorch} for training machine learning models. To realize the quantum computing parts, for initial proof-of-concept realizations and hyperparameter tuning, we resorted to the \texttt{qiskit-torch-module}~\cite{meyer2024qiskit}. For later experiments and the polished version, we switched to \texttt{PennyLane}~\cite{bergholm2018pennylane}, due to the more extensive support of simulating mixed states and quantum channels. The \texttt{VarQEC} module itself consists of $7$ main components, which we will briefly describe in the following:

\paragraph{Training and testing script.} This is the central component of the \texttt{VarQEC} pipeline and integrates all other components. It imports the respective models, circuit ansätze, error channels, and loss functions. It currently makes use of the L-BFGS optimizer~\cite{liu1989limited} with a history size of $100$ and $10$ iterations per training epoch, but can easily be adapted to support other optimizer instances. The training script allows for the isolated training of encoding operations with the distinguishability loss, but also supports the successive training of recovery operations with the fidelity loss. The testing script allows to perform a detailed analysis of the trained models, but also implements several known error correction codes as a baseline. Furthermore, it also allows to evaluate the potential code distance introduced in \cref{subapp:dloss_code_distance}.

\paragraph{Loss functions.} These are one of the main contributions of this paper, and are therefore implemented in a modular function, to enable usage also in other setups. For the fidelity loss, we only provide a simplified implementation based on the QVECTOR approach, as described in \cref{app:learning_recovery}. For the distinguishability loss, we provide two separately finetuned versions, one to reduce memory overhead on the expense of additional time, and one that is adapted the other way around.

\paragraph{Variational Models.} These machine learning models represent either pure encoding circuits or also contain consecutive recovery operations. The implementation offers flexibility regarding model width, i.e.\ the use of data, ancilla, and recovery qubits, but also model depth, i.e.\ the number of involved trainable parameters. It imports the three components of state encoding, trainable circuit ansatz, and error channels. All computations are implemented in a fully differentiable way utilizing the \texttt{autograd}~\cite{paszke2019pytorch} method, which allows for straightforward training with gradient-based optimizers.

\paragraph{State Encoding.} We provide two different methods to prepare the initial states, corresponding to the exact and approximate formulation of the respective loss functions. The first one initializes an adjustable number of Haar-random states. This is done by sampling the respective number of unitaries of size $2^k \times 2^k$ randomly from the unitary group $\mathcal{U}(2^k)$~\cite{mezzadri2006generate}, which conceptually is supported for an arbitrary number of data qubits $k$. The second method prepares states from a two-design. To enhance the experimental efficiency, which is crucial as the loss function is trained on the two-design approximations, we implement the respective state initializations in a parameterized way to allow for batch processing. One single qubit, i.e.\ $k=1$, we place an arbitrary single-qubit rotation $U_3$, which is initialized with the values denoted in \cref{tab:initialization} produces the desired states. We implement the same methodology also for $k=2$ qubits, but we sketched in \cref{subapp:dloss_two_design} how this also could be extended to larger systems.

\begin{table}[t]
    \centering
    \begin{tabular}{c|c|c|c|c|c}
         $\ket{0}$ & $\ket{1}$ & $\ket{+}$ & $\ket{-}$ & $\ket{+i}$ & $\ket{-i}$ \\
         \hline
         $U_3(0.0,0.0,0.0)$ & $~U_3(\pi,0.0,\pi)~$ & $~U_3(\frac{\pi}{2},0.0,\pi)~$ & $U_3(\frac{\pi}{2},-\pi,-\pi)$ & $~U_3(\frac{\pi}{2},\frac{\pi}{2},-\pi)~$ & $U_3(\frac{\pi}{2},-\frac{\pi}{2},-\pi)$ \\
    \end{tabular}
    \caption{How to perform parameterized initialization on single-qubit two-design states. We place an arbitrary rotation gate $U_3(\theta,\phi,\delta)$ on the data qubit, which with the denoted parameters prepares the respective states. This procedure allows for batch-processing of all states involved in the computation of the loss function, which significantly speeds up the training procedure.}
    \label{tab:initialization}
\end{table}

\paragraph{Circuit Ansatz.} The circuit ansatz determines the layout and allocation of trainable parameters that are trained to be the encoding or, respectively, recovery operation. In the \texttt{VarQEC} module, we implement the \gls{rea} detailed in \cref{subapp:randomized_ansatz}, but it is also readily extensible to other custom ansätze. The module enables setting a restricted connectivity as in \cref{subapp:restricted_connectivity_resilience}, which selects target and control pairs only for qubits that are physically connected.

\paragraph{Noise channels.} The noise channels are implemented and simulated using the \texttt{default.mixed} simulator of \texttt{PennyLane}. This includes all the noise channels that were detailed in \cref{subapp:noise_channels}. It is also straightforward to extend the module with custom noise channels, or directly load hardware properties like $T_1$ and $T_2$ times.

\paragraph{Helper Functions.} In addition to the main components detailed above, the \texttt{VarQEC} module also implements several supplementary functionalities. This includes the logging of training progress, different plotting functionalities, and the possibility to configure hyperparameters and the entire setup via the command line and configuration files.

\medskip
\noindent
Apart from this, we also provide a separate module that allows for the deployment of the trained models on quantum hardware. This experimental module is also available as described in the data availability statement. It implements methods to directly connect to the IBM Quantum experience~\cite{IBMQuantum}, but also the IQM Resonance services~\cite{IQMResonance}.


\section{\label{app:experiments_extended}Extended Empirical Results}

In this appendix, we provide additional empirical evaluations to further substantiate the findings presented in \cref{sec:empirical}. We explore the performance of the proposed \gls{varqec} method under various noise channels, assess its scalability to multiple logical qubits, further examine its stability, and analyze the resilience under restricted connectivity.


\subsection{\label{subapp:additional_noise}Additional Noise Channels}

\begin{figure}[t]
  \centering
  \subfigure[\label{subfig:encoding_bit_flip}Bit-flip noise with noise strength $p=0.1$.]{
    \includegraphics[width=0.47\linewidth]{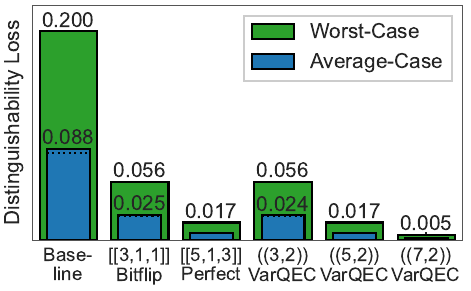}
  }\quad
  \subfigure[\label{subfig:encoding_thermal_relaxation}Thermal relaxation noise of duration $t=10.0$us with $T_1=200$us and $T_2=100$us.]{
    \includegraphics[width=0.47\linewidth]{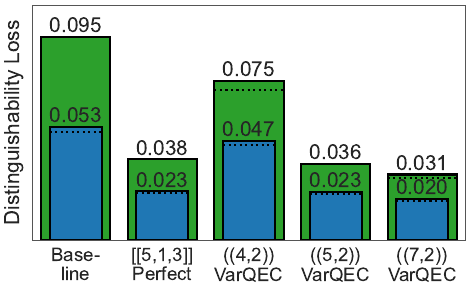}
  }\\
  \subfigure[\label{subfig:encoding_amplitude}Amplitude damping noise with damping parameter $\gamma=0.1$.]{
    \includegraphics[width=0.47\linewidth]{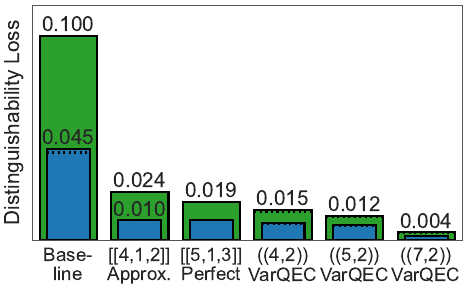}
  }\quad
  \subfigure[\label{subfig:encoding_amplitude_phase}Amplitude and Phase damping noise with damping parameter $\gamma=0.1$.]{
    \includegraphics[width=0.47\linewidth]{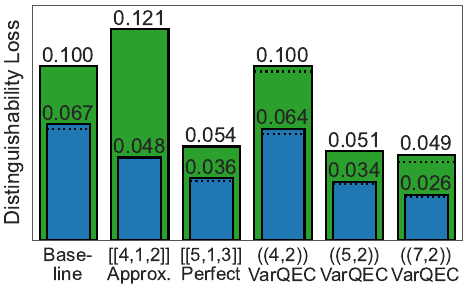}
  }\\
  \caption{\label{fig:encoding_extended}Distinguishability loss for different \gls{qec} codes on (a) bit-flip, (b) thermal relaxation, (c) amplitude damping, and (d) consecutive amplitude and phase damping noise. All bars visualize the average-case loss $\mathcal{D}$ and worst-case loss $\overline{\mathcal{D}}$. Additionally, the two-design approximations $\mathcal{D}_{\mathcal{S}}, \overline{\mathcal{D}}_{\mathcal{S}}$ are visualized as dotted lines if the approximation was not exact. For the noise models, the unencoded baseline, the $[[5,1,3]]$ \emph{perfect} code~\cite{laflamme1996perfect}, optionally the $[[3,1,1]]$ \emph{bit-flip}~\cite{shor1995scheme} and $[[4,1,2]]$ \emph{approximate} \gls{qec} code~\cite{leung1997approximate}, and \gls{varqec} codes with $n=3$ to $n=7$ physical qubit are compared. All \gls{varqec} ansätze are trained for $10$ epochs and contain $(n-1)\cdot(n-2)$ two-qubit blocks. Results on symmetric and asymmetric depolarizing noise can be found in \cref{fig:encoding}.}
\end{figure}

We first extend our analysis to learning encoding operations for additional noise models. The results for bit-flip, thermal relaxation, amplitude damping, and consecutive amplitude and phase damping channels are given in \cref{fig:encoding_extended}. For the bit-flip noise in \cref{subfig:encoding_bit_flip} additional to the $[[5,1,3]]$ perfect code we also consider the $[[3,1,1]]$ bit-flip code~\cite{shor1995scheme}. Confirming our observations from the main part, the $((3,2))$ \gls{varqec} code performs on par with the bit-flip code, and the $((5,2))$ \gls{varqec} code performs on par with the perfect code. Moreover, increasing the number of physical qubits to $n=7$ allows for further reduction of the distinguishability loss. For the thermal relaxation channel in \cref{subfig:encoding_thermal_relaxation}, results are similar to those in depolarizing noise in \cref{subfig:encoding_depolarizing}, but the \gls{varqec} codes manage to have a small edge over the perfect code. A similar connection can be observed for the amplitude and phase damping noise in \cref{subfig:encoding_amplitude_phase}. for isolated amplitude damping noise in \cref{subfig:encoding_amplitude} we additionally compare the $[[4,1,2]]$ \emph{approximate} \gls{qec} code~\cite{leung1997approximate}. Also, here increasing the number of physical qubits in the \gls{varqec} codes significantly reduces the achieved distinguishability loss further.

\begin{figure}[t]
  \centering
  \subfigure[Bitflip noise with noise strength $p=0.1$.]{
    \includegraphics[width=0.47\linewidth]{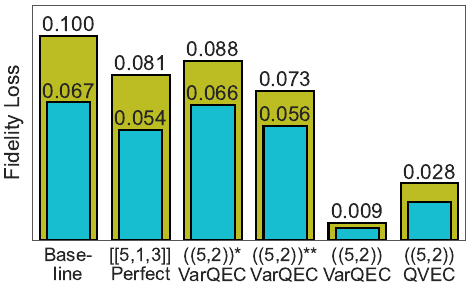}
  }\quad
  \subfigure[Thermal relaxation noise of duration $t=10.0$us with $T_1=200$us, $T_2=100$us.]{
    \includegraphics[width=0.47\linewidth]{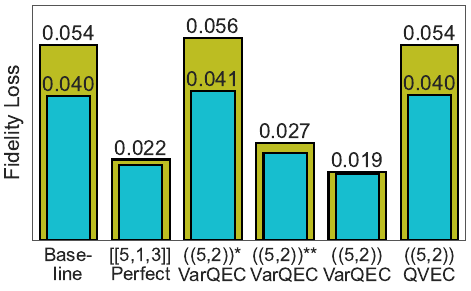}
  }\\
  \subfigure[Amplitude damping noise with damping parameter $\gamma=0.1$.]{
    \includegraphics[width=0.47\linewidth]{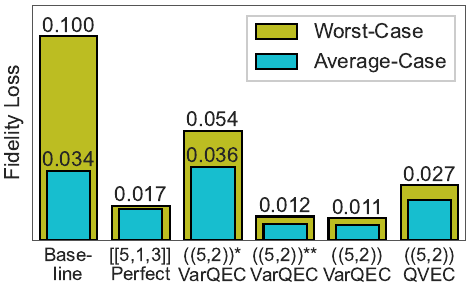}
  }\quad
  \subfigure[Amplitude and Phase damping noise with damping parameter $\gamma=0.1$.]{
    \includegraphics[width=0.47\linewidth]{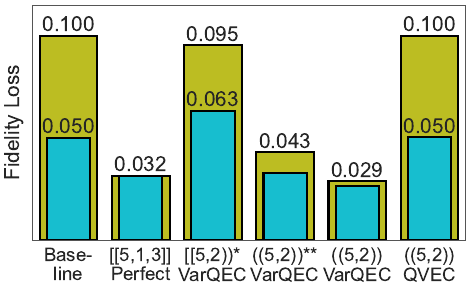}
  }
  \caption{\label{fig:recovery_extended}Fidelity loss for different \gls{qec} codes and respective recovery operations on (a) bit-flip, (b) thermal relaxation, (c) amplitude damping, and (d) consecutive amplitude and phase damping noise. All bars visualize the average-case loss $\mathcal{F}$ and worst-case loss $\overline{\mathcal{F}}$. For the noise models, the unencoded baseline, and the $[[5,1,3]]$ \emph{perfect} code is visualized, optionally extended by the $[[3,1,1]]$ \emph{bit-flip} and $[[4,1,2]]$ \emph{approximate} \gls{qec} code. Additionally, we compare three instances of \gls{varqec} codes, where $((5,2))^{*}$ employs a random encoding, $((5,2))^{**}$ employs a encoding that has converged to a locally optimal distinguishability loss, and the encoding of $((5,2))$ has converged to a global optimum. The recovery operations are successively trained following $\cref{eq:fidelity_loss_simplified}$. For the rightmost model, the encoding and recovery operations have been trained end-to-end with the QVECTOR approach~\cite{johnson2017qvector} as described in \cref{app:learning_recovery}. The recovery ansätze are trained using the fidelity loss for $50$ epochs and contain $200$ two-qubit blocks. As we consider a single-recovery scenario, no separate recovery register was used, i.e.\ the recovery operation acts solely on the $n$-qubit encoded and disturbed state~\cite{johnson2017qvector}. The distinguishability loss for all encodings is summarized in \cref{tab:recovery_dloss_extended}. Results on symmetric and asymmetric depolarizing noise can be found in \cref{fig:recovery}.}
\end{figure}
\begin{table}[t]
    \centering
    \begin{tabular}{cc|cccccc}
         & & Base- & $[[5,1,3]]$ & $((5,2))^*$ & $((5,2))^{**}$ & $((5,2))$ & $((5,2))$ \\
         & & line & Perfect & VarQEC & VarQEC & VarQEC & QVEC
         \\
         \hline
         \multirow{4}{*}{bitflip} & $\overline{\mathcal{D}}=$ & ~~$0.200$~~ & ~~$\pmb{0.017}$~~ & ~~$0.171$~~ & ~~$0.142$~~ & ~~$\pmb{0.017}$~~ & ~~$0.056$~~ \\
         & $\overline{\mathcal{F}} \leq$ & & & ~~$0.168$~~ & ~~$0.263$~~ & ~~$0.034$~~ & \\
         & $\overline{\mathcal{F}} =$ & ~~$0.100$~~ & ~~$0.081$~~ & ~~$0.088$~~ & ~~$0.083$~~ & ~~$\pmb{0.009}$~~ & ~~$0.028$~~ \\
         & $\overline{\mathcal{F}} \geq$ & & & ~~$0.008$~~ & ~~$0.020$~~ & ~~$0.001$~~ & \\
         \hline
         \multirow{4}{*}{thermal relaxation} & $\overline{\mathcal{D}}=$ & ~~$0.095$~~ & ~~$0.038$~~ & ~~$0.095$~~ & ~~$0.050$~~ & ~~$\pmb{0.036}$~~ & ~~$0.095$~~ \\
         & $\overline{\mathcal{F}} \leq$ & & & ~~$0.109$~~ & ~~$0.098$~~ & ~~$0.071$~~ & \\
         & $\overline{\mathcal{F}} =$ & ~~$0.054$~~ & ~~$0.022$~~ & ~~$0.056$~~ & ~~$0.027$~~ & ~~$\pmb{0.019}$~~ & ~~$0.054$~~ \\
         & $\overline{\mathcal{F}} \geq$ & & & ~~$0.003$~~ & ~~$0.003$~~ & ~~$0.001$~~ & \\
         \hline
         \multirow{4}{*}{amplitude damping} & $\overline{\mathcal{D}}=$ & ~~$0.100$~~ & ~~$0.019$~~ & ~~$0.080$~~ & ~~$0.020$~~ & ~~$\pmb{0.012}$~~ & ~~$0.043$~~ \\
         & $\overline{\mathcal{F}} \leq$ & & & ~~$0.104$~~ & ~~$0.040$~~ & ~~$0.024$~~ & \\
         & $\overline{\mathcal{F}} =$ & ~~$0.100$~~ & ~~$0.017$~~ & ~~$0.054$~~ & ~~$0.012$~~ & ~~$\pmb{0.011}$~~ & ~~$0.027$~~ \\
         & $\overline{\mathcal{F}} \geq$ & & & ~~$0.003$~~ & ~~$0.001$~~ & ~~$0.001$~~ & \\
         \hline
         & $\overline{\mathcal{D}}=$ & $0.100$ & $0.054$ & $0.154$ & $0.084$ & $\pmb{0.051}$ & $0.100$ \\ 
         amplitude-phase & $\overline{\mathcal{F}} \leq$ & & & ~~$0.182$~~ & ~~$0.161$~~ & ~~$0.100$~~ & \\
         damping & $\overline{\mathcal{F}} =$ & ~~$0.100$~~ & ~~$0.032$~~ & ~~$0.095$~~ & ~~$0.043$~~ & ~~$\pmb{0.029}$~~ & ~~$0.100$~~ \\
         & $\overline{\mathcal{F}} \geq$ & & & ~~$0.009$~~ & ~~$0.007$~~ & ~~$0.003$~~ & \\
    \end{tabular}
    \caption{    \label{tab:recovery_dloss_extended}Worst-case distinguishability loss $\overline{\mathcal{D}}$ exhibited by the encodings for the models in \cref{fig:recovery_extended}. For the baseline and $[[5,1,3]]$ code, this is evaluated on the unencoded baseline and static encoding operation, respectively. For the \gls{varqec} codes, this corresponds to an untrained encoding for $((5,2))^{*}$, to an local optimum for $((5,2))^{**}$, and to a global optimum for $((5,2))$. For the $((5,2))$ QVECTOR code, the same ansatz with $12$ two-qubit blocks as for the \gls{varqec} codes is employed, but training was only implicitly conducted via the fidelity loss. Additionally, we again report the worst-case fidelity loss $\overline{\mathcal{F}}$ (third line) and the respective upper (second) and lower bounds (fourth line) following \cref{the:lower_bound,the:upper_bound} for the \gls{varqec} models.}
\end{table}

\begin{figure}[t]
  \centering
  \subfigure[\label{subfig:encoding_depolarizing_multi}Depolarizing noise with noise strength $p=0.1$.]{
    \includegraphics[width=0.47\linewidth]{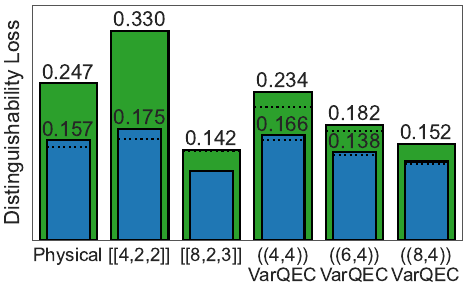}
  }\quad
  \subfigure[\label{subfig:encoding_asymmetric_depolarizing_multi}Asymmetric depolarizing noise with noise strength $p=0.1$ and asymmetry $c=0.5$ following \cref{eq:noise_asymmetry}.]{
    \includegraphics[width=0.47\linewidth]{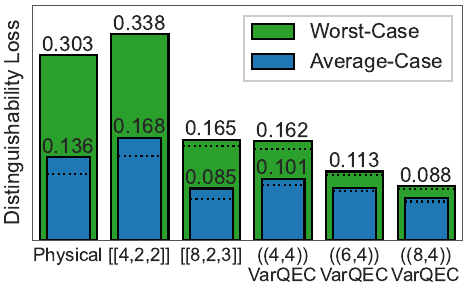}
  }
  \caption{\label{fig:multi_qubit}Distinguishability loss for different \gls{qec} codes with two logical qubits, i.e.\ $k=2$, on (a) depolarizing and (b) asymmetric depolarizing noise. All bars visualize the average-case loss $\mathcal{D}$ and worst-case loss $\overline{\mathcal{D}}$. Additionally, the two-design approximations $\mathcal{D}_{\mathcal{S}}, \overline{\mathcal{D}}_{\mathcal{S}}$ are visualized as dotted lines if the approximation was not exact. For both noise models, the unencoded baseline, the $[[4,2,2]]$ \gls{qec} code~\cite{vaidman1996error}, the $[[8,2,3]]$ \gls{qec} code~\cite{olle2024simultaneous}, and \gls{varqec} codes with $n=4$, $n=6$, and $n=8$ physical qubits are compared. All \gls{varqec} ansätze are trained for $10$ epochs and contain $n\cdot(n-1)$ two-qubit blocks. Results on only a single data qubit can be found in \cref{fig:encoding}.}
\end{figure}

Building upon this, we train respective recovery operations and evaluate the fidelity losses in \cref{fig:recovery_extended}. The distinguishability losses of the underlying encoding circuits and the implied bounds are summarized in \cref{tab:recovery_dloss_extended}. In agreement with the results from the main part, we can observe that a lower distinguishability loss allows for a higher-fidelity recovery operation. On all considered instances, we can observe that the optimal \gls{varqec} codes outperform the unencoded baseline, the $[[5,1,3]]$ perfect code, and also the $((5,2))$ code trained with the QVECTOR approach. In \cref{subapp:dloss_code_distance} we determine the $((5,2))$ \gls{varqec} code trained on amplitude damping noise to exhibit a potential approximate code distance of $d=2$. This aligns with the code distance of the $[[4,1,2]]$ approximate \gls{qec} code, which is known to be especially suited for the amplitude damping channel~\cite{leung1997approximate}. We again want to emphasize that the QVECTOR approach employs the same circuit ansätze as the \gls{varqec} codes. For bit-flip and amplitude damping noise, the $((5,2))$ QVECTOR codes also significantly improve upon the unencoded baseline performance, which is in agreement with the results from Ref.~\cite{johnson2017qvector}. However, already on these and especially on the more complex thermal relaxation noise channel, this end-to-end ansatz produces results that are inferior to the \gls{varqec} procedure.


\subsection{\label{subapp:multiple_qubits}Multiple Data Qubits}

We now assess the extension of the previous experiments to the setup with multi-qubit logical states, in particular $k=2$. For this two-qubit setup, we enumerate elements of a two-design in \cref{subapp:dloss_two_design}. In \cref{fig:multi_qubit}, we report the distinguishability loss for a variety of such setups on both symmetric and asymmetric depolarizing noise. We include the unencoded baseline, which in this context corresponds to a two-qubit state. We additionally compare to the $[[4,2,2]]$ \gls{qec} code~\cite{vaidman1996error} and the $[[8,2,3]]$ \gls{qec} code~\cite{olle2024simultaneous}. Using our \gls{varqec} routine, we train codes with sizes from $n=4$ to $n=8$ physical qubits.

The results in on the symmetric depolarizing channel in \cref{subfig:encoding_depolarizing_multi} indicate that for this setup, the $((8,4))$ \gls{varqec} are not fully competitive with the $[[8,2,3]]$ \gls{qec} code, which, similar to the perfect code on a single logical qubit, is the smallest code with distance $d=3$ on two logical qubits. However, we assume this gap could be eliminated by resorting to a more sophisticated ansatz construction. On the contrary, the results on asymmetric depolarizing noise in \cref{subfig:encoding_asymmetric_depolarizing_multi} demonstrate that the \gls{varqec} codes reduce the distinguishability loss compared to the \emph{standard} \gls{qec} codes.

We recall that in \gls{qec} it is standard to encode only a few logical qubits in a single code patch. Surface codes typically store one logical qubit per patch~\cite{fowler2012surface}, while modern \gls{qldpc} codes can encode around a dozen logical qubits~\cite{bravyi2024high}. Fault-tolerant interactions between such patches are implemented using protocols like lattice surgery~\cite{horsman2012surface}. Developing analogous protocols and operations for the \gls{varqec} procedure is left for future work.


\begin{figure}[t]
  \centering
  \subfigure[\label{subfig:stability_strength_depolarizing}Depolarizing noise with noise strength $p=0.1$.]{
    \includegraphics[width=0.47\linewidth]{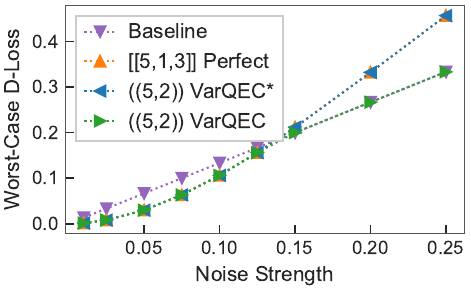}
  }\quad
  \subfigure[\label{subfig:stability_strength_asym_depolarizing}Asymmetric depolarizing noise with noise strength $p=0.1$ and asymmetry $c=0.5$ following \cref{eq:noise_asymmetry}.]{
    \includegraphics[width=0.47\linewidth]{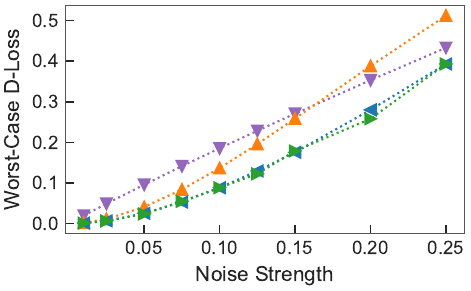}
  }
  \caption{\label{fig:stability_strength}Stability of standard \gls{qec} and \gls{varqec} codes with increasing noise strength for (a) symmetric and (b) asymmetric depolarizing noise. All data points represent the respective worst-case distinguishability loss $\overline{\mathcal{D}}$. The $((5,2))^*$ \gls{varqec} code is always the same \emph{static} one trained on a noise strength $p=0.1$. The other $((5,2))$ \gls{varqec} code is \emph{dynamically} trained on the respective denoted noise configuration.}
\end{figure}

\begin{figure}[t]
  \centering
  \subfigure[\label{subfig:stability_noise_strength}Depolarizing noise with noise strength $p=0.1$ on $n-1$ of the qubits, and with a noise strength as denoted on the $x$-axis on the remaining qubit. For the \gls{qec} codes, the deviating qubit is associated with the original data qubit, while the unencoded baseline employs the best available wire.]{
    \includegraphics[width=0.47\linewidth]{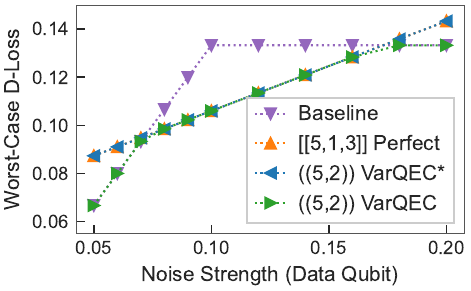}
  }\quad
  \subfigure[\label{subfig:stabilty_type}Transfer of \gls{varqec} code with $n=5$ physical qubits trained on depolarizing noise with strength $p=0.1$ to other noise types. The bitflip noise also features a noise strength of $p=0.1$, while phase and amplitude damping both have a damping parameter of $\gamma=0.1$.]{
    \includegraphics[width=0.47\linewidth]{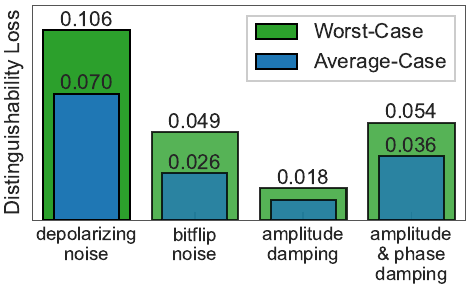}
  }
  \caption{Stability of standard \gls{qec} and \gls{varqec} codes in setups of varying (a) noise strength on different qubits and (b) noise channels. All data points represent the respective worst-case distinguishability loss $\overline{\mathcal{D}}$ and optionally average-case loss $\mathcal{D}$. The $((5,2))^*$ \gls{varqec} code is always the same static one trained on depolarizing noise with noise strength $p=0.1$ on all qubits. The other $((5,2))$ \gls{varqec} code is dynamically trained on the respective noise configuration.}
  \label{fig:stability_extended}
\end{figure}

\subsection{\label{subapp:extended_stability}Extended Stability Analysis}

We now analyze the stability of the \gls{varqec} codes and the adaptability of the \gls{varqec} procedure. Therefore, for both symmetric and asymmetric noise, we vary the noise strength from $p=0.05$ to $p=0.25$. Similar to the main part, we compare the unencoded baseline, the $[[5,1,3]]$ perfect code, a static \gls{varqec} code trained on noise strength of $p=0.1$, and \gls{varqec} codes dynamically trained on the respective noise strengths. For the depolarizing noise in \cref{subfig:stability_strength_depolarizing}, we observe that all \gls{qec} codes perform equally well up to the physical noise threshold of about $p=0.15$. After that, both the $[[5,1,3]]$ code and static \gls{varqec} code perform worse than just a physical qubit without encoding. The dynamically trained \gls{varqec} code at least performs on par with the purely physical setup, but also can not improve upon it. For the asymmetric depolarizing noise in \cref{subfig:stability_strength_asym_depolarizing}, the $[[5,1,3]]$ code requires a physical threshold of again approximately $p=0.15$. However, both \gls{varqec} codes can significantly improve upon this and still work for $p=0.25$. While dynamically tailoring the \gls{varqec} code brings slight advantages for some configurations, the differences are not significant.

We extend this analysis by evaluating the scenario of non-uniform noise strength between the involved qubits in \cref{subfig:stability_noise_strength}, which is typically resembled in actual hardware due to manufacturing quality~\cite{place2021new}. More concretely, we vary the strength of depolarizing noise on the qubit corresponding to the initial data wire from values $p=0.05$ to $p=0.2$. All other involved ancilla qubits exhibit a depolarizing noise of $p=0.1$. To compare to a harder baseline, for the case of a single physical qubit, we select the lower noise strength. As expected, the $[[5,1,3]]$ perfect code outperforms the purely physical setup up until a threshold noise strength of approximately $0.175$. The \emph{static} version of the $((5,2))$ \gls{varqec} code performs on par for all setups. The $((5,2))$ \gls{varqec} \emph{dynamically} trained on the respective noise configurations does not drop beyond the baseline, even above the physical threshold.

In \cref{subfig:stabilty_type} we perform some transferability analysis of the $((5,2))$ \gls{varqec} code that was trained on depolarizing noise of strength $p=0.1$. As evaluated in \cref{subapp:dloss_code_distance}, this code exhibits a potential code distance of $d=3$, and therefore also should apply to other noise channels. Indeed, the transferred code exhibits distinguishability losses that are significantly beyond the unencoded baselines (compare \cref{fig:encoding_extended}). Moreover, for amplitude damping and combined amplitude and phase damping noise, the results are again competitive with the $[[5,1,3]]$ perfect code. We have to highlight that one can further improve upon this by tailoring the codes explicitly to the respective noise channels. However, these results still demonstrate that the learned \gls{varqec} codes partially generalize to other noise models. This might allow for some warm-start parameter initialization approaches~\cite{meyer2024warm}, indicating that models do not always need to be trained from scratch but rather only finetuned.


\subsection{\label{subapp:restricted_connectivity_resilience}Restricting Connectivity Structure and Fault Resilience}

\begin{figure}[t]
    \centering
    \subfigure[``dense'' layout]{
        \includegraphics[width=0.16\textwidth]{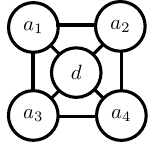}
    }\qquad
    \subfigure[``square'' layout]{
        \includegraphics[width=0.2\textwidth]{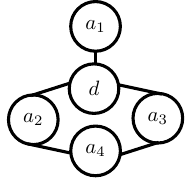}
    }\qquad
    \subfigure[``star'' layout]{
         \includegraphics[width=0.16\textwidth]{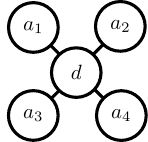}
    }\qquad
    \subfigure[``hexagonal'' layout]{
         \includegraphics[width=0.2\textwidth]{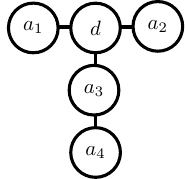}
    }
    \caption{\label{fig:layout}Different layouts for $n=5$ physical qubits satisfying the connectivity restrictions of some potential underlying hardware. Throughout most of this paper, we assumed a ``full'' connectivity, which is not available on most hardware systems. Similarly, the ``dense'' layout might be hard to realize with on most platforms. The ``square'' as well as the ``star'' layout aligns e.g.\ with the grid-based topology of the Google Quantum AI chips~\cite{acharya2024quantum}. The ``hexagonal'' layout is adapted to the \emph{heavy-hex} topology of IBM Quantum devices~\cite{IBMQuantum}.}
\end{figure}

Most quantum hardware platforms do not support all-to-all connectivity, but rather assume some restricted topology. In principle, it is possible to compile a quantum circuit to the respective layout, which, however, typically requires routing with e.g.\ swap gates. We can entirely avoid this for the \gls{varqec} procedure by initially selecting a variational ansatz that is adapted to the hardware connectivity restrictions. Therefore, we now consider the four different layouts in \cref{fig:layout}, i.e., \ ``dense'', ``square'', ``star'', and ``hexagonal''. The \glspl{rea} are configured to inherently apply to these restrictions.

\begin{table}[t]
    \centering
    \begin{tabular}{c|ccccc}
         & \multicolumn{5}{c}{layout} \\
         & ~~~~~``full''~~~~~ & ~~~~``dense''~~~~ & ~~~``square''~~~ & ~~~~~``star''~~~~~ & ``hexagonal'' \\
         \hline
         $[[5,1,3]]$ Perfect & $6$ & $9$ & $15$ & $12$ & $18$ \\
         $((5,2))$ VarQEC & $12$ & $10$ & $10$ & $12$ & $16$ \\
    \end{tabular}
    \caption{\label{tab:layout_depolarizing}Number of two-qubit gates for $[[5,1,3]]$ perfect code compiled to the different layouts in \cref{fig:layout} and smallest number of two-qubit blocks for which a $((5,2))$ \gls{varqec} code achieves the optimal worst-case distinguishability loss of $\overline{\mathcal{D}} \approx 0.106$ for depolarizing noise with $p=0.1$.}
\end{table}

For the results in \cref{tab:layout_depolarizing}, we trained $((5,2))$ \gls{varqec} codes with $6$ to $20$ two-qubit blocks on depolarizing noise. For each layout, we report the respective smallest code instance that produces the optimal distinguishability loss of $\overline{\mathcal{D}}\approx0.106$. In addition to that, we compile the $[[5,1,3]]$ perfect encoding circuit~\cite{xu2021variational} for the respective layout, using the \texttt{qiskit.transpile} method~\cite{javadi2024quantum}, and report the respective number of two-qubit gates. This significantly increases the circuit depth, while for the \gls{varqec} codes, the circuit complexity only needs to increase slightly with lower connectivity density.

\begin{table}[t]
    \centering
    \begin{tabular}{c|cccccccc}
         & \multicolumn{8}{c}{number of blocks} \\
         & 6 & 8 & 10 & 12 & 14 & 16 & 18 & 20 \\
         \hline
         ``Full'' layout & $0.0615$ & $0.0459$ & $0.0366$ & $\pmb{0.0349}$ & $\pmb{0.0348}$ & $\pmb{0.0347}$ & $\pmb{0.0347}$ & $\pmb{0.0348}$ \\
         ``Dense'' layout & $0.0573$ & $0.0363$ & $0.0352$ & $0.0359$ & $0.0358$ & $\pmb{0.0348}$ & $\pmb{0.0347}$ & $\pmb{0.0348}$ \\
         ``Square'' layout & $0.0619$ & $0.0502$ & $0.0359$ & $\pmb{0.0347}$ & $0.0366$ & $\pmb{0.0348}$ & $\pmb{0.0350}$ & $\pmb{0.0348}$ \\
         ``Star'' layout & $0.0611$ & $0.0453$ & $0.0359$ & $0.0359$ & $0.0359$ & $\pmb{0.0348}$ & $\pmb{0.0347}$ & $\pmb{0.0348}$ \\
         ``Hexagonal'' layout & $0.0502$ & $0.0502$ & $0.0367$ & $0.0383$ & $0.0359$ & $0.0367$ & $0.0354$ & $\pmb{0.0347}$ \\ 
    \end{tabular}
    \caption{Worst-case distinguishability loss $\overline{\mathcal{D}}$ for thermal relaxation noise with duration $t=10.0\mu\text{s}$ with $T_1=200\mu\text{s}$ and $T_2=100\mu\text{s}$ using an increasing number of two-qubit blocks in $((5,2))$ \gls{varqec} encoding circuits.}
    \label{tab:layout_thermal_relaxation}
\end{table}

\begin{figure}[t]
  \centering
  \includegraphics[width=\linewidth]{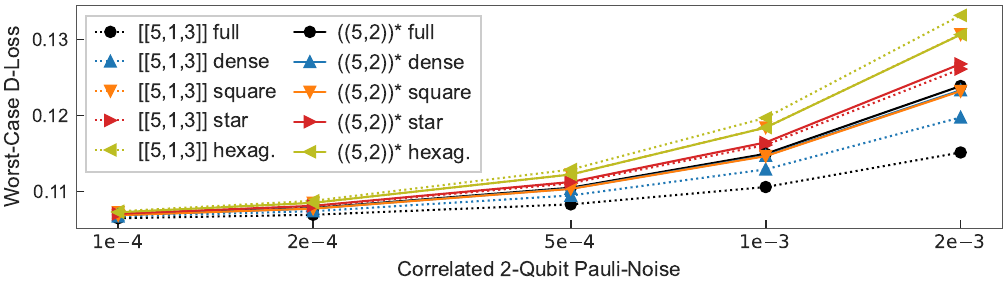}
  \caption{  \label{fig:resilience_extended}Resilience of standard \gls{qec} and \gls{varqec} codes in setups of non-ideal encoding operations under symmetric depolarizing noise of strength $p=0.1$. All data points represent the respective worst-case distinguishability loss $\overline{\mathcal{D}}$. The solid lines represent static \gls{varqec} codes trained under the assumption of ideal encoding operations, adapted to the respective layouts. The dashed lines correspond to the $[[5,1,3]]$ \gls{qec} code, compiled for the respective connectivity structure. The unencoded baseline, which does not include any two-qubit operations, is at approximately $0.133$. The encoding circuit depths of the respective codes are summarized in \cref{tab:layout_depolarizing}.}
\end{figure}

For realistically modeling idle qubits, one typically resorts to the thermal relaxation noise channel. Therefore, we repeat the above analysis on the \gls{varqec} circuits for this setup with $T_1=200\mu\text{s}$, $T_2=100\mu\text{s}$, $t=10.0\mu\text{s}$, and report results in \cref{tab:layout_thermal_relaxation}. The results underline that the \gls{varqec} can easily adapt to sparse connectivity, with just a slight increase in circuit depth. 

Last but not least, we consider the noise-resilience of the encodings under the assumption of restricted connectivity \cref{fig:resilience_extended}. Similar to the main part, we evaluate this by adding correlated depolarizing noise of increasing strength after each two-qubit gate. We consider symmetric depolarizing noise, i.e.\ under the assumption of ideal encoding operations, the $[[5,1,3]]$ code is optimal. For ``full'', ``dense'', and ``'square' connectivity layouts, the compiled codes are more resilient than the \gls{varqec} codes, mainly due to the circuits being shallower. However, the \gls{varqec} codes shine under sparse connectivity structures like especially ``hexagonal'', where this trend reverses. We want to emphasize that, e.g.\ for asymmetric depolarizing noise, the \gls{varqec} codes would exhibit superior noise-resilience on all layouts, as already under full-connectivity assumptions, there was quite some performance gap as analyzed in~\cref{subfig:stability_noise_resilience}.

\medskip
\noindent
This extended analysis of the \gls{varqec} procedure reinforces the conclusion from the main part, in particular that the \gls{varqec} codes are adaptable to a wide range of noise models. The codes themselves, furthermore, are stable over a wide range of noise fluctuations. Moreover, the approach itself is perfectly suited for developing codes under restricted connectivity.



\section{\label{app:hardware_experiments}Closer Look at the Hardware Experiments}

In this appendix, we provide additional details on the hardware experiments conducted to validate the practical applicability of our proposed \gls{varqec} codes. By deploying our models on different quantum computing platforms, we aim to demonstrate the adaptability and resilience of our approach under realistic conditions.


\subsection{\label{subapp:hardware_setup}General Experimental Setup}

\begin{figure}[t]
  \centering
  \subfigure[\label{subfig:experiment_simulation_ibmq}Thermal relaxation noise with $T_1=180\mu\text{s}$ and $T_2=120\mu\text{s}$ for $t=10.0\mu\text{s}$ as encountered in the median on the \texttt{ibmq\_marrakesh} device.]{
    \includegraphics[width=0.47\linewidth]{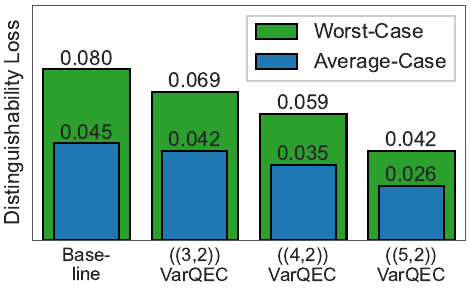}
  }\quad
  \subfigure[Thermal relaxation noise with $T_1=48\mu\text{s}$ and $T_2=20\mu\text{s}$ for $t=2.0\mu\text{s}$ as encountered in the median on the \texttt{IQM Garnet} device.]{
    \includegraphics[width=0.47\linewidth]{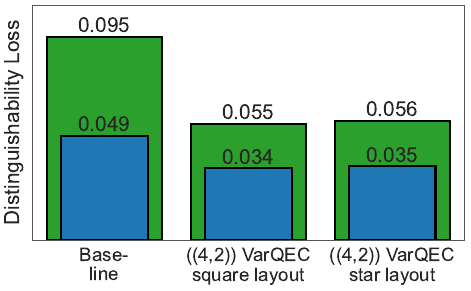}
  }
  \caption{  \label{fig:experiment_simulation}Distinguishability loss for different \gls{qec} codes on thermal relaxation noise modeling the median error properties for (a) an IBMQ device and (b) an IQM device. All bars visualize the average-case loss $\mathcal{D}$ and worst-case loss $\overline{\mathcal{D}}$. For the IBMQ noise model, we compare the unencoded baseline and \gls{varqec} codes with $n=3$ to $n=5$ physical qubits. The $((4,2))$ \gls{varqec} code for this setup is plotted in \cref{fig:ansatz_ibmq}. For the IQM noise model, we compare the unencoded baseline to the results of $((4,2))$ \gls{varqec} codes with different layout alignments. The codes are trained in classical simulation and afterwards deployed on quantum hardware, as denoted by the experiments in this section.}
\end{figure}

We utilized the \gls{varqec} framework for training the codes in simulation. We provide a separate implementation for submitting the results to IBM and IQM hardware, as well as for retrieving and analyzing the results, as indicated in the data availability statement. Our implementation builds upon the \texttt{qiskit-experiments} module~\cite{kanazawa2023qiskit}, which we modified to support cluster re-sampling~\cite{cameron2008bootstrap,meyer2025benchmarking} of the data. The quantum states were reconstructed using mitigated state tomography, offering a proof-of-concept demonstration of our approach despite the scalability limitations of state tomography.

For the IBM Quantum device, we simulated idling by adding \texttt{qiskit.delay} operations of specific durations to the involved qubits. On the IQM device, which does not natively support delay operations, we approximated idling by adding several identity gates equivalent to the desired delay, assuming that each single-gate execution takes approximately $30.0n\text{s}$~\cite{abdurakhimov2024technology}. To efficiently utilize hardware resources and reduce access costs, we executed multiple logical patches in parallel. Specifically, we deployed $8$ logical patches on the $156$-qubit IBM device and $3$ logical patches on the 20-qubit IQM device. State tomography was also executed in parallel, and we verified that this parallel execution did not significantly degrade the quality of the results compared to sequential execution.

\begin{figure}[t]
  \centering
  \subfigure[Alignment of physical qubits to logical patches following the ``star'' shape. For the four-qubit setup, this is also compatible with the ``hexagonal'' layout.]{
    \includegraphics[width=\linewidth]{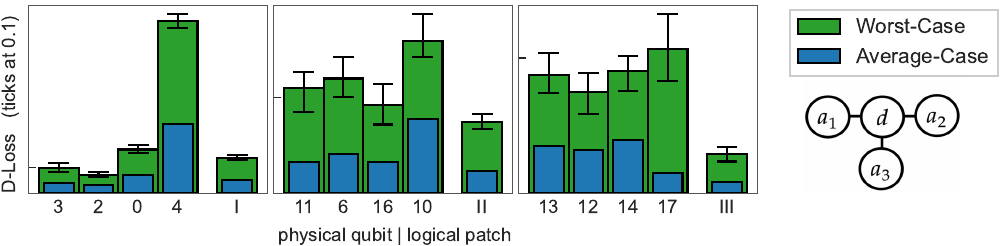}
  }\\
  \subfigure[Alignment of physical qubits to logical patches following the ``square'' shape.]{
    \includegraphics[width=\linewidth]{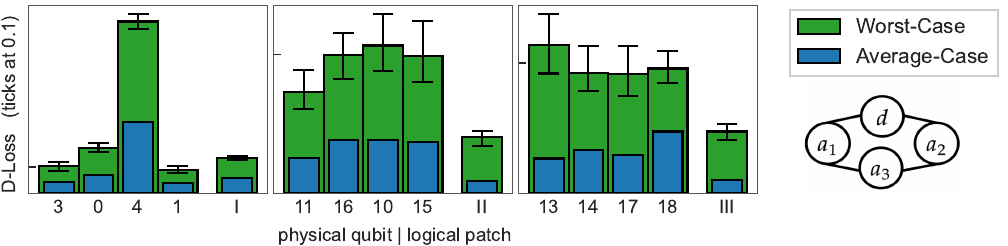}
  }
  \caption{\label{fig:experiment_iqm}Reconstructed distinguishability loss for the $3$ patches of the experiment with the $((4,2))$ \gls{varqec} code on the \texttt{IQM Garnet} device. The qubits have been idled for $1.0\mu\text{s}$ for this experiment. We consider two different physical qubit to logical patch alignment strategies, in particular (a) ``star'' and (b) ``square'' alignment.
  For each patch, the first four bars indicate the unencoded baseline losses for the respective physical qubits, while the fifth one denotes the loss for the entire encoded patch. The losses were estimated with mitigated state tomography using $1000$ shots per circuit execution, and the error bars denote 5th and 95th percentile ranges estimated by cluster re-sampling.}
\end{figure}

We compared the hardware results to the training predictions on the distinguishability loss, as depicted in \cref{fig:experiment_simulation}. While the hardware measurements reported in \cref{fig:experiment_ibmq,fig:experiment_iqm} exhibit quantitatively higher distinguishability loss than the simulation predictions, they still demonstrate beyond break-even performance as we discuss in the following, corroborating the observations from the main text in \cref{sec:hardware}.


\subsection{\label{subapp:hardware_ibmq}Extended Results on IBMQ Devices}

\begin{figure}[t]
  \centering
  \includegraphics[width=\linewidth]{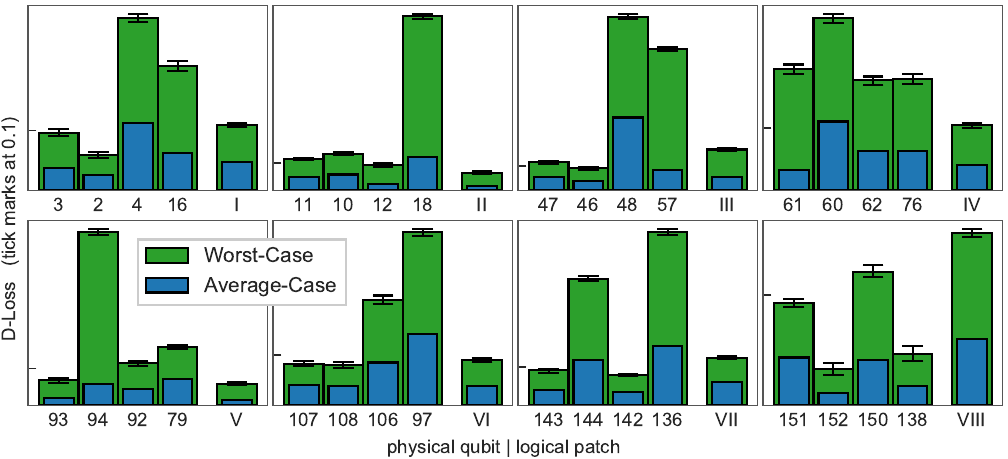}
  \caption{\label{fig:experiment_ibmq}Reconstructed distinguishability loss for the $8$ patches of the experiment with the $((4,2))$ \gls{varqec} code on the \texttt{ibm\_marrakesh} device. The qubits have been idled for $5.0\mu\text{s}$ for this experiment. For each patch, the first four bars indicate the unencoded baseline losses for the respective physical qubits, while the fifth one denotes the loss for the entire encoded patch. The losses were estimated with mitigated state tomography using $10000$ shots per circuit execution, and the error bars denote 5th and 95th percentile ranges estimated by cluster re-sampling. The alignment and layout of physical wires to logical patch is also summarized in \cref{tab:alignment_ibmq}, and the circuit realizing the encoding is given in \cref{fig:ansatz_ibmq}.}
\end{figure}

\begin{table}[t]
    \centering
    \begin{tabular}{cc|cccccccccc}
         & & \multicolumn{8}{c}{logical patch} & & \multirow{7}{*}{  
            \tikzset{every picture/.style={line width=0.75pt}} 
            \begin{tikzpicture}[x=0.9pt,y=0.9pt,yscale=-0.7,xscale=.7]
            \draw [line width=1.5]    (126.2,50.6) -- (134.2,50.6) ;
            \draw [line width=1.5]    (166.2,50.6) -- (174.2,50.6) ;
            \draw [color={rgb, 255:red, 74; green, 74; blue, 74 }  ,draw opacity=1 ][line width=1.5]    (150.2,67.6) -- (150.2,75.6) ;
            \draw [color={rgb, 255:red, 128; green, 128; blue, 128 }  ,draw opacity=1 ][line width=1.5]    (150.2,106.6) -- (150.2,114.6) ;
            \draw  [line width=1.5]   (150.48, 51.2) circle [x radius= 15.91, y radius= 15.91]   ;
            \draw (150.48,51.2) node    {$\ d\ $};
            \draw  [line width=1.5]   (190.5, 50.19) circle [x radius= 15.95, y radius= 15.95]   ;
            \draw (190.5,50.19) node    {$a_{2}$};
            \draw  [line width=1.5]   (110.5, 51.19) circle [x radius= 15.95, y radius= 15.95]   ;
            \draw (110.5,51.19) node    {$a_{1}$};
            \draw  [color={rgb, 255:red, 74; green, 74; blue, 74 }  ,draw opacity=1 ][line width=1.5]   (149.5, 91.19) circle [x radius= 15.95, y radius= 15.95]   ;
            \draw (149.5,91.19) node  [color={rgb, 255:red, 74; green, 74; blue, 74 }  ,opacity=1 ]  {$a_{3}$};
            \draw  [color={rgb, 255:red, 128; green, 128; blue, 128 }  ,draw opacity=1 ][line width=1.5]   (150.5, 131.19) circle [x radius= 15.95, y radius= 15.95]   ;
            \draw (150.5,131.19) node  [color={rgb, 255:red, 128; green, 128; blue, 128 }  ,opacity=1 ]  {$a_{4}$};
            \end{tikzpicture}
         } \\
         & & I & II & III & IV & V & VI & VII & VIII \\
         \cline{1-10}
         & $d$ & 3 & 11 & 47 & 61 & 93 & 107 & 143 & 151 \\
         & $a_1$ & 2 & 10 & 46 & 60 & 94 & 108 & 144 & 152 \\
         wire & $a_2$ & 4 & 12 & 48 & 62 & 92 & 106 & 142 & 150 \\
         & $a_3$ & 16 & 18 & 57 & 76 & 79 & 97 & 136 & 138 \\
         & $a_4$ & ~23~ & ~31~ & ~67~ & ~81~ & ~73~ & ~87~ & 123 & 131
    \end{tabular}
    \caption{Physical wire allocation of the encoding patches for the experiment on the IBMQ device \texttt{ibm\_marrakesh} depicted in \cref{fig:experiment_ibmq}. The device uses a \texttt{Heron r2} chip with 156 qubits, the physical wire indices refer to the numbering in the IBMQ documentation~\cite{IBMQuantum}. The connectivity is adapted to the hexagonal connectivity structure, each single 5-qubit encoding patch is allocated as depicted on the right. Hereby, $d$ denotes the initial physical data qubit to be protected, and $a_1$, $a_2$, $a_3$, and $a_4$ are ancilla qubits. For the experiments with encodings on 4 and 3 wires, the ancilla qubits $a_4$ and $a_3$ are dropped, respectively.}
    \label{tab:alignment_ibmq}
\end{table}

\begin{figure}[t]
  \centering
  \includegraphics[width=\linewidth]{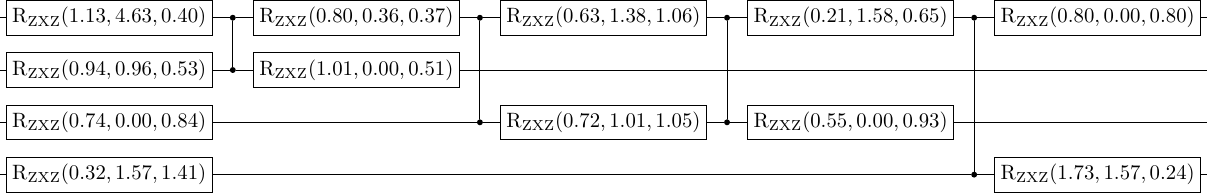}
  \caption{\label{fig:ansatz_ibmq}Circuit ansatz implementing the $((4,2))$ \gls{varqec} code. The parameters were trained on the noise setup for the \texttt{ibm\_marrakesh} experiment that is analyzed in \cref{fig:experiment_ibmq}. The circuit is an instance of a \gls{rea} (see \cref{subapp:randomized_ansatz}) restricted to the ``hexagonal'' four-qubit layout with the basis gate set of the IBM devices. It contains four two-qubit blocks, with overall $36$ parameters. The uppermost wire corresponds to the initial data qubit.}
\end{figure}

We conducted experiments on the \texttt{ibm\_marrakesh} device~\cite{IBMQuantum}, which features a $156$-qubit \texttt{Heron r2} chip. The median properties of the qubits were identified as $T_1=180\mu\text{s}$ and $T_2=120\mu\text{s}$. Due to additional sources of error, such as noise introduced during the encoding operations and measurement delays, we trained the encoding with a noise duration of $t=10.0\mu\text{s}$ and deployed it on the hardware for $t=5.0\mu\text{s}$ (compare also \cref{sec:hardware}).

We trained and deployed \gls{varqec} codes with $n=3$ to $n=5$ physical qubits. The experiments required approximately $12$ QPU minutes for the $3$-qubit code, $25$ QPU minutes for the $4$-qubit code, and $72$ QPU minutes for the $5$-qubit code, accessed via the IBM Quantum premium plan. The logical qubit allocations for the $((3,2))$, $((4,2))$ and $((5,2))$ \gls{varqec} codes are summarized in \cref{tab:alignment_ibmq}. The gates used in our ansätze were aligned with the native gate set of the device, employing CZ gates for two-qubit operations and parameterized single-qubit rotations implemented as $R_{zxz}(\theta,\phi,\xi) = R_z(\xi)R_x(\phi)R_z(\theta)$. The results, depicted in \cref{subfig:experiment_wires} of the main text, indicate a decrease in the distinguishability loss with an increasing number of physical qubits, thus demonstrating improved error correction performance with larger codes.

Detailed results for the $8$ individual patches of the $((4,2))$ \gls{varqec} code are provided in \cref{fig:experiment_ibmq}. We note both worst-case and average-case distinguishability losses, with error bars indicating the 5th and 95th percentile ranges evaluated by cluster re-sampling. The experiments were conducted with $10000$ shots for each circuit in the mitigated state tomography process. The concrete ansatz for the $((4,2))$ \gls{varqec} code, consisting of four two-qubit blocks, is illustrated in \cref{fig:ansatz_ibmq}. We observed that while the \gls{varqec} code does not outperform the best physical qubit in every individual patch, especially in patches with significant qubit quality fluctuations, the overall results indicate a reduction in the distinguishability loss on average. This is particularly pronounced in patches where the qubit quality is more uniform (e.g., patch IV). In some patches, the logical results did not outperform the best physical qubit, which we attribute to variances in qubit quality and, in one case, a defective two-qubit connection (patch VIII).


\subsection{\label{subapp:hardware_iqm}Results on IQM Devices}

We also deployed our VarQEC codes on the $20$-qubit \texttt{IQM Garnet} device with a crystal topology, accessed via the IQM Resonance services~\cite{IQMResonance}. The median coherence times were $T_1=48\mu\text{s}$ and $T_2=20\mu\text{s}$~\cite{abdurakhimov2024technology}. Given these limited coherence times, we reduced the idling time to $t=2.0\mu\text{s}$ during training and further to $t=1.0\mu\text{s}$ for hardware deployment. Due to access constraints, we limited the number of shots per circuit to $1000$. The ansatz gates were adapted to the native gate set of the device, utilizing CZ gates for two-qubit operations and parameterized single-qubit gates implemented as $PR_x(\theta,\phi) = R_z(\phi)R_x(\theta)R_z(-\phi)$.

The denser grid connectivity of the IQM device allowed us to experiment with different ansatz topologies, specifically the ``star'' and ``square'' layouts. However, the results do not show a significant difference between the two layouts, which we expect to change for larger codes, as analyzed in \cref{subapp:restricted_connectivity_resilience}. The increased span of the error bars can be attributed to the reduction of the shot count by a factor of $10$. We observed that in both layouts, qubit $4$ was defective during our experiments, leading to outlier results in patch I. In patches II and III, we achieved a clear net gain in performance, as the logical distinguishability loss was significantly lower than that of the best physical qubit. The smaller variance in qubit quality on the IQM device facilitated better evaluation of our code's functionality, given that our training assumed uniform noise levels across qubits.

\medskip

\noindent
The hardware experiments conducted on both IBM Quantum and IQM devices demonstrate the practical viability and adaptability of our \gls{varqec} approach under realistic conditions. Despite the inherent limitations of current quantum hardware, such as qubit decoherence and gate imperfections, our codes showed improved performance in preserving quantum information. Our results underline the effectiveness of tailoring error correction codes to specific hardware platforms and noise characteristics. The \gls{varqec} procedure's adaptability allows for optimizing encodings that are robust against the particular noise profiles and connectivity constraints of different quantum devices. Future work could further enhance performance by incorporating more sophisticated device-specific noise models directly into the training process.

\renewcommand*{\bibfont}{\footnotesize}
\setlength{\bibsep}{0pt}
\bibliographystyle{plain}
\bibliography{mybibliography}

\end{document}